\documentclass[longauth]{aaEC}

\usepackage{graphicx}
\usepackage{natbib}
\usepackage{scalerel}
\usepackage{comment}

\usepackage[table]{xcolor}

\usepackage{txfonts}
\usepackage[pdfencoding=auto,psdextra]{hyperref}
\hypersetup{
    colorlinks=true,
    linkcolor=blue,
    filecolor=magenta,      
    urlcolor=blue,
    citecolor=blue
}
\makeatletter
\renewcommand*\aa@pageof{, page \thepage{} of \pageref*{LastPage}}
\makeatother

\usepackage[utf8]{inputenc}

\usepackage[switch, modulo]{lineno}

\usepackage{euclid}
\defcitealias{Q1-SP013}{EC:LM25}
\defcitealias{bisigello}{B25}

\begin{document}
%
%

\title{\Euclid: Quick Data Release (Q1) -- The connection between galaxy close encounters and radio activity\thanks{This paper is published on behalf of the Euclid Consortium.}}    
\newcommand{\orcid}[1]{} 

\author{M.~Magliocchetti\orcid{0000-0001-9158-4838}\thanks{\email{manuela.magliocchetti@inaf.it}}\inst{\ref{aff1}}
\and A.~La~Marca\orcid{0000-0002-7217-5120}\inst{\ref{aff2},\ref{aff3}}
\and L.~Bisigello\orcid{0000-0003-0492-4924}\inst{\ref{aff4}}
\and M.~Bondi\orcid{0000-0002-9553-7999}\inst{\ref{aff5}}
\and F.~Ricci\orcid{0000-0001-5742-5980}\inst{\ref{aff6},\ref{aff7}}
\and S.~Fotopoulou\orcid{0000-0002-9686-254X}\inst{\ref{aff8}}
\and L.~Wang\orcid{0000-0002-6736-9158}\inst{\ref{aff2},\ref{aff3}}
\and R.~Scaramella\orcid{0000-0003-2229-193X}\inst{\ref{aff7},\ref{aff9}}
\and L.~Pentericci\inst{\ref{aff7}}
\and I.~Prandoni\orcid{0000-0001-9680-7092}\inst{\ref{aff5}}
\and J.~G.~Sorce\orcid{0000-0002-2307-2432}\inst{\ref{aff10},\ref{aff11}}
\and H.~J.~A.~Rottgering\orcid{0000-0001-8887-2257}\inst{\ref{aff12}}
\and M.~J.~Hardcastle\orcid{0000-0003-4223-1117}\inst{\ref{aff13}}
\and J.~Petley\orcid{0000-0002-4496-0754}\inst{\ref{aff12}}
\and F.~La~Franca\orcid{0000-0002-1239-2721}\inst{\ref{aff6},\ref{aff7}}
\and K.~Rubinur\orcid{0000-0001-5574-5104}\inst{\ref{aff14}}
\and Y.~Toba\orcid{0000-0002-3531-7863}\inst{\ref{aff15},\ref{aff16}}
\and Y.~Zhong\orcid{0009-0001-3910-2288}\inst{\ref{aff17},\ref{aff6}}
\and M.~Mezcua\orcid{0000-0003-4440-259X}\inst{\ref{aff18},\ref{aff19}}
\and G.~Zamorani\orcid{0000-0002-2318-301X}\inst{\ref{aff20}}
\and F.~Shankar\orcid{0000-0001-8973-5051}\inst{\ref{aff21}}
\and B.~Altieri\orcid{0000-0003-3936-0284}\inst{\ref{aff22}}
\and S.~Andreon\orcid{0000-0002-2041-8784}\inst{\ref{aff23}}
\and N.~Auricchio\orcid{0000-0003-4444-8651}\inst{\ref{aff20}}
\and C.~Baccigalupi\orcid{0000-0002-8211-1630}\inst{\ref{aff24},\ref{aff25},\ref{aff26},\ref{aff27}}
\and M.~Baldi\orcid{0000-0003-4145-1943}\inst{\ref{aff28},\ref{aff20},\ref{aff29}}
\and S.~Bardelli\orcid{0000-0002-8900-0298}\inst{\ref{aff20}}
\and A.~Biviano\orcid{0000-0002-0857-0732}\inst{\ref{aff25},\ref{aff24}}
\and E.~Branchini\orcid{0000-0002-0808-6908}\inst{\ref{aff30},\ref{aff31},\ref{aff23}}
\and M.~Brescia\orcid{0000-0001-9506-5680}\inst{\ref{aff32},\ref{aff33}}
\and J.~Brinchmann\orcid{0000-0003-4359-8797}\inst{\ref{aff34},\ref{aff35},\ref{aff36}}
\and S.~Camera\orcid{0000-0003-3399-3574}\inst{\ref{aff37},\ref{aff38},\ref{aff39}}
\and G.~Ca\~nas-Herrera\orcid{0000-0003-2796-2149}\inst{\ref{aff40},\ref{aff12}}
\and V.~Capobianco\orcid{0000-0002-3309-7692}\inst{\ref{aff39}}
\and C.~Carbone\orcid{0000-0003-0125-3563}\inst{\ref{aff41}}
\and J.~Carretero\orcid{0000-0002-3130-0204}\inst{\ref{aff42},\ref{aff43}}
\and M.~Castellano\orcid{0000-0001-9875-8263}\inst{\ref{aff7}}
\and G.~Castignani\orcid{0000-0001-6831-0687}\inst{\ref{aff20}}
\and S.~Cavuoti\orcid{0000-0002-3787-4196}\inst{\ref{aff33},\ref{aff44}}
\and K.~C.~Chambers\orcid{0000-0001-6965-7789}\inst{\ref{aff45}}
\and A.~Cimatti\inst{\ref{aff46}}
\and C.~Colodro-Conde\inst{\ref{aff47}}
\and G.~Congedo\orcid{0000-0003-2508-0046}\inst{\ref{aff48}}
\and C.~J.~Conselice\orcid{0000-0003-1949-7638}\inst{\ref{aff49}}
\and L.~Conversi\orcid{0000-0002-6710-8476}\inst{\ref{aff50},\ref{aff22}}
\and Y.~Copin\orcid{0000-0002-5317-7518}\inst{\ref{aff51}}
\and A.~Costille\inst{\ref{aff52}}
\and F.~Courbin\orcid{0000-0003-0758-6510}\inst{\ref{aff53},\ref{aff54},\ref{aff55}}
\and H.~M.~Courtois\orcid{0000-0003-0509-1776}\inst{\ref{aff56}}
\and M.~Cropper\orcid{0000-0003-4571-9468}\inst{\ref{aff57}}
\and A.~Da~Silva\orcid{0000-0002-6385-1609}\inst{\ref{aff58},\ref{aff59}}
\and H.~Degaudenzi\orcid{0000-0002-5887-6799}\inst{\ref{aff60}}
\and G.~De~Lucia\orcid{0000-0002-6220-9104}\inst{\ref{aff25}}
\and A.~M.~Di~Giorgio\orcid{0000-0002-4767-2360}\inst{\ref{aff1}}
\and H.~Dole\orcid{0000-0002-9767-3839}\inst{\ref{aff11}}
\and F.~Dubath\orcid{0000-0002-6533-2810}\inst{\ref{aff60}}
\and C.~A.~J.~Duncan\orcid{0009-0003-3573-0791}\inst{\ref{aff48}}
\and X.~Dupac\inst{\ref{aff22}}
\and S.~Dusini\orcid{0000-0002-1128-0664}\inst{\ref{aff61}}
\and S.~Escoffier\orcid{0000-0002-2847-7498}\inst{\ref{aff62}}
\and M.~Farina\orcid{0000-0002-3089-7846}\inst{\ref{aff1}}
\and R.~Farinelli\inst{\ref{aff20}}
\and F.~Faustini\orcid{0000-0001-6274-5145}\inst{\ref{aff7},\ref{aff63}}
\and S.~Ferriol\inst{\ref{aff51}}
\and F.~Finelli\orcid{0000-0002-6694-3269}\inst{\ref{aff20},\ref{aff64}}
\and M.~Frailis\orcid{0000-0002-7400-2135}\inst{\ref{aff25}}
\and E.~Franceschi\orcid{0000-0002-0585-6591}\inst{\ref{aff20}}
\and P.~Franzetti\inst{\ref{aff41}}
\and M.~Fumana\orcid{0000-0001-6787-5950}\inst{\ref{aff41}}
\and S.~Galeotta\orcid{0000-0002-3748-5115}\inst{\ref{aff25}}
\and K.~George\orcid{0000-0002-1734-8455}\inst{\ref{aff65}}
\and B.~Gillis\orcid{0000-0002-4478-1270}\inst{\ref{aff48}}
\and C.~Giocoli\orcid{0000-0002-9590-7961}\inst{\ref{aff20},\ref{aff29}}
\and J.~Gracia-Carpio\inst{\ref{aff66}}
\and A.~Grazian\orcid{0000-0002-5688-0663}\inst{\ref{aff4}}
\and F.~Grupp\inst{\ref{aff66},\ref{aff67}}
\and S.~V.~H.~Haugan\orcid{0000-0001-9648-7260}\inst{\ref{aff14}}
\and J.~Hoar\inst{\ref{aff22}}
\and W.~Holmes\inst{\ref{aff68}}
\and I.~M.~Hook\orcid{0000-0002-2960-978X}\inst{\ref{aff69}}
\and F.~Hormuth\inst{\ref{aff70}}
\and A.~Hornstrup\orcid{0000-0002-3363-0936}\inst{\ref{aff71},\ref{aff72}}
\and K.~Jahnke\orcid{0000-0003-3804-2137}\inst{\ref{aff73}}
\and M.~Jhabvala\inst{\ref{aff74}}
\and B.~Joachimi\orcid{0000-0001-7494-1303}\inst{\ref{aff75}}
\and E.~Keih\"anen\orcid{0000-0003-1804-7715}\inst{\ref{aff76}}
\and S.~Kermiche\orcid{0000-0002-0302-5735}\inst{\ref{aff62}}
\and A.~Kiessling\orcid{0000-0002-2590-1273}\inst{\ref{aff68}}
\and B.~Kubik\orcid{0009-0006-5823-4880}\inst{\ref{aff51}}
\and M.~K\"ummel\orcid{0000-0003-2791-2117}\inst{\ref{aff67}}
\and H.~Kurki-Suonio\orcid{0000-0002-4618-3063}\inst{\ref{aff77},\ref{aff78}}
\and A.~M.~C.~Le~Brun\orcid{0000-0002-0936-4594}\inst{\ref{aff79}}
\and S.~Ligori\orcid{0000-0003-4172-4606}\inst{\ref{aff39}}
\and P.~B.~Lilje\orcid{0000-0003-4324-7794}\inst{\ref{aff14}}
\and V.~Lindholm\orcid{0000-0003-2317-5471}\inst{\ref{aff77},\ref{aff78}}
\and I.~Lloro\orcid{0000-0001-5966-1434}\inst{\ref{aff80}}
\and G.~Mainetti\orcid{0000-0003-2384-2377}\inst{\ref{aff81}}
\and D.~Maino\inst{\ref{aff82},\ref{aff41},\ref{aff83}}
\and E.~Maiorano\orcid{0000-0003-2593-4355}\inst{\ref{aff20}}
\and O.~Mansutti\orcid{0000-0001-5758-4658}\inst{\ref{aff25}}
\and O.~Marggraf\orcid{0000-0001-7242-3852}\inst{\ref{aff84}}
\and M.~Martinelli\orcid{0000-0002-6943-7732}\inst{\ref{aff7},\ref{aff9}}
\and N.~Martinet\orcid{0000-0003-2786-7790}\inst{\ref{aff52}}
\and F.~Marulli\orcid{0000-0002-8850-0303}\inst{\ref{aff85},\ref{aff20},\ref{aff29}}
\and R.~J.~Massey\orcid{0000-0002-6085-3780}\inst{\ref{aff86}}
\and E.~Medinaceli\orcid{0000-0002-4040-7783}\inst{\ref{aff20}}
\and S.~Mei\orcid{0000-0002-2849-559X}\inst{\ref{aff87},\ref{aff88}}
\and Y.~Mellier\inst{\ref{aff89},\ref{aff90}}
\and M.~Meneghetti\orcid{0000-0003-1225-7084}\inst{\ref{aff20},\ref{aff29}}
\and E.~Merlin\orcid{0000-0001-6870-8900}\inst{\ref{aff7}}
\and G.~Meylan\inst{\ref{aff91}}
\and A.~Mora\orcid{0000-0002-1922-8529}\inst{\ref{aff92}}
\and M.~Moresco\orcid{0000-0002-7616-7136}\inst{\ref{aff85},\ref{aff20}}
\and L.~Moscardini\orcid{0000-0002-3473-6716}\inst{\ref{aff85},\ref{aff20},\ref{aff29}}
\and R.~Nakajima\orcid{0009-0009-1213-7040}\inst{\ref{aff84}}
\and C.~Neissner\orcid{0000-0001-8524-4968}\inst{\ref{aff93},\ref{aff43}}
\and R.~C.~Nichol\orcid{0000-0003-0939-6518}\inst{\ref{aff94}}
\and S.-M.~Niemi\orcid{0009-0005-0247-0086}\inst{\ref{aff40}}
\and C.~Padilla\orcid{0000-0001-7951-0166}\inst{\ref{aff93}}
\and S.~Paltani\orcid{0000-0002-8108-9179}\inst{\ref{aff60}}
\and F.~Pasian\orcid{0000-0002-4869-3227}\inst{\ref{aff25}}
\and K.~Pedersen\inst{\ref{aff95}}
\and W.~J.~Percival\orcid{0000-0002-0644-5727}\inst{\ref{aff96},\ref{aff97},\ref{aff98}}
\and V.~Pettorino\orcid{0000-0002-4203-9320}\inst{\ref{aff40}}
\and S.~Pires\orcid{0000-0002-0249-2104}\inst{\ref{aff99}}
\and G.~Polenta\orcid{0000-0003-4067-9196}\inst{\ref{aff63}}
\and M.~Poncet\inst{\ref{aff100}}
\and L.~A.~Popa\inst{\ref{aff101}}
\and L.~Pozzetti\orcid{0000-0001-7085-0412}\inst{\ref{aff20}}
\and F.~Raison\orcid{0000-0002-7819-6918}\inst{\ref{aff66}}
\and A.~Renzi\orcid{0000-0001-9856-1970}\inst{\ref{aff102},\ref{aff61}}
\and J.~Rhodes\orcid{0000-0002-4485-8549}\inst{\ref{aff68}}
\and G.~Riccio\inst{\ref{aff33}}
\and E.~Romelli\orcid{0000-0003-3069-9222}\inst{\ref{aff25}}
\and M.~Roncarelli\orcid{0000-0001-9587-7822}\inst{\ref{aff20}}
\and R.~Saglia\orcid{0000-0003-0378-7032}\inst{\ref{aff67},\ref{aff66}}
\and Z.~Sakr\orcid{0000-0002-4823-3757}\inst{\ref{aff103},\ref{aff104},\ref{aff105}}
\and D.~Sapone\orcid{0000-0001-7089-4503}\inst{\ref{aff106}}
\and B.~Sartoris\orcid{0000-0003-1337-5269}\inst{\ref{aff67},\ref{aff25}}
\and M.~Schirmer\orcid{0000-0003-2568-9994}\inst{\ref{aff73}}
\and P.~Schneider\orcid{0000-0001-8561-2679}\inst{\ref{aff84}}
\and T.~Schrabback\orcid{0000-0002-6987-7834}\inst{\ref{aff107}}
\and A.~Secroun\orcid{0000-0003-0505-3710}\inst{\ref{aff62}}
\and G.~Seidel\orcid{0000-0003-2907-353X}\inst{\ref{aff73}}
\and S.~Serrano\orcid{0000-0002-0211-2861}\inst{\ref{aff19},\ref{aff108},\ref{aff18}}
\and P.~Simon\inst{\ref{aff84}}
\and C.~Sirignano\orcid{0000-0002-0995-7146}\inst{\ref{aff102},\ref{aff61}}
\and G.~Sirri\orcid{0000-0003-2626-2853}\inst{\ref{aff29}}
\and L.~Stanco\orcid{0000-0002-9706-5104}\inst{\ref{aff61}}
\and J.~Steinwagner\orcid{0000-0001-7443-1047}\inst{\ref{aff66}}
\and P.~Tallada-Cresp\'{i}\orcid{0000-0002-1336-8328}\inst{\ref{aff42},\ref{aff43}}
\and A.~N.~Taylor\inst{\ref{aff48}}
\and I.~Tereno\orcid{0000-0002-4537-6218}\inst{\ref{aff58},\ref{aff109}}
\and N.~Tessore\orcid{0000-0002-9696-7931}\inst{\ref{aff75},\ref{aff57}}
\and S.~Toft\orcid{0000-0003-3631-7176}\inst{\ref{aff110},\ref{aff111}}
\and R.~Toledo-Moreo\orcid{0000-0002-2997-4859}\inst{\ref{aff112}}
\and F.~Torradeflot\orcid{0000-0003-1160-1517}\inst{\ref{aff43},\ref{aff42}}
\and I.~Tutusaus\orcid{0000-0002-3199-0399}\inst{\ref{aff18},\ref{aff19},\ref{aff104}}
\and L.~Valenziano\orcid{0000-0002-1170-0104}\inst{\ref{aff20},\ref{aff64}}
\and J.~Valiviita\orcid{0000-0001-6225-3693}\inst{\ref{aff77},\ref{aff78}}
\and T.~Vassallo\orcid{0000-0001-6512-6358}\inst{\ref{aff25},\ref{aff65}}
\and G.~Verdoes~Kleijn\orcid{0000-0001-5803-2580}\inst{\ref{aff3}}
\and A.~Veropalumbo\orcid{0000-0003-2387-1194}\inst{\ref{aff23},\ref{aff31},\ref{aff30}}
\and Y.~Wang\orcid{0000-0002-4749-2984}\inst{\ref{aff113}}
\and J.~Weller\orcid{0000-0002-8282-2010}\inst{\ref{aff67},\ref{aff66}}
\and E.~Zucca\orcid{0000-0002-5845-8132}\inst{\ref{aff20}}
\and M.~Huertas-Company\orcid{0000-0002-1416-8483}\inst{\ref{aff47},\ref{aff114},\ref{aff115},\ref{aff116}}
\and V.~Scottez\orcid{0009-0008-3864-940X}\inst{\ref{aff89},\ref{aff117}}}
										   
\institute{INAF-Istituto di Astrofisica e Planetologia Spaziali, via del Fosso del Cavaliere, 100, 00100 Roma, Italy\label{aff1}
\and
SRON Netherlands Institute for Space Research, Landleven 12, 9747 AD, Groningen, The Netherlands\label{aff2}
\and
Kapteyn Astronomical Institute, University of Groningen, PO Box 800, 9700 AV Groningen, The Netherlands\label{aff3}
\and
INAF-Osservatorio Astronomico di Padova, Via dell'Osservatorio 5, 35122 Padova, Italy\label{aff4}
\and
INAF, Istituto di Radioastronomia, Via Piero Gobetti 101, 40129 Bologna, Italy\label{aff5}
\and
Department of Mathematics and Physics, Roma Tre University, Via della Vasca Navale 84, 00146 Rome, Italy\label{aff6}
\and
INAF-Osservatorio Astronomico di Roma, Via Frascati 33, 00078 Monteporzio Catone, Italy\label{aff7}
\and
School of Physics, HH Wills Physics Laboratory, University of Bristol, Tyndall Avenue, Bristol, BS8 1TL, UK\label{aff8}
\and
INFN-Sezione di Roma, Piazzale Aldo Moro, 2 - c/o Dipartimento di Fisica, Edificio G. Marconi, 00185 Roma, Italy\label{aff9}
\and
Univ. Lille, CNRS, Centrale Lille, UMR 9189 CRIStAL, 59000 Lille, France\label{aff10}
\and
Universit\'e Paris-Saclay, CNRS, Institut d'astrophysique spatiale, 91405, Orsay, France\label{aff11}
\and
Leiden Observatory, Leiden University, Einsteinweg 55, 2333 CC Leiden, The Netherlands\label{aff12}
\and
Department of Physics, Astronomy and Mathematics, University of Hertfordshire, College Lane, Hatfield AL10 9AB, UK\label{aff13}
\and
Institute of Theoretical Astrophysics, University of Oslo, P.O. Box 1029 Blindern, 0315 Oslo, Norway\label{aff14}
\and
Department of Physical Sciences, Ritsumeikan University, Kusatsu, Shiga 525-8577, Japan\label{aff15}
\and
Academia Sinica Institute of Astronomy and Astrophysics (ASIAA), 11F of ASMAB, No.~1, Section 4, Roosevelt Road, Taipei 10617, Taiwan\label{aff16}
\and
Department of Physics, School of Advanced Science and Engineering, Faculty of Science and Engineering, Waseda University, 3-4-1 Okubo, Shinjuku, 169-8555 Tokyo, Japan\label{aff17}
\and
Institute of Space Sciences (ICE, CSIC), Campus UAB, Carrer de Can Magrans, s/n, 08193 Barcelona, Spain\label{aff18}
\and
Institut d'Estudis Espacials de Catalunya (IEEC),  Edifici RDIT, Campus UPC, 08860 Castelldefels, Barcelona, Spain\label{aff19}
\and
INAF-Osservatorio di Astrofisica e Scienza dello Spazio di Bologna, Via Piero Gobetti 93/3, 40129 Bologna, Italy\label{aff20}
\and
School of Physics \& Astronomy, University of Southampton, Highfield Campus, Southampton SO17 1BJ, UK\label{aff21}
\and
ESAC/ESA, Camino Bajo del Castillo, s/n., Urb. Villafranca del Castillo, 28692 Villanueva de la Ca\~nada, Madrid, Spain\label{aff22}
\and
INAF-Osservatorio Astronomico di Brera, Via Brera 28, 20122 Milano, Italy\label{aff23}
\and
IFPU, Institute for Fundamental Physics of the Universe, via Beirut 2, 34151 Trieste, Italy\label{aff24}
\and
INAF-Osservatorio Astronomico di Trieste, Via G. B. Tiepolo 11, 34143 Trieste, Italy\label{aff25}
\and
INFN, Sezione di Trieste, Via Valerio 2, 34127 Trieste TS, Italy\label{aff26}
\and
SISSA, International School for Advanced Studies, Via Bonomea 265, 34136 Trieste TS, Italy\label{aff27}
\and
Dipartimento di Fisica e Astronomia, Universit\`a di Bologna, Via Gobetti 93/2, 40129 Bologna, Italy\label{aff28}
\and
INFN-Sezione di Bologna, Viale Berti Pichat 6/2, 40127 Bologna, Italy\label{aff29}
\and
Dipartimento di Fisica, Universit\`a di Genova, Via Dodecaneso 33, 16146, Genova, Italy\label{aff30}
\and
INFN-Sezione di Genova, Via Dodecaneso 33, 16146, Genova, Italy\label{aff31}
\and
Department of Physics "E. Pancini", University Federico II, Via Cinthia 6, 80126, Napoli, Italy\label{aff32}
\and
INAF-Osservatorio Astronomico di Capodimonte, Via Moiariello 16, 80131 Napoli, Italy\label{aff33}
\and
Instituto de Astrof\'isica e Ci\^encias do Espa\c{c}o, Universidade do Porto, CAUP, Rua das Estrelas, PT4150-762 Porto, Portugal\label{aff34}
\and
Faculdade de Ci\^encias da Universidade do Porto, Rua do Campo de Alegre, 4150-007 Porto, Portugal\label{aff35}
\and
European Southern Observatory, Karl-Schwarzschild-Str.~2, 85748 Garching, Germany\label{aff36}
\and
Dipartimento di Fisica, Universit\`a degli Studi di Torino, Via P. Giuria 1, 10125 Torino, Italy\label{aff37}
\and
INFN-Sezione di Torino, Via P. Giuria 1, 10125 Torino, Italy\label{aff38}
\and
INAF-Osservatorio Astrofisico di Torino, Via Osservatorio 20, 10025 Pino Torinese (TO), Italy\label{aff39}
\and
European Space Agency/ESTEC, Keplerlaan 1, 2201 AZ Noordwijk, The Netherlands\label{aff40}
\and
INAF-IASF Milano, Via Alfonso Corti 12, 20133 Milano, Italy\label{aff41}
\and
Centro de Investigaciones Energ\'eticas, Medioambientales y Tecnol\'ogicas (CIEMAT), Avenida Complutense 40, 28040 Madrid, Spain\label{aff42}
\and
Port d'Informaci\'{o} Cient\'{i}fica, Campus UAB, C. Albareda s/n, 08193 Bellaterra (Barcelona), Spain\label{aff43}
\and
INFN section of Naples, Via Cinthia 6, 80126, Napoli, Italy\label{aff44}
\and
Institute for Astronomy, University of Hawaii, 2680 Woodlawn Drive, Honolulu, HI 96822, USA\label{aff45}
\and
Dipartimento di Fisica e Astronomia "Augusto Righi" - Alma Mater Studiorum Universit\`a di Bologna, Viale Berti Pichat 6/2, 40127 Bologna, Italy\label{aff46}
\and
Instituto de Astrof\'{\i}sica de Canarias, V\'{\i}a L\'actea, 38205 La Laguna, Tenerife, Spain\label{aff47}
\and
Institute for Astronomy, University of Edinburgh, Royal Observatory, Blackford Hill, Edinburgh EH9 3HJ, UK\label{aff48}
\and
Jodrell Bank Centre for Astrophysics, Department of Physics and Astronomy, University of Manchester, Oxford Road, Manchester M13 9PL, UK\label{aff49}
\and
European Space Agency/ESRIN, Largo Galileo Galilei 1, 00044 Frascati, Roma, Italy\label{aff50}
\and
Universit\'e Claude Bernard Lyon 1, CNRS/IN2P3, IP2I Lyon, UMR 5822, Villeurbanne, F-69100, France\label{aff51}
\and
Aix-Marseille Universit\'e, CNRS, CNES, LAM, Marseille, France\label{aff52}
\and
Institut de Ci\`{e}ncies del Cosmos (ICCUB), Universitat de Barcelona (IEEC-UB), Mart\'{i} i Franqu\`{e}s 1, 08028 Barcelona, Spain\label{aff53}
\and
Instituci\'o Catalana de Recerca i Estudis Avan\c{c}ats (ICREA), Passeig de Llu\'{\i}s Companys 23, 08010 Barcelona, Spain\label{aff54}
\and
Institut de Ciencies de l'Espai (IEEC-CSIC), Campus UAB, Carrer de Can Magrans, s/n Cerdanyola del Vall\'es, 08193 Barcelona, Spain\label{aff55}
\and
UCB Lyon 1, CNRS/IN2P3, IUF, IP2I Lyon, 4 rue Enrico Fermi, 69622 Villeurbanne, France\label{aff56}
\and
Mullard Space Science Laboratory, University College London, Holmbury St Mary, Dorking, Surrey RH5 6NT, UK\label{aff57}
\and
Departamento de F\'isica, Faculdade de Ci\^encias, Universidade de Lisboa, Edif\'icio C8, Campo Grande, PT1749-016 Lisboa, Portugal\label{aff58}
\and
Instituto de Astrof\'isica e Ci\^encias do Espa\c{c}o, Faculdade de Ci\^encias, Universidade de Lisboa, Campo Grande, 1749-016 Lisboa, Portugal\label{aff59}
\and
Department of Astronomy, University of Geneva, ch. d'Ecogia 16, 1290 Versoix, Switzerland\label{aff60}
\and
INFN-Padova, Via Marzolo 8, 35131 Padova, Italy\label{aff61}
\and
Aix-Marseille Universit\'e, CNRS/IN2P3, CPPM, Marseille, France\label{aff62}
\and
Space Science Data Center, Italian Space Agency, via del Politecnico snc, 00133 Roma, Italy\label{aff63}
\and
INFN-Bologna, Via Irnerio 46, 40126 Bologna, Italy\label{aff64}
\and
University Observatory, LMU Faculty of Physics, Scheinerstrasse 1, 81679 Munich, Germany\label{aff65}
\and
Max Planck Institute for Extraterrestrial Physics, Giessenbachstr. 1, 85748 Garching, Germany\label{aff66}
\and
Universit\"ats-Sternwarte M\"unchen, Fakult\"at f\"ur Physik, Ludwig-Maximilians-Universit\"at M\"unchen, Scheinerstrasse 1, 81679 M\"unchen, Germany\label{aff67}
\and
Jet Propulsion Laboratory, California Institute of Technology, 4800 Oak Grove Drive, Pasadena, CA, 91109, USA\label{aff68}
\and
Department of Physics, Lancaster University, Lancaster, LA1 4YB, UK\label{aff69}
\and
Felix Hormuth Engineering, Goethestr. 17, 69181 Leimen, Germany\label{aff70}
\and
Technical University of Denmark, Elektrovej 327, 2800 Kgs. Lyngby, Denmark\label{aff71}
\and
Cosmic Dawn Center (DAWN), Denmark\label{aff72}
\and
Max-Planck-Institut f\"ur Astronomie, K\"onigstuhl 17, 69117 Heidelberg, Germany\label{aff73}
\and
NASA Goddard Space Flight Center, Greenbelt, MD 20771, USA\label{aff74}
\and
Department of Physics and Astronomy, University College London, Gower Street, London WC1E 6BT, UK\label{aff75}
\and
Department of Physics and Helsinki Institute of Physics, Gustaf H\"allstr\"omin katu 2, University of Helsinki, 00014 Helsinki, Finland\label{aff76}
\and
Department of Physics, P.O. Box 64, University of Helsinki, 00014 Helsinki, Finland\label{aff77}
\and
Helsinki Institute of Physics, Gustaf H{\"a}llstr{\"o}min katu 2, University of Helsinki, 00014 Helsinki, Finland\label{aff78}
\and
Laboratoire d'etude de l'Univers et des phenomenes eXtremes, Observatoire de Paris, Universit\'e PSL, Sorbonne Universit\'e, CNRS, 92190 Meudon, France\label{aff79}
\and
SKAO, Jodrell Bank, Lower Withington, Macclesfield SK11 9FT, United Kingdom\label{aff80}
\and
Centre de Calcul de l'IN2P3/CNRS, 21 avenue Pierre de Coubertin 69627 Villeurbanne Cedex, France\label{aff81}
\and
Dipartimento di Fisica "Aldo Pontremoli", Universit\`a degli Studi di Milano, Via Celoria 16, 20133 Milano, Italy\label{aff82}
\and
INFN-Sezione di Milano, Via Celoria 16, 20133 Milano, Italy\label{aff83}
\and
Universit\"at Bonn, Argelander-Institut f\"ur Astronomie, Auf dem H\"ugel 71, 53121 Bonn, Germany\label{aff84}
\and
Dipartimento di Fisica e Astronomia "Augusto Righi" - Alma Mater Studiorum Universit\`a di Bologna, via Piero Gobetti 93/2, 40129 Bologna, Italy\label{aff85}
\and
Department of Physics, Institute for Computational Cosmology, Durham University, South Road, Durham, DH1 3LE, UK\label{aff86}
\and
Universit\'e Paris Cit\'e, CNRS, Astroparticule et Cosmologie, 75013 Paris, France\label{aff87}
\and
CNRS-UCB International Research Laboratory, Centre Pierre Bin\'etruy, IRL2007, CPB-IN2P3, Berkeley, USA\label{aff88}
\and
Institut d'Astrophysique de Paris, 98bis Boulevard Arago, 75014, Paris, France\label{aff89}
\and
Institut d'Astrophysique de Paris, UMR 7095, CNRS, and Sorbonne Universit\'e, 98 bis boulevard Arago, 75014 Paris, France\label{aff90}
\and
Institute of Physics, Laboratory of Astrophysics, Ecole Polytechnique F\'ed\'erale de Lausanne (EPFL), Observatoire de Sauverny, 1290 Versoix, Switzerland\label{aff91}
\and
Telespazio UK S.L. for European Space Agency (ESA), Camino bajo del Castillo, s/n, Urbanizacion Villafranca del Castillo, Villanueva de la Ca\~nada, 28692 Madrid, Spain\label{aff92}
\and
Institut de F\'{i}sica d'Altes Energies (IFAE), The Barcelona Institute of Science and Technology, Campus UAB, 08193 Bellaterra (Barcelona), Spain\label{aff93}
\and
School of Mathematics and Physics, University of Surrey, Guildford, Surrey, GU2 7XH, UK\label{aff94}
\and
DARK, Niels Bohr Institute, University of Copenhagen, Jagtvej 155, 2200 Copenhagen, Denmark\label{aff95}
\and
Waterloo Centre for Astrophysics, University of Waterloo, Waterloo, Ontario N2L 3G1, Canada\label{aff96}
\and
Department of Physics and Astronomy, University of Waterloo, Waterloo, Ontario N2L 3G1, Canada\label{aff97}
\and
Perimeter Institute for Theoretical Physics, Waterloo, Ontario N2L 2Y5, Canada\label{aff98}
\and
Universit\'e Paris-Saclay, Universit\'e Paris Cit\'e, CEA, CNRS, AIM, 91191, Gif-sur-Yvette, France\label{aff99}
\and
Centre National d'Etudes Spatiales -- Centre spatial de Toulouse, 18 avenue Edouard Belin, 31401 Toulouse Cedex 9, France\label{aff100}
\and
Institute of Space Science, Str. Atomistilor, nr. 409 M\u{a}gurele, Ilfov, 077125, Romania\label{aff101}
\and
Dipartimento di Fisica e Astronomia "G. Galilei", Universit\`a di Padova, Via Marzolo 8, 35131 Padova, Italy\label{aff102}
\and
Institut f\"ur Theoretische Physik, University of Heidelberg, Philosophenweg 16, 69120 Heidelberg, Germany\label{aff103}
\and
Institut de Recherche en Astrophysique et Plan\'etologie (IRAP), Universit\'e de Toulouse, CNRS, UPS, CNES, 14 Av. Edouard Belin, 31400 Toulouse, France\label{aff104}
\and
Universit\'e St Joseph; Faculty of Sciences, Beirut, Lebanon\label{aff105}
\and
Departamento de F\'isica, FCFM, Universidad de Chile, Blanco Encalada 2008, Santiago, Chile\label{aff106}
\and
Universit\"at Innsbruck, Institut f\"ur Astro- und Teilchenphysik, Technikerstr. 25/8, 6020 Innsbruck, Austria\label{aff107}
\and
Satlantis, University Science Park, Sede Bld 48940, Leioa-Bilbao, Spain\label{aff108}
\and
Instituto de Astrof\'isica e Ci\^encias do Espa\c{c}o, Faculdade de Ci\^encias, Universidade de Lisboa, Tapada da Ajuda, 1349-018 Lisboa, Portugal\label{aff109}
\and
Cosmic Dawn Center (DAWN)\label{aff110}
\and
Niels Bohr Institute, University of Copenhagen, Jagtvej 128, 2200 Copenhagen, Denmark\label{aff111}
\and
Universidad Polit\'ecnica de Cartagena, Departamento de Electr\'onica y Tecnolog\'ia de Computadoras,  Plaza del Hospital 1, 30202 Cartagena, Spain\label{aff112}
\and
Infrared Processing and Analysis Center, California Institute of Technology, Pasadena, CA 91125, USA\label{aff113}
\and
Instituto de Astrof\'isica de Canarias (IAC); Departamento de Astrof\'isica, Universidad de La Laguna (ULL), 38200, La Laguna, Tenerife, Spain\label{aff114}
\and
Universit\'e PSL, Observatoire de Paris, Sorbonne Universit\'e, CNRS, LERMA, 75014, Paris, France\label{aff115}
\and
Universit\'e Paris-Cit\'e, 5 Rue Thomas Mann, 75013, Paris, France\label{aff116}
\and
ICL, Junia, Universit\'e Catholique de Lille, LITL, 59000 Lille, France\label{aff117}}    



%
%
 \abstract{Using the large statistics provided by both \Euclid and the LOFAR surveys, we present the first large-scale study
of the connection between radio emission, its morphology, and the merging properties of the hosts of radio sources up to $z\sim 2$. By dividing the radio sample into active galactic nuclei (AGN) and star-forming galaxies, we find that radio-emitting AGN show a clear preference to reside within galaxies undergoing a merging event. This is more significant for AGN that present extended and/or complex radio emission: indeed, about half of them are associated with merging systems, while only $\SI {\sim 15}\percent$ are hosted by an isolated galaxy. The observed trend is primarily driven by AGN residing at $z<1$, especially in the case of high  -- $P_{144\,\rm MHz}>10^{24}$\,W\, Hz$^{-1}$\,sr$^{-1}$ -- radio luminosities ($\SI {\sim 60}\percent$ in mergers versus $\SI {\sim 10}\percent$ isolated regardless of radio appearance). On the other hand, this preference seems to disappear at higher redshifts, where only bright AGN with extended radio emission still prefer galaxies undergoing a merging event. The situation is reversed in the case of radio-emitting star-forming galaxies, which are preferentially associated with isolated systems. This is more significant as we move towards low radio-luminosity/star-formation objects ($P_{144\, \rm MHz}<10^{23}$\,W\,Hz$^{-1}$\,sr$^{-1}$) for which we find $\SI {\sim 40}\percent$ in isolated systems versus $\SI {\sim 20}\percent$ in mergers. These values hold regardless of redshift. We interpret the above result for AGN with their need to accrete outer gas from local encounters in order to trigger (radio) activity, especially in the case of extended radio emission such as hot-spots and lobes. This is mostly observed at $z<1$, since in the local Universe galaxies are more gas deprived than their higher-redshift counterparts. Internal gas reservoirs instead seem sufficient to trigger star formation within the majority of galaxies, which indeed prefer to be associated with isolated systems at all redshifts probed.}

%
%
%
%
\titlerunning{\Euclid\/: close encounters and radio activity}
   \authorrunning{M.Magliocchetti et al.}
   
   \maketitle
%
%
%
%
   
\section{\label{sc:Intro}Introduction}

In recent years there has been a rising interest in understanding the possible connection between AGN activity and close galaxy-galaxy encounters such as merging events. This possibility has been mainly investigated for AGN selected at optical, X-ray and mid-IR wavelengths (e.g., \citealt{Ellison19, Pierce23, tanaka, bickley, lamarca24,Q1-SP013} just to mention few of the most recent works). In general there seems to be a consensus for a positive link between the two phenomena (i.e. AGN activity being more frequently associated with merging systems), even though some studies claim this to be confined to only mid-IR selected AGN, and to be entirely lost in specific galaxy populations such as dwarfs (e.g., \citealt{bichang,villforth19,erostegui}, see also \citealt{villforth23} for a recent review). Other unsolved issues concern possible dependencies of the observed frequency of galaxy mergers in AGN hosts on AGN luminosity (e.g. \citealt{comerford}), dust obscuration (e.g., \citealt{ricci}), and/or environment (e.g., \citealt{koulouridis}).

Even more uncertain is the situation for radio-selected AGN, which constitute a relatively small fraction of the general AGN population observed at other wavelengths (e.g., \citealt{hickox}, but also see \citealt{calistro2024} for more recent results based on deeper radio observations), even though their incidence is observed to largely increase with the stellar mass of the host galaxy (e.g., \citealt{sabater19, maglio18}). Indeed, relatively little information can be found in the literature on possible effects of galaxy mergers on radio activity of AGN origin. Moreover, in general the results are hampered by small sample statistics. 

\cite{heckman} analysed 43 powerful radio AGN at $z<0.3$, finding disturbed optical morphologies explainable with galaxy encounters for about $\SI {44}\percent$ of them, less so (about $\SI {30}\percent$) if only limiting to a subsample of 23 sources complete in radio luminosity.
These disturbances were more frequently observed for objects belonging to the class of Fanaroff--Riley (FR; \citealt{FR}) II with strong emission lines in their optical spectra (High Excitation Radio Galaxies; HERGs) indicating efficient accretion onto the central black hole.
\cite{ramos11} considered 46 radio-bright, $0.05<z<0.7$ sources from the 2Jy sample (\citealt{tadhunter}), mainly FR\,II-HERGs, and for this population reported a frequency of mergers of about $\SI {94}\percent$. For the 11 sources belonging to the FR\,I class and showing weak or no emission lines in their optical spectra (Low Excitation Radio Galaxies; LERGs) indicative of inefficient accretion, they instead found a value of $\SI {\sim 27}\percent$. 
\cite{chiaberge} instead presented their analysis for 19 type 2 radio-loud AGN at $1<z<2.5$ spanning 5 decades in radio power. The 11 objects taken from the 3CR catalogue (\citealt{spinrad}) were all FR\,IIs, and for these they found that they were all associated with merging events. The same was true for $\SI {88}\percent$ of the eight fainter radio-loud AGN considered in the analysis. These results were then compared with lower redshift samples for which the reported merging fraction was $\SI {\sim 70}\percent$, making the authors conclude that not only are virtually all radio-loud AGN associated with merging systems, but also that such a phenomenon is independent of both redshift and radio luminosity. 

Results similar to those of \cite{chiaberge} were presented by \cite{breiding} for 28 radio-loud quasars at $1<z<2$ from the 3CR catalogue, while, on the other hand, \cite{pierce19} found for a local sample of 30 intermediate radio-luminosity HERGs that the fraction of hosts with signatures of merging was $\SI {\sim 53}\percent$, not only dependent on radio luminosity ($\SI {\sim 67}\percent$ versus $\SI {\sim 40}\percent$ for the brighter vs. fainter half), but also much lower than what was reported for brighter 2Jy sources by \cite{ramos11}. This luminosity dependence has been more recently confirmed by \cite{pierce22} for a larger sample of 155 radio AGN at $z<0.3$. Indeed, according to the above work, the fraction of HERGs associated with galaxies that exhibit optical signatures of disturbance goes from $\SI {37}\percent$ to $\SI {66}\percent$ in their radio-brightest sample. 

Different findings have instead been obtained recently in the case of LERGs, which -- in agreement with the work of \cite{ramos11} -- do not seem to show any large enhancement in the frequency of galaxy close interactions when compared to the general galaxy population ($\SI {\sim 27}\percent$, e.g., \citealt{gordon} and \citealt{gao}, see also \citealt{ellison}), at least in the local, $z\simeq 0$, Universe.
A recent study by \cite{wang} reports an excess of close ($<18$~kpc) companions around $z\simeq 3.5$ luminous radio-loud AGN. However, the sample is limited to just four sources. 

From the above discussion it follows that, although there seems to be some convergence on a scenario that envisages a connection between merging events and the triggering of (radio) AGN activity (mainly in the case of efficient gas accretion onto the central black hole that is thought to produce the HERG population, e.g., \citealt{hardcastle}), 
no unbiased and statistically sound conclusion has been reached so far, especially regarding the possible dependence of this connection on radio luminosity, radio morphology, or cosmic evolution (i.e., high redshift vs. local Universe).

The present paper aims to fill this gap, thanks to the exquisite statistical power provided by the LOw-Frequency ARray (LOFAR; \citealt{van}) and \Euclid surveys (\citealt{EuclidSkyOverview}), which in both cases combine sensitivity with large probed areas. In particular, our analysis concentrates on 10 deg$^2$ centred on the so-called Euclid Deep Field North (EDF-N) which are simultaneously covered by deep LOFAR (\citealt{bondi}) and  \cite{Q1cite} (Q1; \citealt{Q1-TP001}) observations. Very high resolution images (mean full width at half maximum of $0\arcsecf158$ with a standard deviation of $0\arcsecf001$) captured by the VIS instrument (\citealt{EuclidSkyVIS}) onboard of \Euclid have been obtained for all galaxies down to a completeness limit of $\IE=25.5$ (\citealt{Q1-TP002}). This has allowed the investigation of the morphological properties of \Euclid galaxies, at least for the bright tail of the distribution (e.g. \citealt{Q1-SP040, Q1-SP047, Q1-SP043,Q1-SP049,Q1-SP027, Q1-SP009}), and also the classification of objects on the basis of their optical appearance as isolated or undergoing a merging event (\citealt{Q1-SP013}, hereafter \citetalias{Q1-SP013}). 

For the present work, we incorporate the information provided by \citetalias{Q1-SP013} into the description of the LOFAR sources presented in \cite{bondi}, which are then divided into AGN and star-forming galaxies (SFG) following the method proposed by \cite{maglio13}. For both populations we will combine radio emission and optical appearance in order to investigate possible links between these two observables. Particular attention will be devoted to those sources that present signatures for complex or extended radio emission such as lobes, jets, or more diffuse patterns. 

The paper is structured as follows. In Sect.~\ref{radio} we introduce the radio catalogue and the method adopted to distinguish between AGN and SFG, while in Sect.~\ref{combined} the \Euclid sample obtained from VIS observations and the combined catalogue used for the subsequent analysis are outlined. Section~\ref{results} presents our results for radio-selected AGN and SFG. Here we will also discuss and take care of possible systematics in our analysis. Lastly, Sect.~\ref{conclusions} investigates our findings in a broader context and draws the conclusions. Some radio and optical cutouts for LOFAR sources with an extended radio morphology are shown in Appendix \ref{images}. Throughout the paper we assume a $\Lambda$CDM cosmology with $H_0 = 70$\,km\,s$^{-1}$\,Mpc$^{-1}$,  $\Omega_{\rm m}= 0.3$, and $\Omega_\Lambda=0.7$.

\begin{table*}
\begin{center}
\caption{Properties of the radio AGN sample.}
\small{\begin{tabular}{lllllllllll}
\hline\hline
 & & & &  & & & & & &\\[-9pt]
 &AGN & Matched& Mer & Iso  & S & C& S+& S+& C+& C+\\
&&&(optical)&(optical)&(radio)&(radio)&Mer&Iso&Mer&Iso\\
\hline
 & & & &  & & & & & &\\[-9pt]
 $0.5<z<2$& 1793 &819  &308   & 201 &  667   & 152 &239& 180& 69 &21\\
$0.5<z<1$& 870 & 521 &    214&      125&     413&    108&  162&     112&           52&        13\\      
$1<z<2$& 923 & 298  &   94   &   76    & 254 & 44 & 77	& 68	     & 17&	8\\
$P<10^{24}$ \& $0.5<z<2$&1086&  554   &   194    & 154   &522  &  32  &  183&	   149&	  11    &      5 \\
$P>10^{24}$ \& $0.5<z<2$&707  &    265 &    114 &   47  &  145  &  120 &  56   &  31   & 58  &       16\\
$10^{23.35}<P<10^{24.2}$ and $0.5<z<1$&453 &   260  &	96   & 67 &  213 &  47  & 78  &    60&	   18  	&      7  \\
$10^{23.35}<P<10^{24.2}$ and $1<z<2$&511  &  171   &   52  &   47  &  170   &  1	&   51  &    47&	   1	&      0\\
$P>10^{24.2}$ and $0.5<z<1$&126 &  83	& 46  &   9  &   24  &  59	&   13	&    3	 & 33    &  6\\
$P>10^{24.2}$ and $1<z<2$&412&   127	& 42	&29  &  84  &  43  &  26   &    21&	  16	 &     8\\
\hline
\label{table1}
\end{tabular}}
\tablefoot{The second column reports the number of all LOFAR AGN obtained from the work of \citetalias{bisigello}, the third the number of those that are also present in the catalogue of \citetalias{Q1-SP013}, the fourth the number of radio AGN with optical counterparts classified as mergers (Mer), while the fifth is for radio AGN with optical counterparts classified as isolated galaxies (Iso). The sixth column provides the number of LOFAR AGN with a compact radio morphology (S), the seventh the number of those that instead show complex radio morphologies (C). Finally, columns 8--11 present the number of radio AGN with combined cases of radio and optical morphologies. Different rows are for various redshift and 144\,MHz radio-luminosity (P) cuts expressed in W\,Hz$^{-1}$\,sr$^{-1}$, as indicated in the first column.}
\end{center}
\end{table*}

\begin{table*}
\begin{center}
\caption{Properties of the radio SFG sample.}
\small
\begin{tabular}{lllllllllll}
\hline\hline
 & & & &  & & & & & &\\[-9pt]
 &SFG& Matched& Mer & Iso  & S & C& S+& S+& C+& C+\\
&& &(optical)&(optical)&(radio)&(radio)&Mer&Iso&Mer&Iso\\
\hline
 & & & &  & & & & & &\\[-9pt]
$0.5<z<2$& 9299 & 3679 & 865 &  1510&  3678 &  1   & 865   &  1510& 0 &0\\
$0.5<z<1$& 4275 & 2455  & 570  &   1110  & 2454 & 1  & 570&    1110  &0 &  0\\
$1<z<2$& 5024&  1224  & 295  &   400 & 1224&  0  & 295 &   400 & 0 & 0 \\
$P<10^{23}$ \& $0.5<z<2$ &3473 & 1985   &442 &  918   & 1985  &   0  &   442 &     918 &      0  &        0\\
$P>10^{23}$ \& $0.5<z<2$&5826 & 1694 &  423  &592   &1693&  1 &  423	 &     592&	 0  	&    0\\
$10^{22.7}<P<10^{23}$ and $0.5<z<1$&1844 &1061 &  255&   464&    1061   &  0 &   255 &      464  &     0   &       0\\
$10^{22.7}<P<10^{23.3}$ and $1<z<2$&1417  &   554  &  120   & 207   &  554 & 0  &   120     &  207 &      0   &       0\\
$P>10^{23}$ and $0.5<z<1$&968 & 561 &  149  &   219  &560    & 1  & 149   &  219 &     0     &    0\\
$P>10^{23.3}$ and $1<z<2$&3607 & 670   & 175  &  193 & 670 &  0  &  175   &    193   &    0     &     0\\
\hline
\label{table2}
\end{tabular}
\tablefoot{As in Table~\ref{table1}, but for the sub-population of radio-selected star-forming galaxies.}
\end{center}
\end{table*}

\section{Radio Selection \label{radio}}
\subsection{Radio data \label{data}}
\cite{bondi} presented the first deep (\SI{72}{\hour} of observations) radio image of the EDF-N region obtained with the LOFAR High Band Antenna (HBA) at $144\,\rm MHz$. The observations produced a 6$\arcsec$ resolution image with a central rms noise of $32\,\mu \rm Jy\, \rm beam^{-1}$. A catalogue of $23\,333$ radio sources above a signal-to-noise ratio (SNR) threshold of 5 was extracted from the inner circular 10 \si{\deg\squared}
region. As the result of visual inspection of the radio morphologies and
with the support of deep IRAC images from the unWISE 5-year catalogue (\citealt{schlafly}), radio sources were split into two main categories: those whose radio emission could be fitted
by PyBDSF (the Python Blob Detector and Source Finder) with a single Gaussian component, and those that required multiple
Gaussian components. In this paper we will refer to the first
category as compact radio sources and to the second one as extended
radio sources. These latter sources constitute $\SI {7}\percent$ of the total radio counts.


\cite{bisigello} (hereafter \citetalias{bisigello}) provide optical and near-infrared counterparts to the \cite{bondi} radio sources, obtained by following a robust identification strategy that combines the statistical power of the likelihood ratio (LR, e.g., \citealt{sutherland}) method -- including both colour and magnitude information -- with targeted visual inspection.
After masking regions close to stars and with unreliable photometry in the optical or near-infrared, the
final catalogue includes 19\,550 radio sources, out of which 19\,401 (corresponding to a remarkable identification rate of $\SI {99.2}\percent$) have a reliable counterpart. Photometric redshifts were then derived using optical-to-radio data by following two different methods available from the {\tt CIGALE} (\citealt{buat}) SED-fitting routine: the first one corresponds to
the value of the best spectral energy distribution (SED) template (i.e., the one with the minimum $\chi^2$), while the second is derived with a Bayesian-like approach (see \citealt{noll} for more details). Although none of the two methods returns very precise redshift determinations (fraction of outliers $\SI {18.5}\percent$ in the first case and $\SI {22.8}\percent$ in the second -- cf.
\citetalias{bisigello}), these are good enough to provide a redshift distribution for the LOFAR sources in the EDF-N which is in agreement with the distributions presented by \cite{Duncan} for the other three LOFAR deep fields (Bo\"otes, Lockman Hole, and ELAIS-N1), and therefore are also good enough for the purposes of the present work, which uses redshifts only for statistical purposes and -- as will be more clear in the next section -- excludes objects with a poor redshift determination. We note that about $\SI {20}\percent$
of the radio sources in the \citetalias{bisigello} catalogue have a spectroscopic redshift determination from the Dark Energy Spectroscopic Instrument -- DESI  (\citealt{desi}).

We stress that we prefer to adopt the photometric redshifts presented in \citetalias{bisigello} with respect to those derived in \citet{Q1-TP005}, as the latter analysis does not include IRAC (\citealt{moneti}) and available radio data in the fit, and neither does it consider AGN templates, which are relevant to the present work. This is also the reason why we do not include information on host galaxy masses in our analysis, since for consistency with the adopted redshifts, we should have estimated them from the \citetalias{bisigello} work, but these are not yet available.

The $144\,\rm MHz$ radio luminosities for the \citetalias{bisigello} sample were then derived for the adopted cosmology and Bayesian redshifts by assuming an average radio spectral slope $\alpha=0.7$ independent of redshift, which holds with good approximation for the overwhelming majority of relatively low-luminosity radio sources, both AGN and star-forming galaxies (e.g., \citealt{maglio14, toba, calistro}). We note that considering a spectral slope of 0.67 or 0.73, as derived by \cite{calistro} separately for AGN and star-forming galaxies, would produce a change in radio luminosity of less than $\SI {6}\percent$. The radio luminosity distribution as a function of redshift for the \citetalias{bisigello} sample is presented in the left-hand panel of Fig.~\ref{fig:L_z}.

\subsection{AGN or star-forming galaxy? \label{AGNvsSF}}
A tricky step in the process of identifying extragalactic sources observed in monochromatic radio surveys is distinguishing between radio emission of AGN origin and that instead due to star-forming processes (e.g., \citealt{smolcic2017}; \citealt{whittam2022}; \citealt{best23}).
In the local Universe things are rather straightforward, since the majority of radio-loud sources are associated with massive elliptical galaxies with little or no ongoing star-formation activity (for a more refined analysis of the local population of hosts see e.g., \citealt{Janssen2012}). They are generally "red and dead'', present radio-to-optical flux ratios\footnote{defined as $q_R=F_{1.4\, \rm GHz}\,10^{(R-12.5)/2.5}$, where $F$ is expressed in mJy and $R$ is the $R$-band magnitude} larger than $30$ and show a remarkable tightness in the $K$-band magnitude versus redshift relation, at least out to $z\simeq 1$ (e.g. \citealt{lilly1}; \citealt{best2005}; {\citealt{capetti}}). On the other hand, star-forming galaxies present a very tight correlation between their far-IR and radio luminosities (e.g. \citealt{delhaize2017}; \citealt{calistro}), supposedly due to the presence of massive stars which on the one hand heat dust, therefore producing IR emission, while on the other hand generate supernovae events that accelerate cosmic rays, producing radio synchrotron radiation (e.g. \citealt{condon}).
However, at high redshifts,  discerning between these two populations becomes more uncertain (cf. \citealt{maglio22} for a review). 
This is why in this work, as in previous ones (\citealt{maglio14,maglio16,maglio15,maglio18}), we decided to rely on the method introduced in \cite{maglio13}. 

The approach is based 
on the work of \cite{mcalpine}, which provides luminosity functions at 1.4\,GHz (hereafter RLF) separately for the two classes of AGN and SFG up to redshifts $z\simeq 2.5$.  Their results show that the 1.4\,GHz radio luminosity $P_{\rm cross}$ beyond which AGN-powered galaxies become the dominant radio population scales with redshift roughly as
\begin{eqnarray}
 {\rm log}_{10} \left [ P_{\rm cross}(z)/P_{0,\rm cross}\right]=z,
\label{eq:P}
\end{eqnarray}
up to $z\sim 1.8$, and then remains constant to $P_{\rm cross}\simeq 10^{23.5}$\,W\,Hz$^{-1}$\,sr$^{-1}$ at least up to $z\simeq 2.5$. The value $P_{0,\rm cross}=10^{21.7}$\,W\,Hz$^{-1}$\,sr$^{-1}$ is what is found in the local Universe and roughly coincides with the break in the radio luminosity function of SFG 
(cf. \citealt{maglio2002, mauch}). Beyond this luminosity, the RLF of star-forming galaxies steeply declines, and the contribution of this population to the total radio counts is drastically reduced to a negligible percentage. 
The same trend is true at higher redshifts, and given the steepness of the RLF of star-forming galaxies beyond the luminosity break (as opposed to the flatter trend of AGN at all $z$), the adopted method has the important
advantage of producing a clean sample of radio-selected AGN (cf. \citealt{maglio14}), while only necessitating redshift determinations instead of expensive spectroscopy or multi-wavelength information for the sources being considered.

In order to proceed with the above method, we first rescaled the LOFAR luminosities to 1.4\,GHz by assuming an average radio spectral index $\alpha=0.7$ (see Sect.~\ref{data}).
We then distinguished between radio emission from AGN and star-forming galaxies by means of Eq.~(\ref{eq:P}) for $z\le 1.8$ and by fixing $P_{\rm cross}(z)=10^{23.5}\,$W$\,$Hz$^{-1}\,$sr$^{-1}$ at higher redshifts. This procedure identifies 4011 AGN (3946 if using the redshifts derived with the minimum $\chi^2$ method, cf. \citetalias{bisigello}) and 15\,390 sources classified as SFG (15\,455 if using redshifts derived with the minimum $\chi^2$ method). We stress here (cf. \citealt{maglio22}) that, while the AGN selection is clean and subject to very little contamination from the population of star-forming galaxies, this is not true for this latter class, since SFG selected as in Eq.~(\ref{eq:P}) are a mixed bag of all those sources that emit in the radio band as a result of processes connected to star formation. These include genuine SFG as well as radio-quiet AGN whose host galaxies are actively forming stars (e.g., \citealt{padovani}). Indeed the fraction of radio-emitting AGN obtained as in Eq.~(\ref{eq:P}) is remarkably similar to the $\SI {\sim 18}\percent$ derived by \cite{best23} on the other three LOFAR deep fields once their LERGs and HERGs are combined together, since our method is agnostic to this distinction. That of SFG is instead higher ($\SI {\sim 81}\percent$ versus the $\SI {\sim 68}\percent$ of \citealt{best23}). However, as expected from the above discussion, this discrepancy is largely reduced if in the \cite{best23} sample we add together SFG and the $\SI {\sim 9}\percent$ contribution from radio-quiet AGN.

\begin{figure*}
\begin{center}
\vspace{0 cm}  
\includegraphics[scale=0.32, viewport=90 135 600 760] {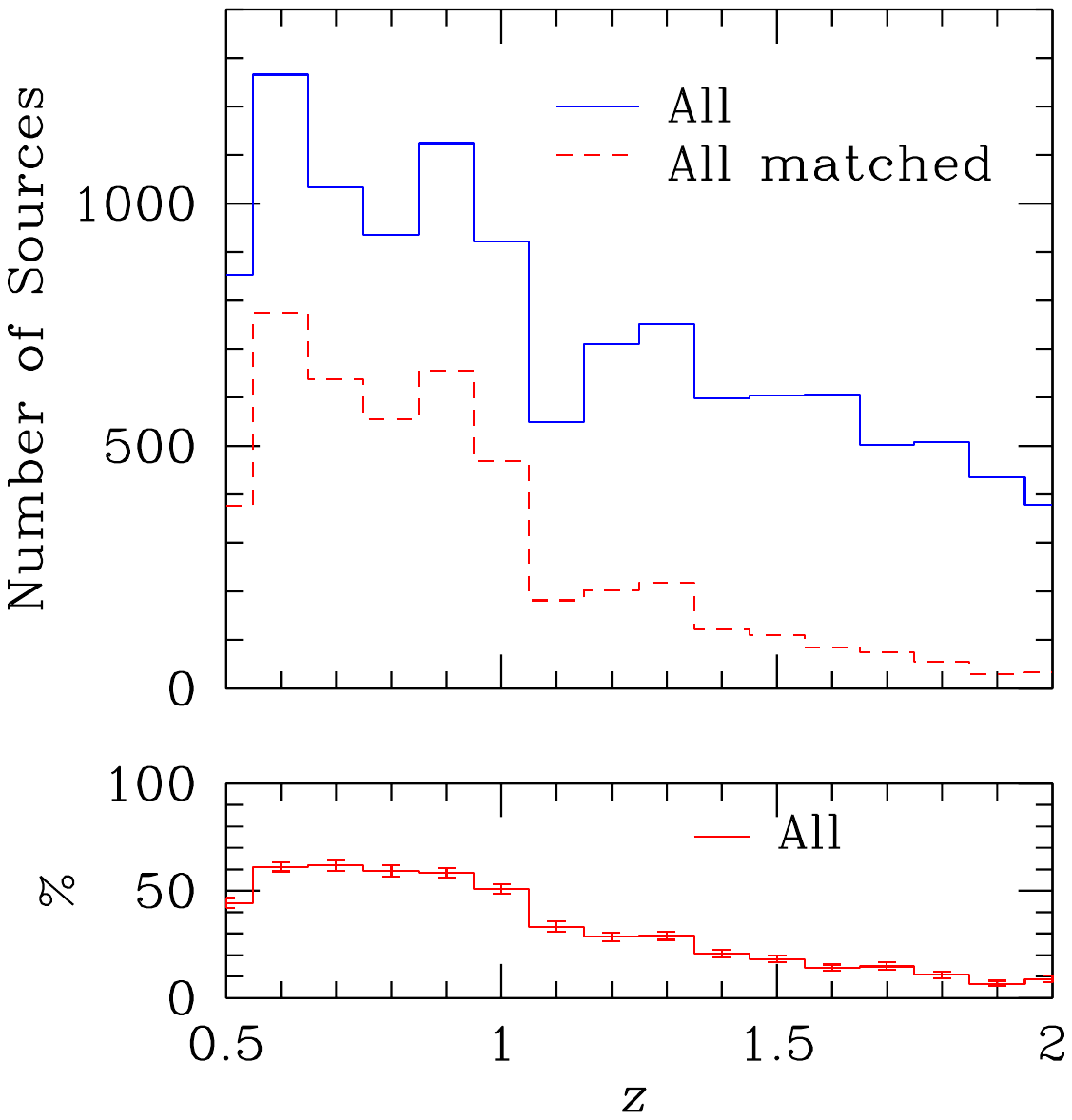}
\includegraphics[scale=0.32, viewport=80 135 600 760] {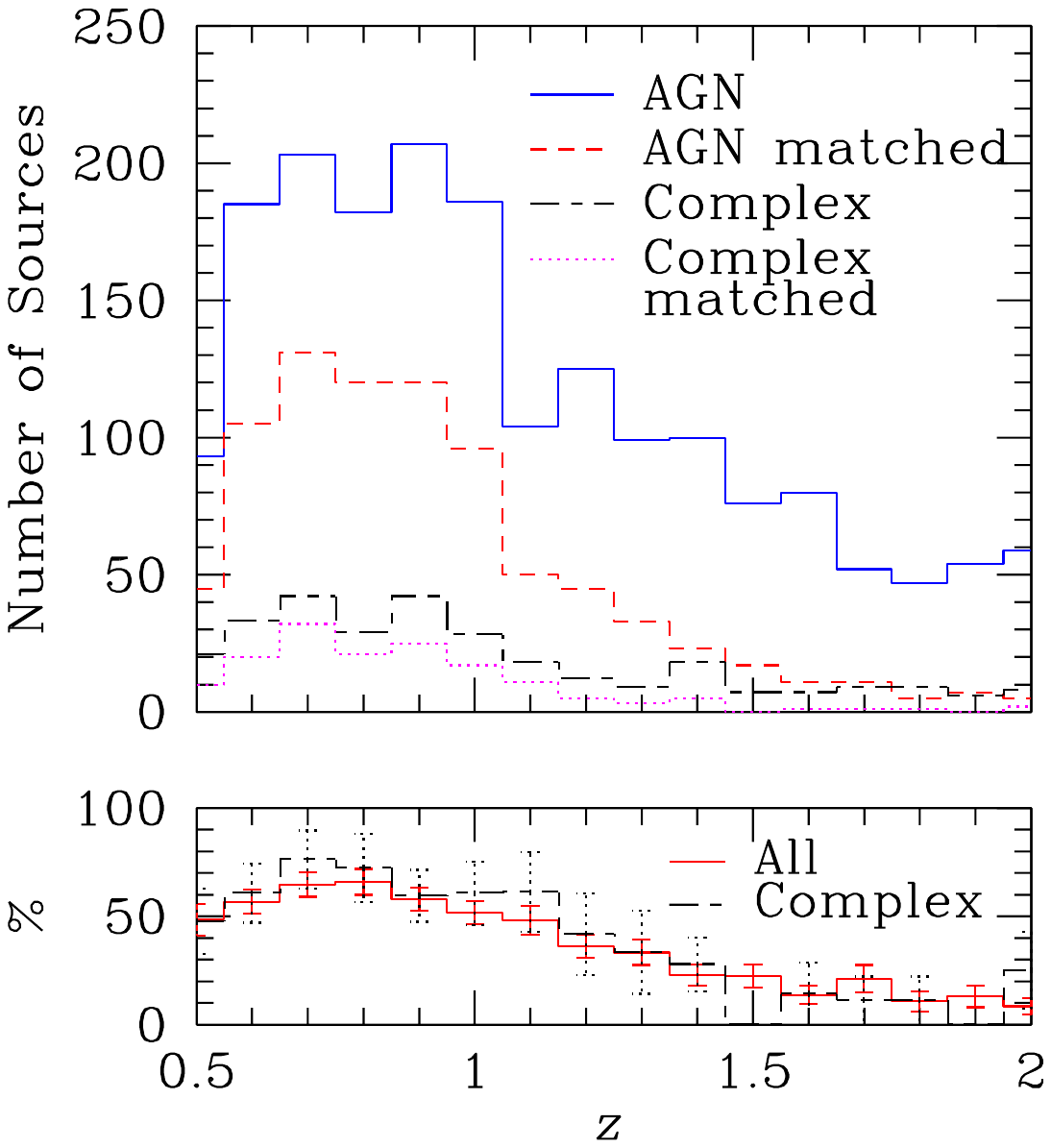}
\includegraphics[scale=0.32, viewport=70 135 600 760] {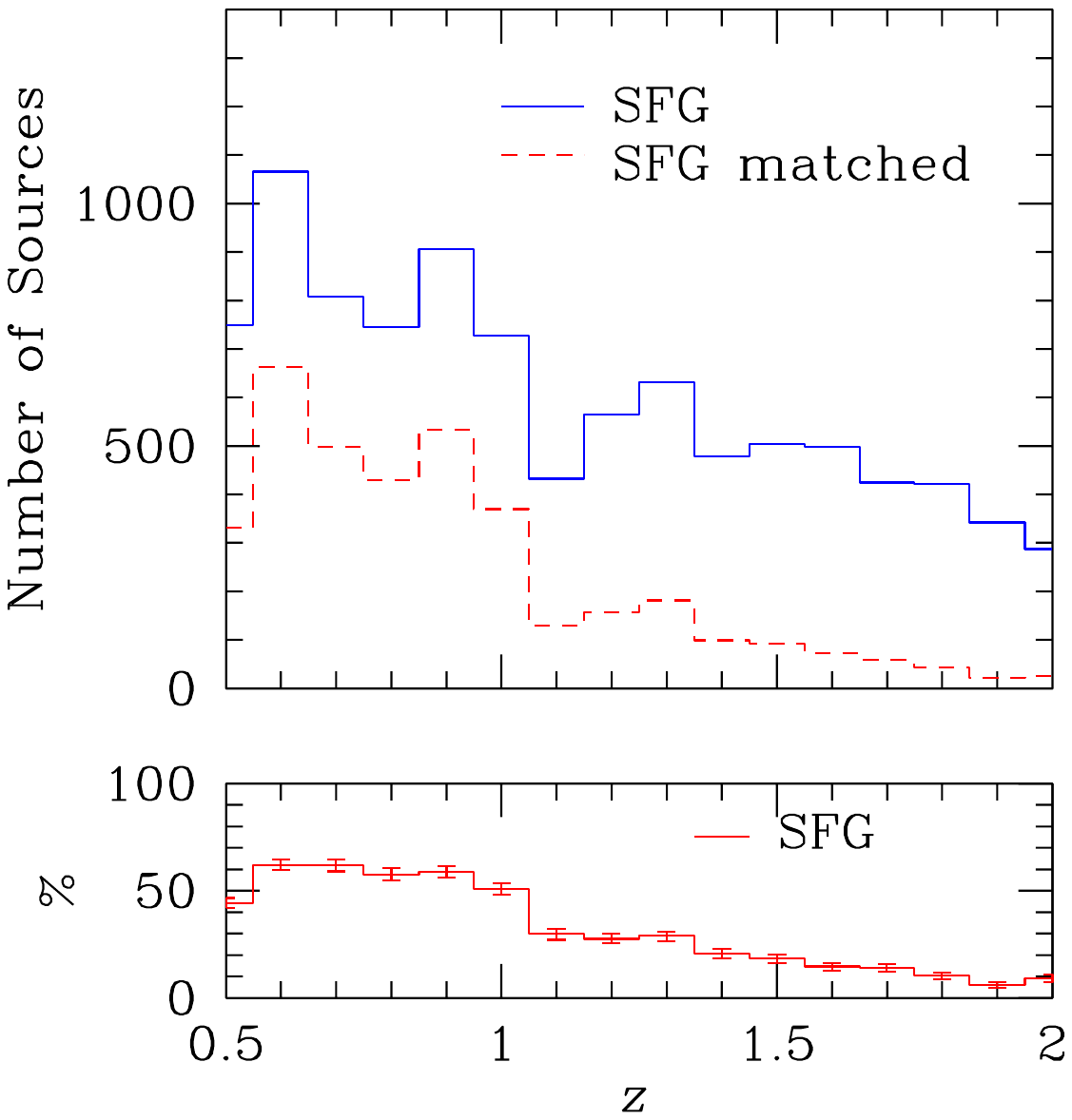}
\caption{Redshift distribution of LOFAR sources in the EDF-N region in the redshift interval $0.5< z < 2$. The left-hand panel refers to the whole sample, the middle panel to the sub-class of radio AGN, and the right-hand panel to star-forming galaxies. In each panel the solid line refers to the parent radio sample from the work of \citetalias{bisigello}, while the dashed histograms correspond to those sources with optical counterparts from the work of \citetalias{Q1-SP013}. The bottom panels show the percentages obtained from the ratios between the above quantities, together with the associated (Poissonian) errors. The middle panel additionally presents the trends for AGN with complex radio morphology, short-long dashed lines for the parent sample and dotted lines for the matched sample.
\label{fig:hist_z}}
\end{center}
\end{figure*}

\begin{figure*}
\begin{center}
\vspace{0 cm}  
\includegraphics[scale=0.32, viewport=300 135 600 760] {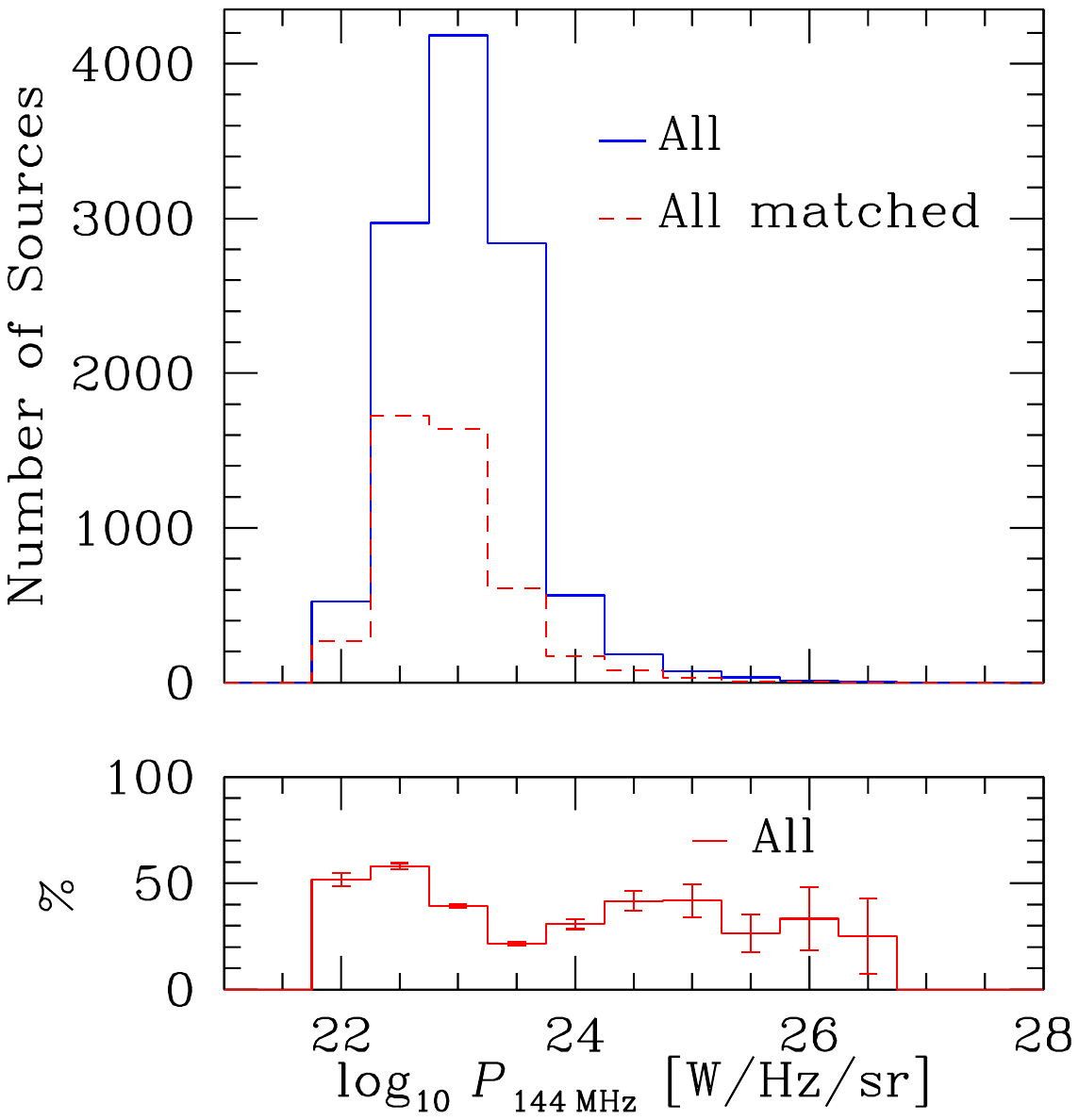}
\includegraphics[scale=0.32, viewport=80 135 600 760] {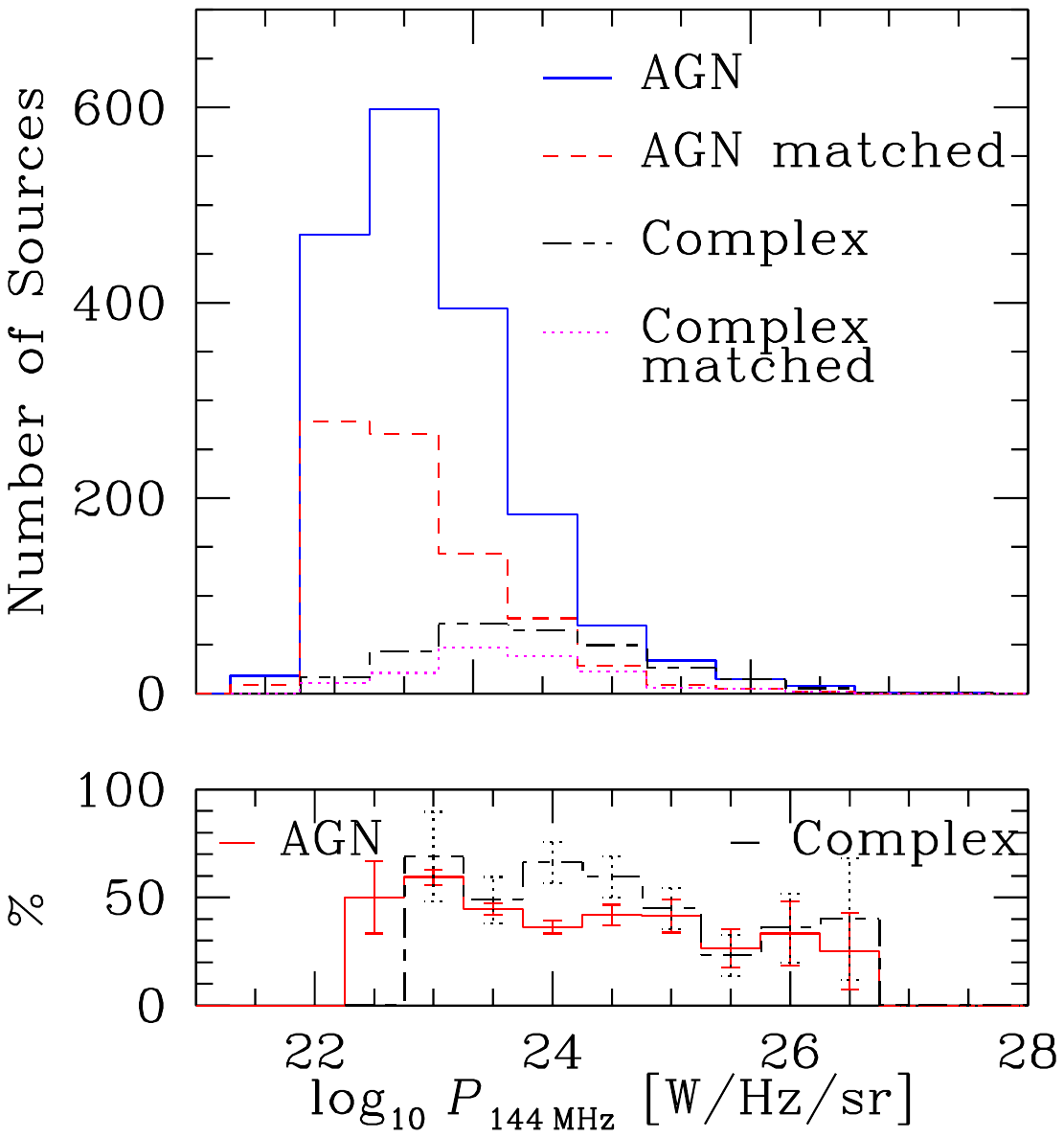}
\includegraphics[scale=0.32, viewport=60 135 400 360] {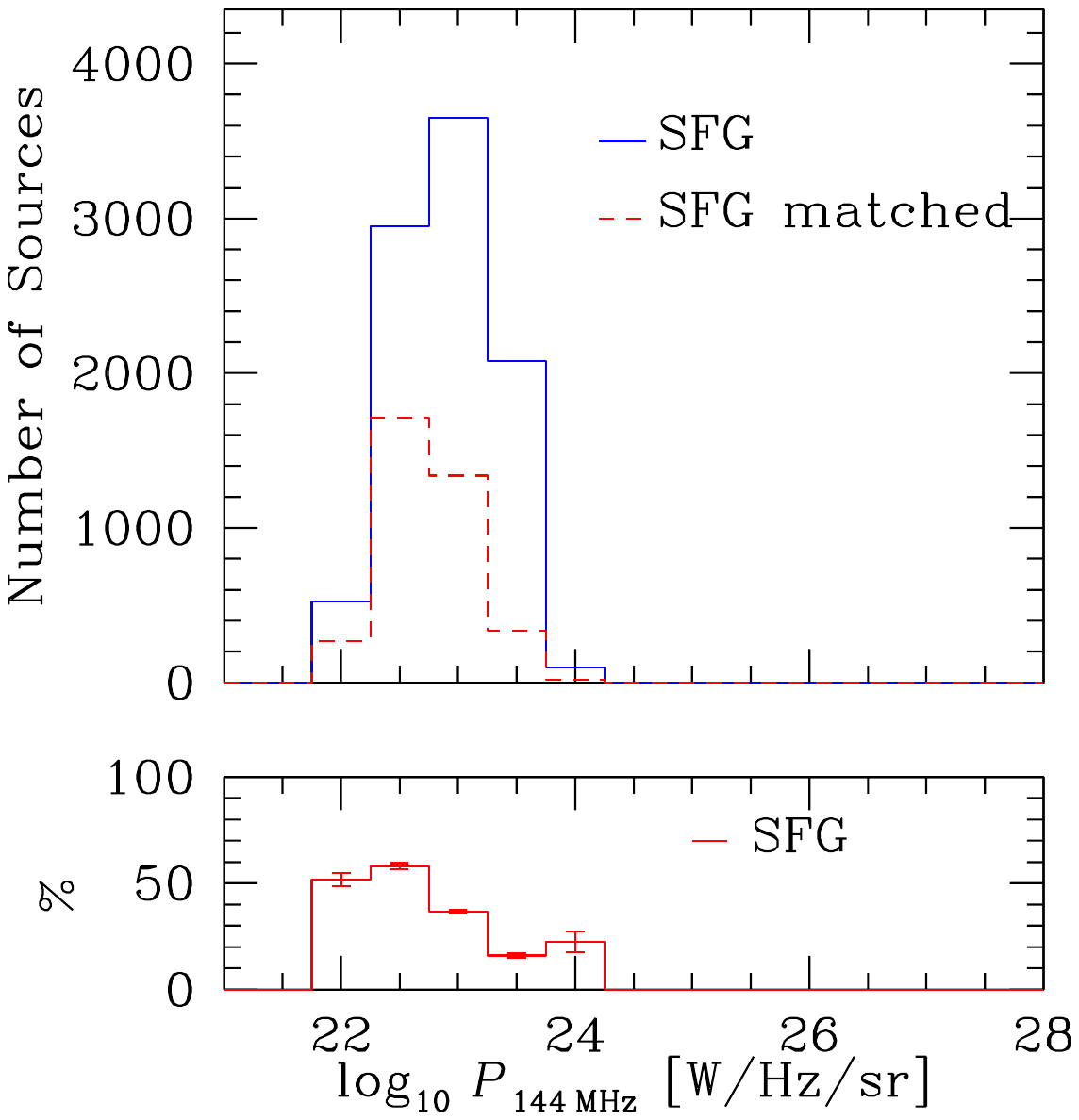}
\caption{Similar to Fig.~\ref{fig:hist_z}, but for the 144\,MHz luminosity distribution of LOFAR sources in the EDF-N belonging to the redshift interval $0.5<z<2$. 
\label{fig:hist_RL}}
\end{center}
\end{figure*}

In spite of the mixed nature of the star-forming sample ($\SI {\sim 85}\percent$ {\it bona-fide} SFG and $\SI {\sim 11}\percent$ radio-quiet AGN according to \citealt{best23}),  since interesting information can also be obtained for these objects, in the following we continue to label them with the cumulative name of SFG and proceed with their analysis, warning the reader to keep the above caveat in mind.

As a final step, for the creation of our master-catalogue, we removed from the AGN and star-forming samples obtained as in Eq.~(\ref{eq:P}) those 1045 objects ($\SI {5}\percent$ of the original sample) that had a different classification when using Bayesian versus minimum $\chi^2$ redshifts. This was done to ensure that all the sources with dubious redshift determinations were excluded. Indeed, since the adopted selection method only depends on redshift once the radio flux is given, different classifications for a single object imply a large discrepancy between photometric redshifts derived with the two different approaches. 

After this final cleaning we end up with 14\,900 SFG (3340 with spectroscopic redshifts) and 3456 AGN (647 with spectroscopic redshifts), out of which 438 (i.e., $\SI {\sim 13}\percent$) have a complex or extended morphology. This is true only for 26 SFG (corresponding to $\SI {0.17}\percent$ of the parent sample), and provides a reassuring check for the adopted classification criterion, since we do not expect processes associated with star formation to produce complex radio morphologies. We note that in principle the adoption of the full probability distribution functions (pdf) for the photometric redshifts might have an effect on some classifications as such pdf distributions can be broad (c.f. Hale et al. in preparation). However, 
given the cleaning procedures explained above, and the fact that many (especially $z<1$) sources are endowed with spectroscopic redshifts, we do not expect this effect to introduce appreciable variations to our samples.

The redshift distributions of both AGN and SFG in the redshift range $0.5<z<2$ relevant to the following analysis (cf. Sect.~\ref{combined}) are shown in the middle and right-hand panels of Fig.~\ref{fig:hist_z}, while their luminosity distributions in the same redshift range are provided in the middle and right-hand panels of Fig.~\ref{fig:hist_RL}. The left-hand panels of both figures reproduce the redshift and luminosity distributions of the parent radio sample. The middle panels of Figs.~\ref{fig:hist_z} and \ref{fig:hist_RL} additionally present the trends for the sub-population of complex/extended radio AGN. It can be seen that their redshift distribution closely follows that of the whole radio AGN population, while -- as expected -- their radio luminosities are shifted towards the bright end of the distribution.

\begin{figure*}
\begin{center}
\vspace{0 cm}  
\includegraphics[scale=0.40, viewport=-30 150 600 760] {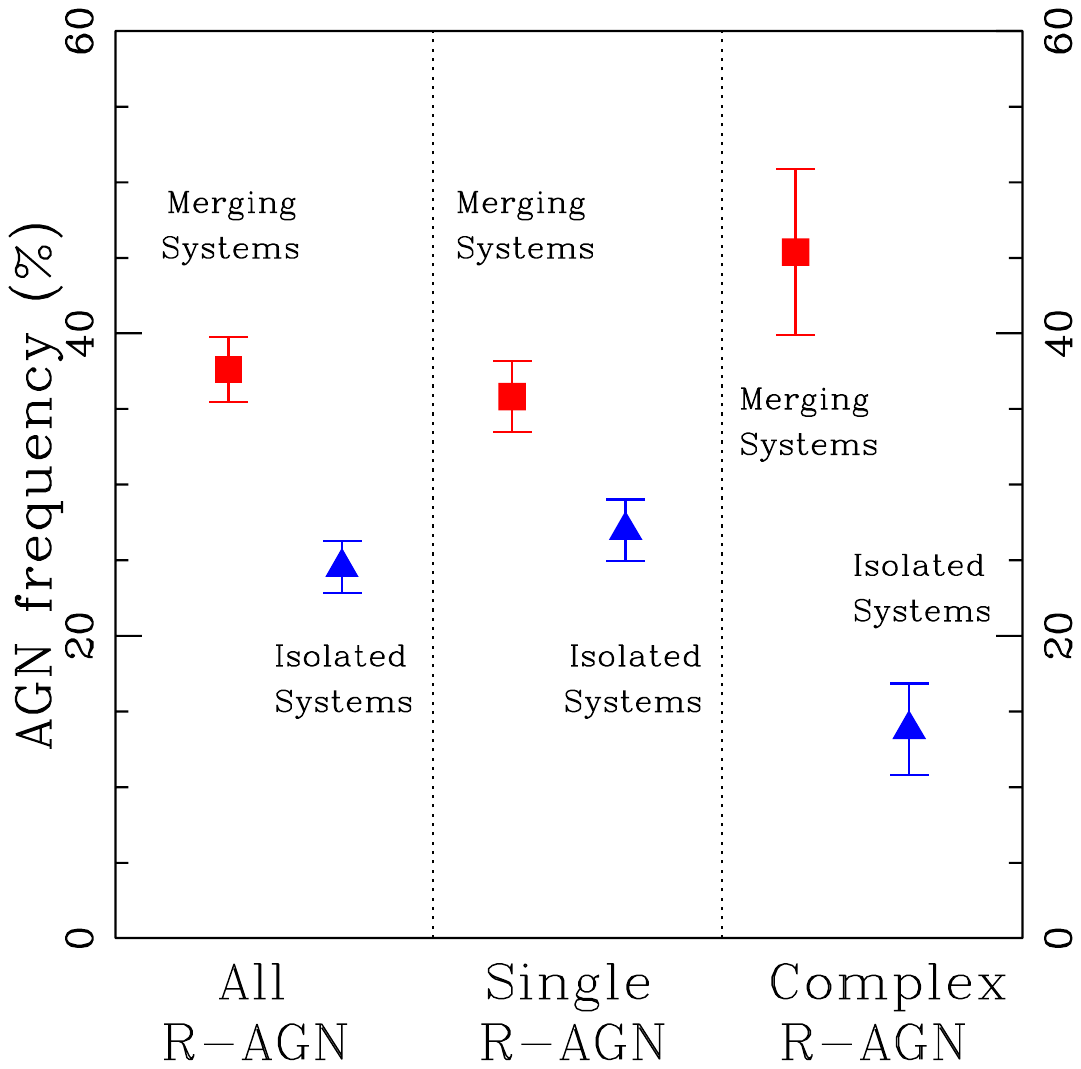}
\includegraphics[scale=0.40, viewport=-20 150 600 760] {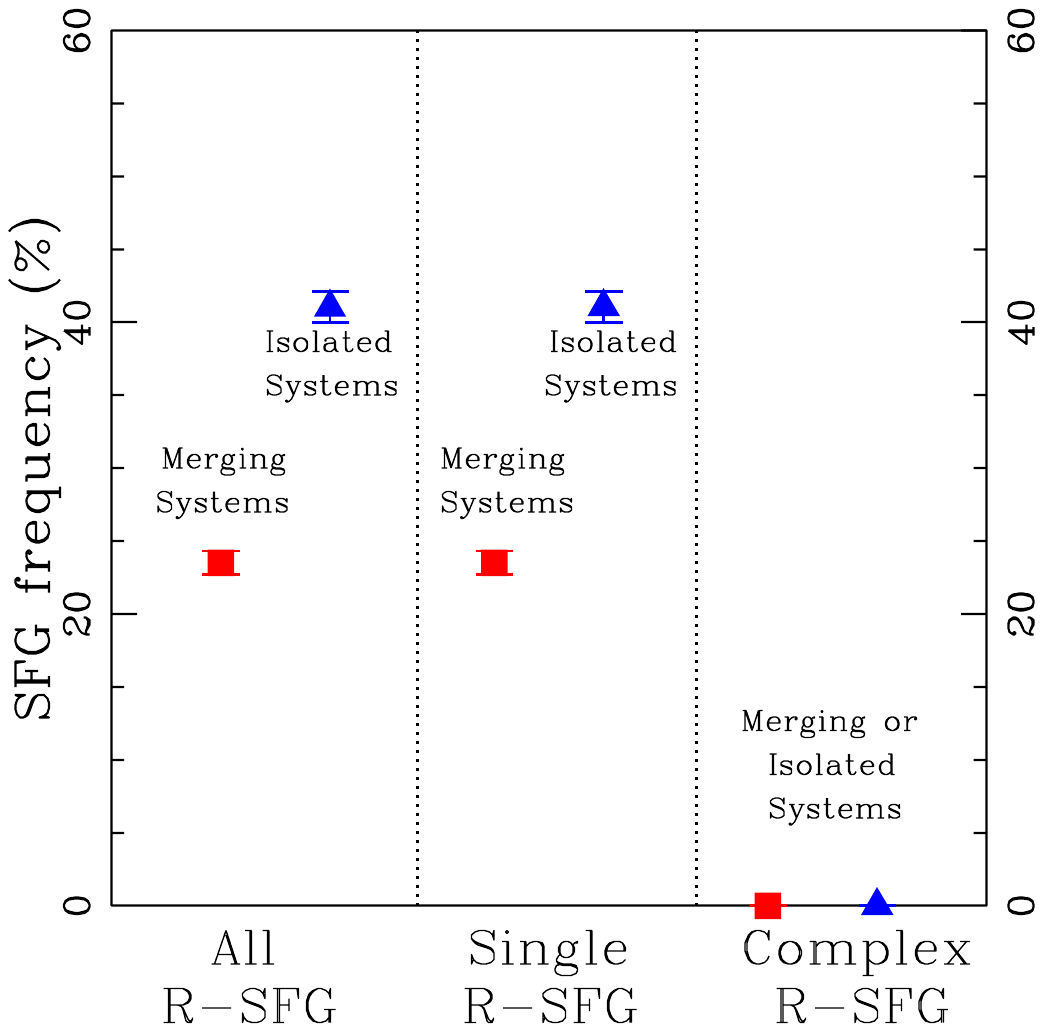}
\caption{Percentage of radio-selected AGN (left-hand panel) and SFG (right-hand panel) in the range $0.5<z<2$ associated with either isolated galaxies (\lq Isolated Systems\rq -- blue triangles) or galaxy-galaxy mergers (\lq Merging Systems\rq -- red squares). Each panel is subdivided into three sections, where the leftmost one shows the case for the whole sample of radio-selected AGN or SFG with a counterpart from the work of \citetalias{Q1-SP013}, the middle one is for radio sources with a compact radio structure, while the rightmost one is for objects presenting complex or extended radio morphologies. Error-bars represent Poissonian uncertainties.
\label{fig:hist_mer}}
\end{center}
\end{figure*}
\begin{figure*}
\begin{center}
\vspace{0 cm}  
\includegraphics[scale=0.4, viewport=-20 130 600 760] {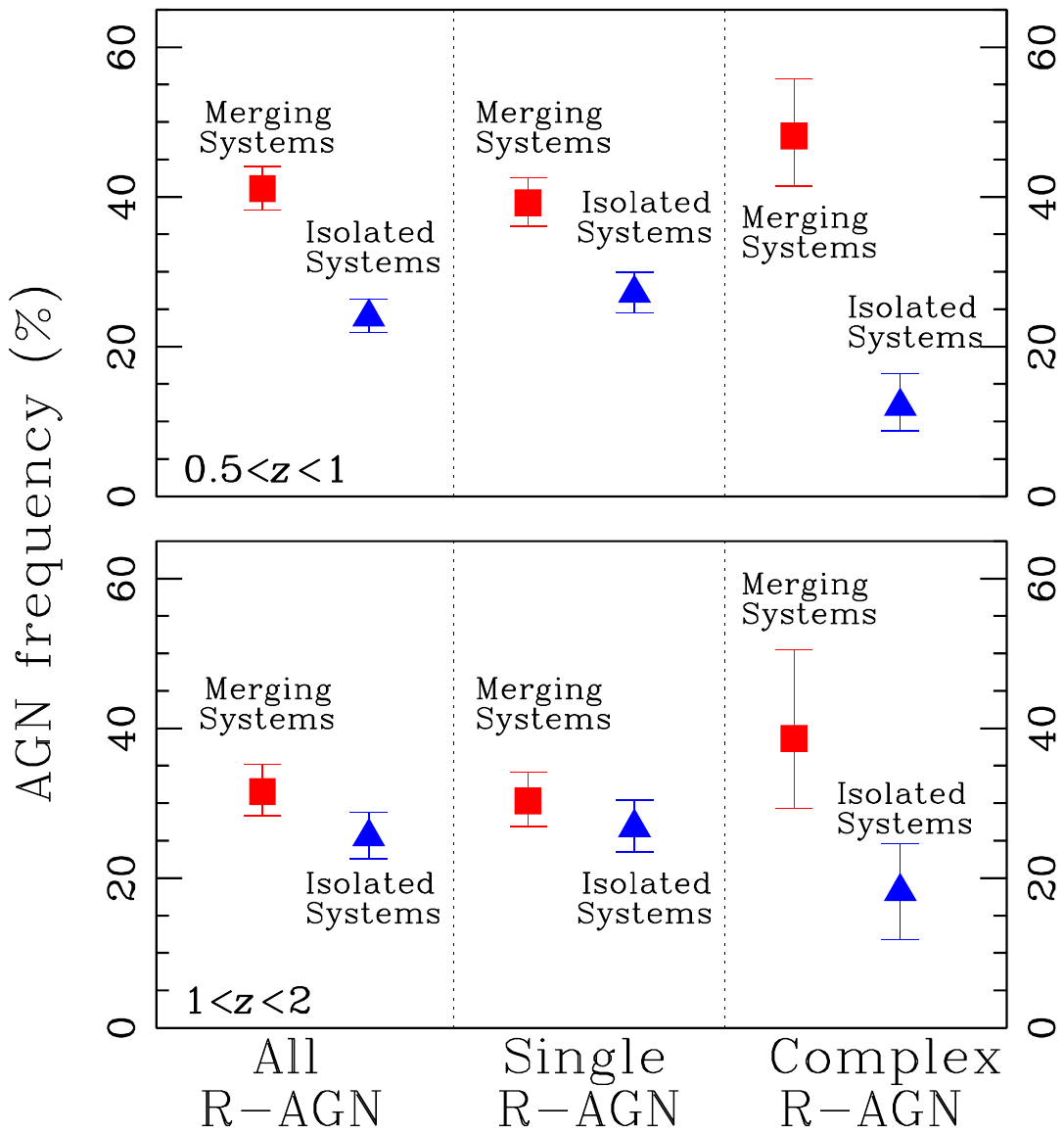}
\includegraphics[scale=0.4, viewport=-20 130 600 760] {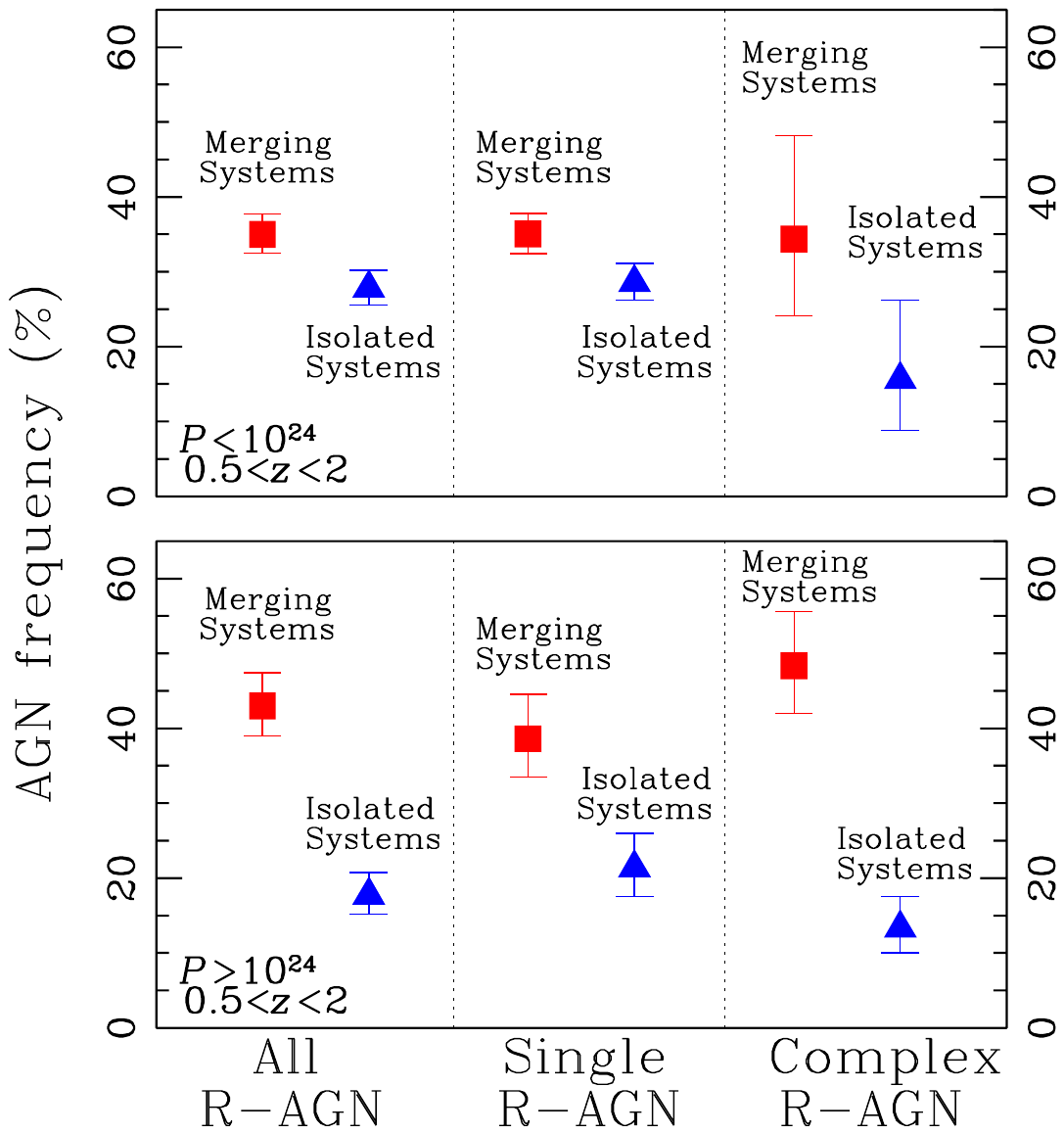}
\caption{Percentage of radio-selected AGN associated with either isolated galaxies or galaxy-galaxy mergers. The subdivision of each panel and the colour/marker styles are as in Fig.~3. The left-hand panel presents the two cases for sources in the $0.5<z<1$ (top) and $1<z<2$ (bottom) range, while the right-hand panel considers objects of different radio luminosities, respectively $P_{144\, \rm MHz}<$ (top) and $>$ (bottom) $10^{24}$\,W\,Hz$^{-1}$\,sr$^{-1}$. In this case error-bars represent Poissonian uncertainties derived according to \cite{gehrels}.
\label{fig:hist_AGN_mer}}
\end{center}
\end{figure*}
\begin{figure*}
\begin{center}
\vspace{-0.5 cm}  
\includegraphics[scale=0.4, viewport=-20 130 600 760] {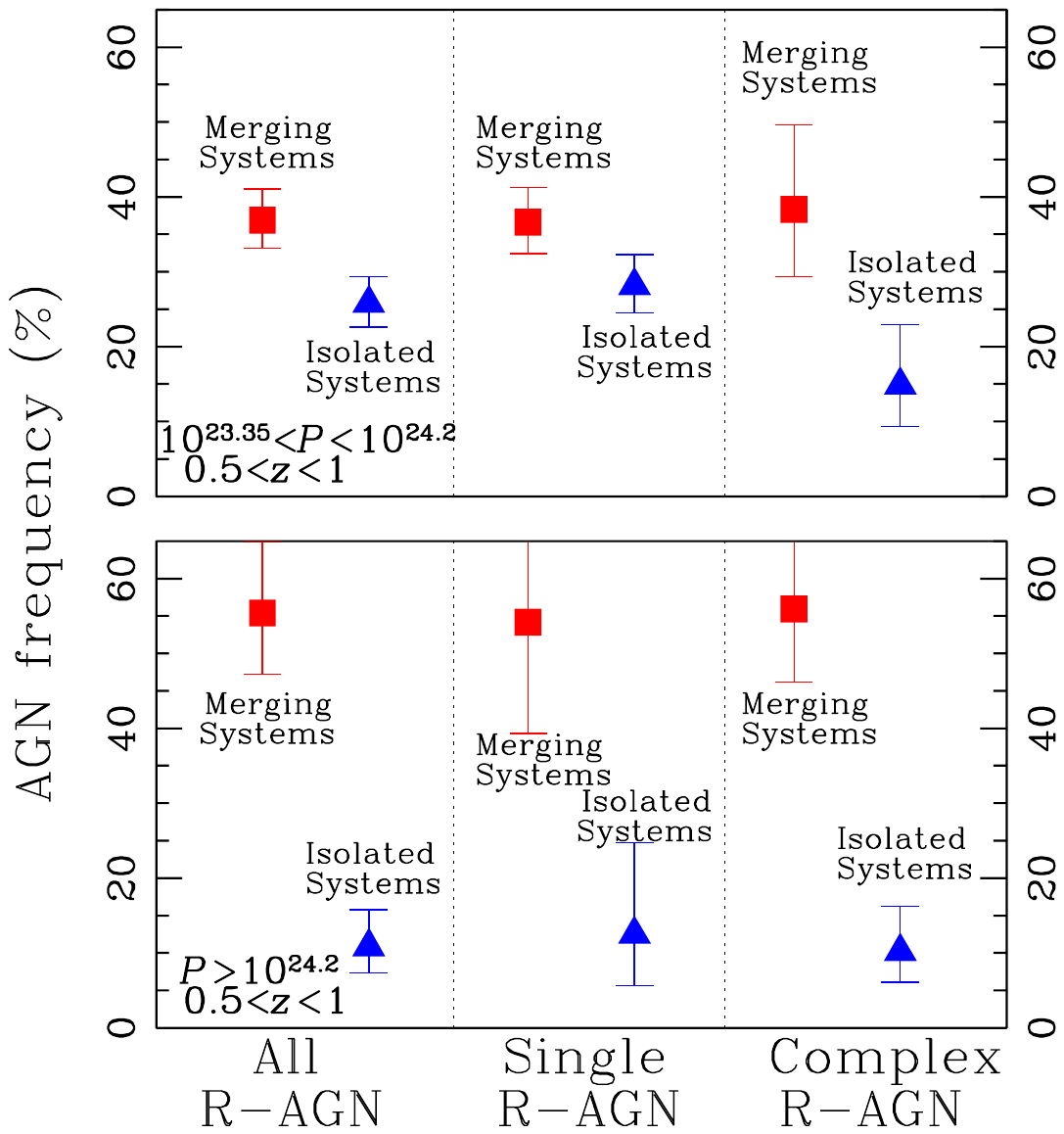}
\includegraphics[scale=0.4, viewport=-20 130 600 760] {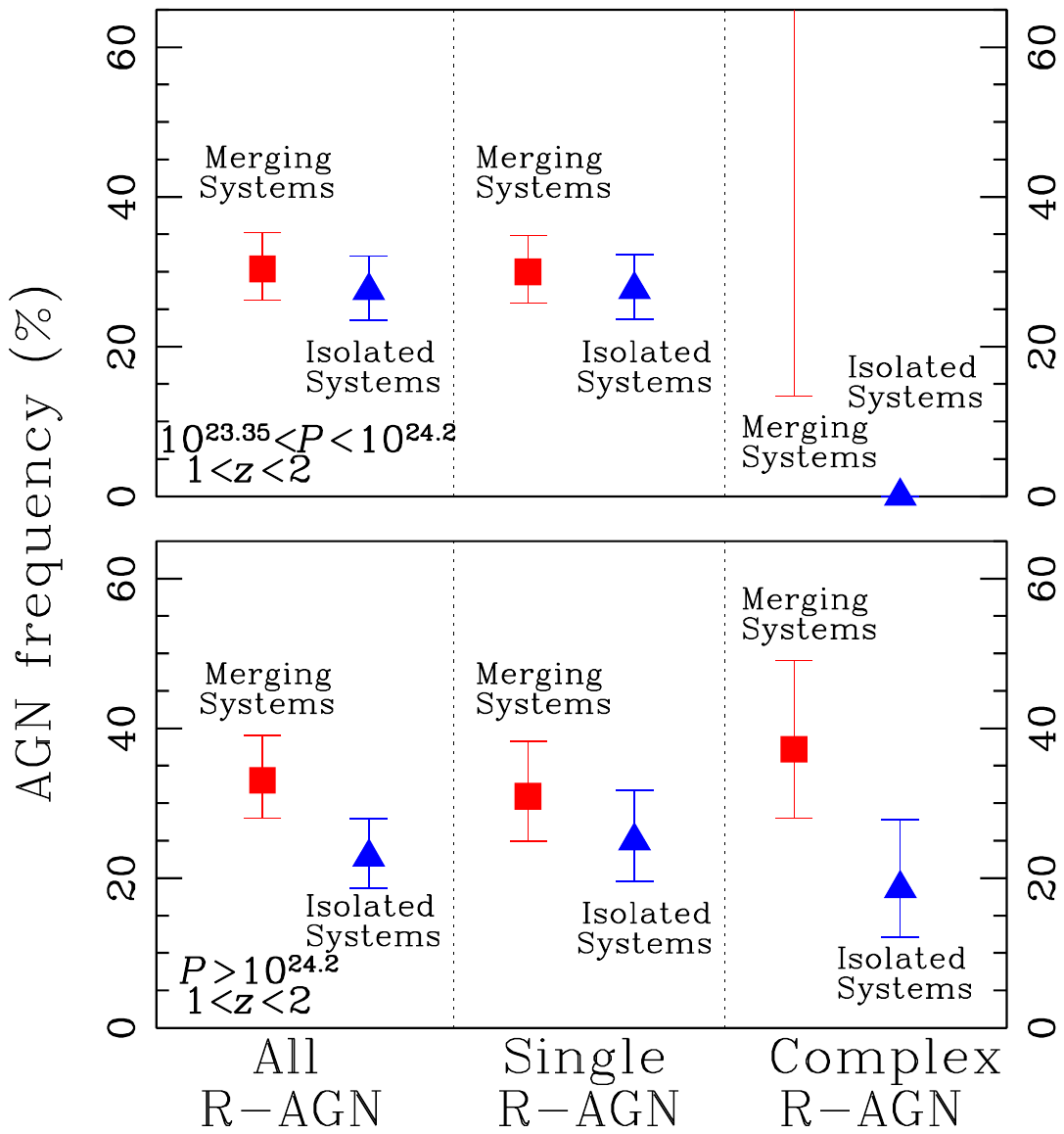}
\caption{Similar to Fig.~4, except that now different panels present different redshift/radio-luminosity (expressed in W\,Hz$^{-1}$\,sr$^{-1}$ units) combinations as shown (see text for details). The point in the top-right part of the top-right panel is off the scale and refers to one single object hosted by a merging system.
\label{fig:hist_AGN_mer_z_P}}
\end{center}
\end{figure*}

\section{The combined catalogue \label{combined}}
For the following analysis, we make use of the catalogue presented by \citetalias{Q1-SP013}. 
In brief, the authors provide the first classification of \Euclid galaxies divided into isolated versus mergers based on their VIS images in the \Euclid Deep Fields. The work processed $\IE$ observations with a convolutional neural network (CNN) trained on cosmological hydrodynamical simulations to classify galaxies as mergers and non-mergers. A redshift interval $0.5<z<2$ was chosen to ensure that the adopted $8\arcsec\times 8\arcsec$ thumbnail centred on each source corresponded to a roughly constant physical scale of 50 kpc $\times$ 50 kpc. Furthermore, in order to ensure as clean a catalogue as possible, \citetalias{Q1-SP013} limited their analysis to galaxies with $\texttt{DET\_QUALITY\_FLAG}<4$ to filter out contaminants, $\texttt{SPURIOUS\_FLAG}=0$ to exclude spurious sources, and $\texttt{MUMAX\_MINUS\_MAG>-2.6}$ to filter out point-like sources. Finally, a cut at $\IE=23.9-2.5\: \rm log_{10} \texttt{ (FLUX\_DETECTION\_TOTAL)} =23.5$ was considered to exclude faint objects for which morphological classification was less robust given the depth of the survey and the resolution of the images. This procedure returned 941\,730 galaxies, respectively recognized as mergers ($\SI {\sim 18}\percent$ of the parent sample), non-mergers ($\SI {\sim 45}\percent$ of the parent sample) and sources for which the classification was uncertain (the remaining $\SI {\sim 37}\percent$) according to the score of the classifier ($\geq 0.59$ for mergers, $<0.35$ for non-mergers, and between 0.35 and 0.59 for unclassified sources (see \citetalias{Q1-SP013} for further information).

We then considered the positions of the optical counterparts of LOFAR sources from the work of \citetalias{bisigello} (see Sect.~\ref{radio}). These have been matched to the catalogue provided by \citetalias{Q1-SP013} by simply requiring offsets $<0\arcsecf3$. With this choice, the chances for spurious associations are basically null. Within the redshift range $0.5<z<2$ we have 1793 radio AGN and 9299 SFG. Of these, 819 AGN (corresponding to $\SI {\sim 47}\percent$ of the total) and 3679 SFG (corresponding to $\SI {\sim 40}\percent$) are associated with a \Euclid galaxy from the work of \citetalias{Q1-SP013} (312 AGN and 604 SFG if only restricting to sources with a spectroscopic redshift determination). The properties of these two samples in the case of associations with  galaxies classified by \citetalias{Q1-SP013} as either interacting/merging or isolated systems are summarised in Tables \ref{table1} and \ref{table2}. Unclassified systems will be discussed in detail in Sect.~\ref{sys}. For what concerns sources with a spectroscopic redshift determination, there are 186 SFG associated with merging systems and 218 SFG hosted by isolated galaxies (157 and 192 for $0.5<z<1$). There are instead 139 AGN in merging systems and 63 AGN in isolated galaxies (122 and 54 for $0.5<z<1$). Comparing the above numbers with those provided in Table~\ref{table1} implies that about $\SI {\sim 52}\percent$ of the $0.5<z<1$ AGN in the matched/classified catalogue have a secure spectroscopic redshift determination.

The dashed histograms (dotted in the case of complex radio morphologies) in Fig.~\ref{fig:hist_z} illustrate the redshift distributions of the LOFAR sources in the matched catalogue. As can be seen from the bottom panels of the same figure, the fraction of radio sources with a counterpart from the \citetalias{Q1-SP013} work is about $\SI {\sim 60}\percent$, approximately constant between $0.5<z<1$ in all the considered cases (whole population, radio AGN including those with complex morphologies, and SFG), and then presents a smooth decline between $z=1$ and 2. As shown in the left panel of Fig.~\ref{fig:morph}, this is due to the $\IE=23.5$ magnitude cut applied to the whole sample of VIS-selected \Euclid galaxies (see above in this section). 
Indeed, we find that while below $z=1$, $\SI {\sim 37.5}\percent$ of \Euclid galaxies are fainter than $\IE=23.5$, this percentage increases to $\SI {\sim 85.7}\percent$ at $z=2$. We note that these values are in perfect agreement with those found in the case of LOFAR sources, independent of their classification (cf. Fig.~\ref{fig:hist_z}). This implies an absence of redshift biases in our combined radio/optical sample.  

Figure~\ref{fig:hist_RL} instead shows the radio luminosity distribution of LOFAR sources belonging to the various sub-categories (all, radio AGN, and SFG) found within the range $0.5<z<2$. Solid histograms (short-long dashed for complex radio morphologies) correspond to the parent radio sample obtained from the work of \citetalias{bisigello}, while dashed histograms (dotted for complex radio morphologies) correspond to those objects with a counterpart from the \citetalias{Q1-SP013} work, regardless of the optical morphological classification of the host galaxy. The luminosity distribution of radio AGN in the matched catalogue closely follows that of the parent (AGN) sample also in the case of AGN with extended morphologies. As further shown in the middle-bottom panel of Fig.~\ref{fig:hist_RL}, this implies that the matched radio AGN sample is free from radio luminosity biases, independent of the extension of radio emission. The same does not hold for the population of matched radio-selected star-forming galaxies, whose luminosity distribution  
presents a peak between $10^{21.5}< P_{144\, \rm MHz}/[{\rm W\,Hz^{-1} \,sr^{-1}}]< 10^{22.5}$ followed by a rather sharp decline at higher luminosities. The net result of the sum of these two trends (constancy of the fractional number of AGN with radio luminosity and peaked distribution in the case of SFG) is the behaviour observed in the left-hand panel of Fig.~\ref{fig:hist_RL}.

\begin{figure*}
\begin{center}
\vspace{-1 cm}  
\includegraphics[scale=0.4, viewport=-20 130 600 760] 
{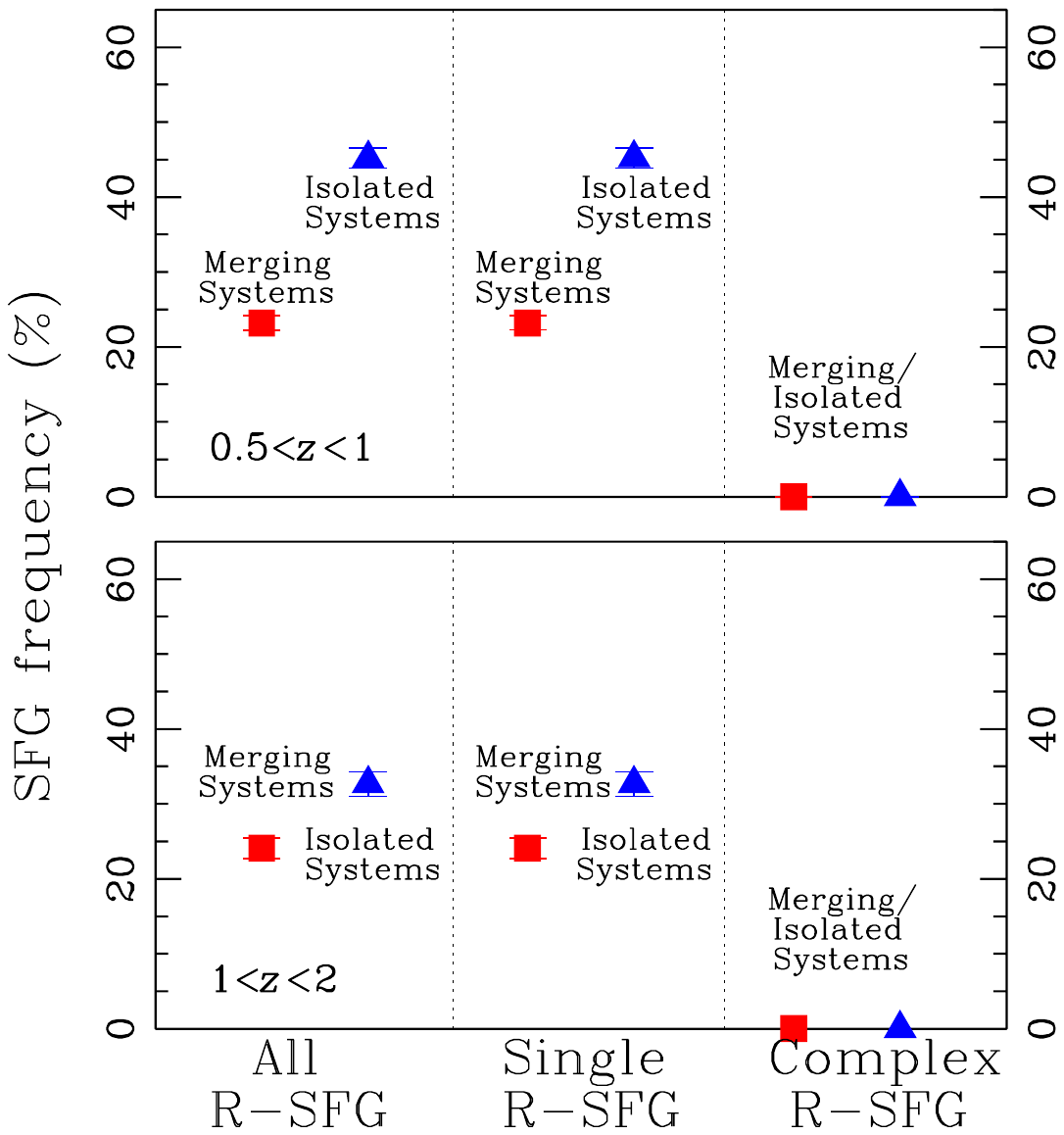}
\includegraphics[scale=0.4, viewport=-20 130 600 760]
{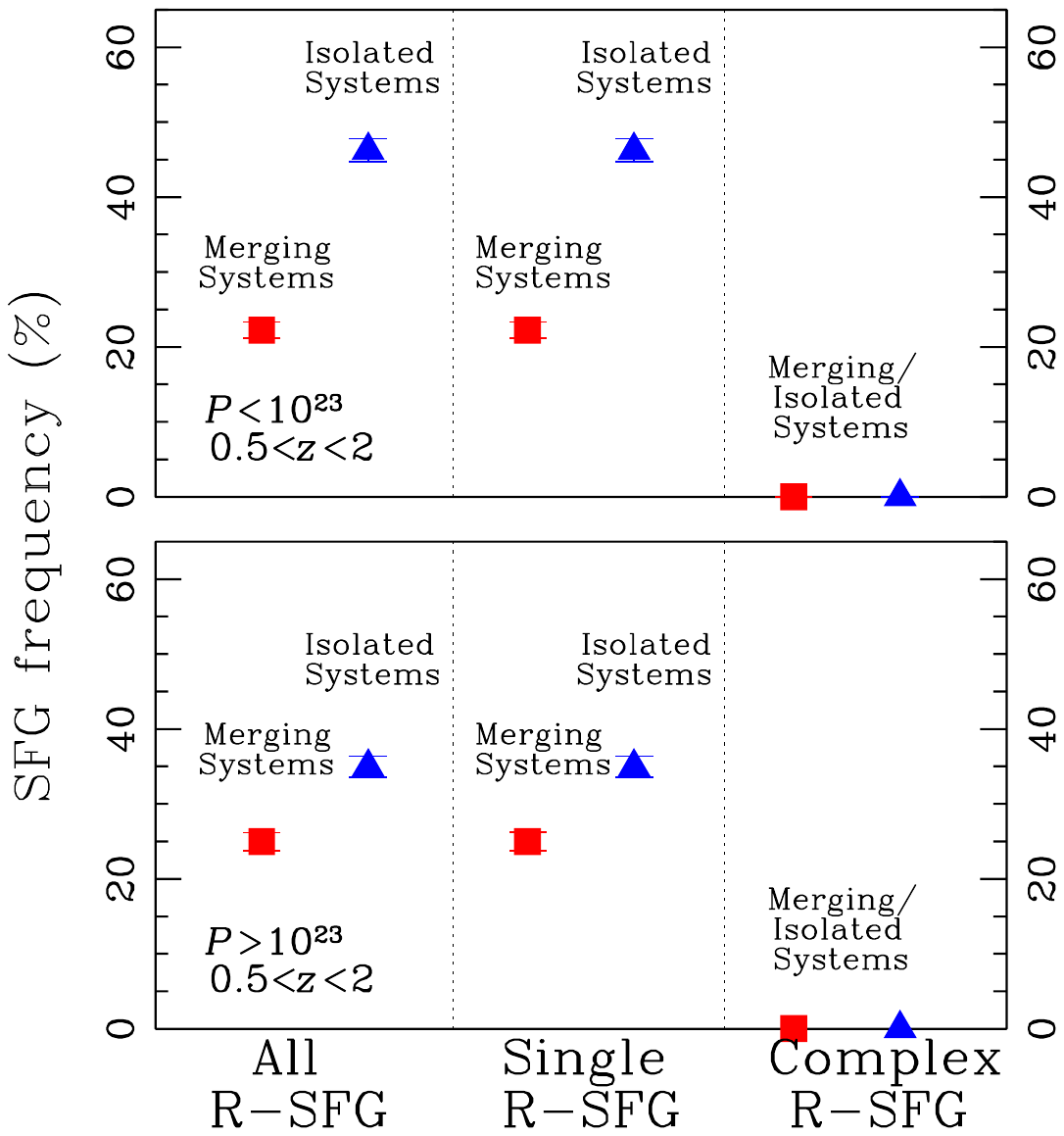}
\caption{Percentage of radio-selected star-forming galaxies (SFG) associated with either isolated galaxies or galaxy-galaxy mergers. The subdivision of each panel and the colour/marker styles are as in Fig.~3. The left-hand panel presents the two cases for sources in the $0.5<z<1$ (top) and $1<z<2$ (bottom) range, while the right-hand panel considers objects of different radio luminosities, respectively $P_{144\, \rm MHz}<$ (top) and $>$ (bottom) $10^{23}$\,W\,Hz$^{-1}$\,sr$^{-1}$. Error-bars represent Poissonian uncertainties.
\label{fig:hist_SF_mer}}
\end{center}
\end{figure*}
\begin{figure*}
\begin{center}
\vspace{-0.5 cm}  
\includegraphics[scale=0.40, viewport=-20 130 600 760] {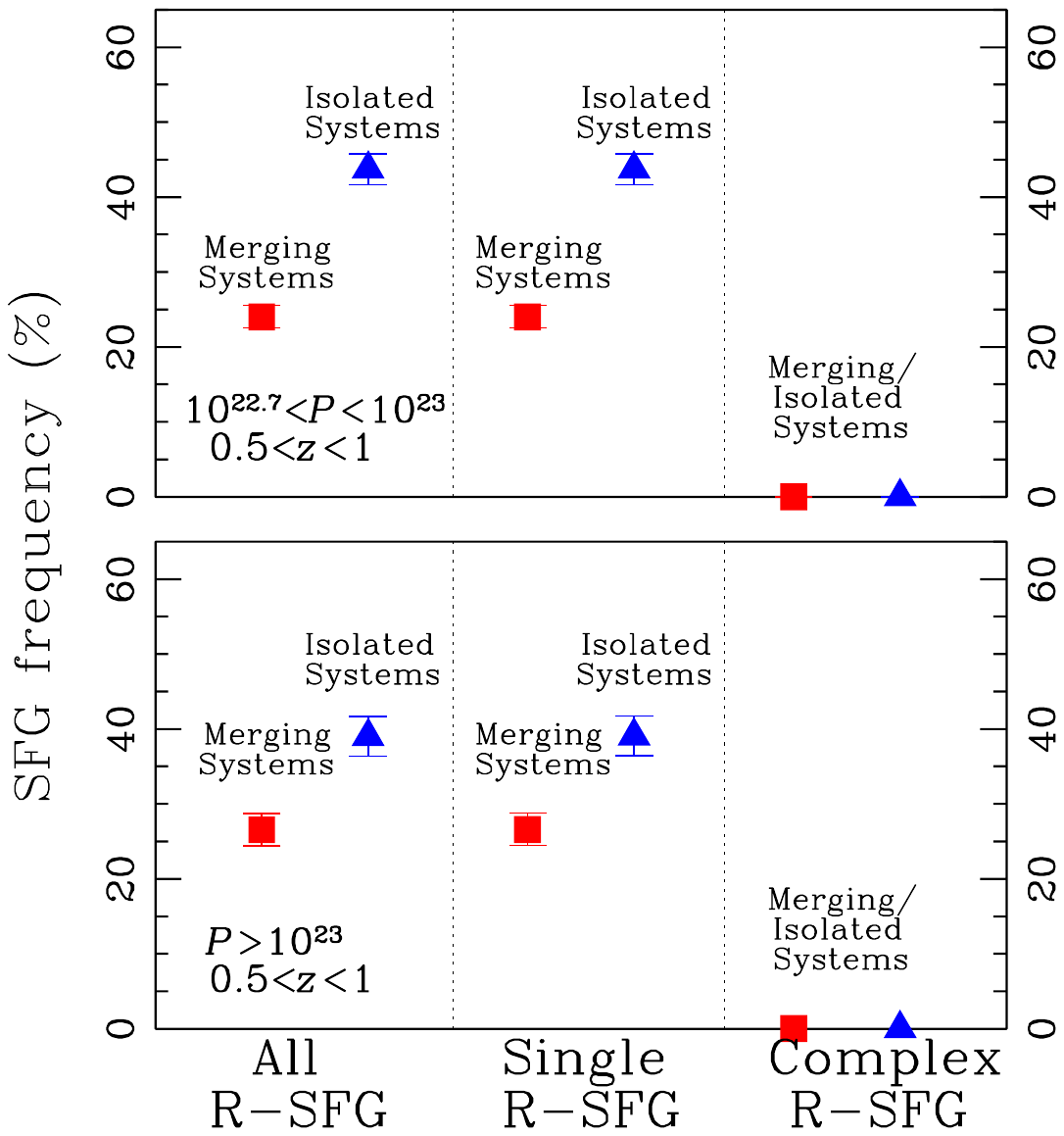}
\includegraphics[scale=0.4, viewport=-20 130 600 760] {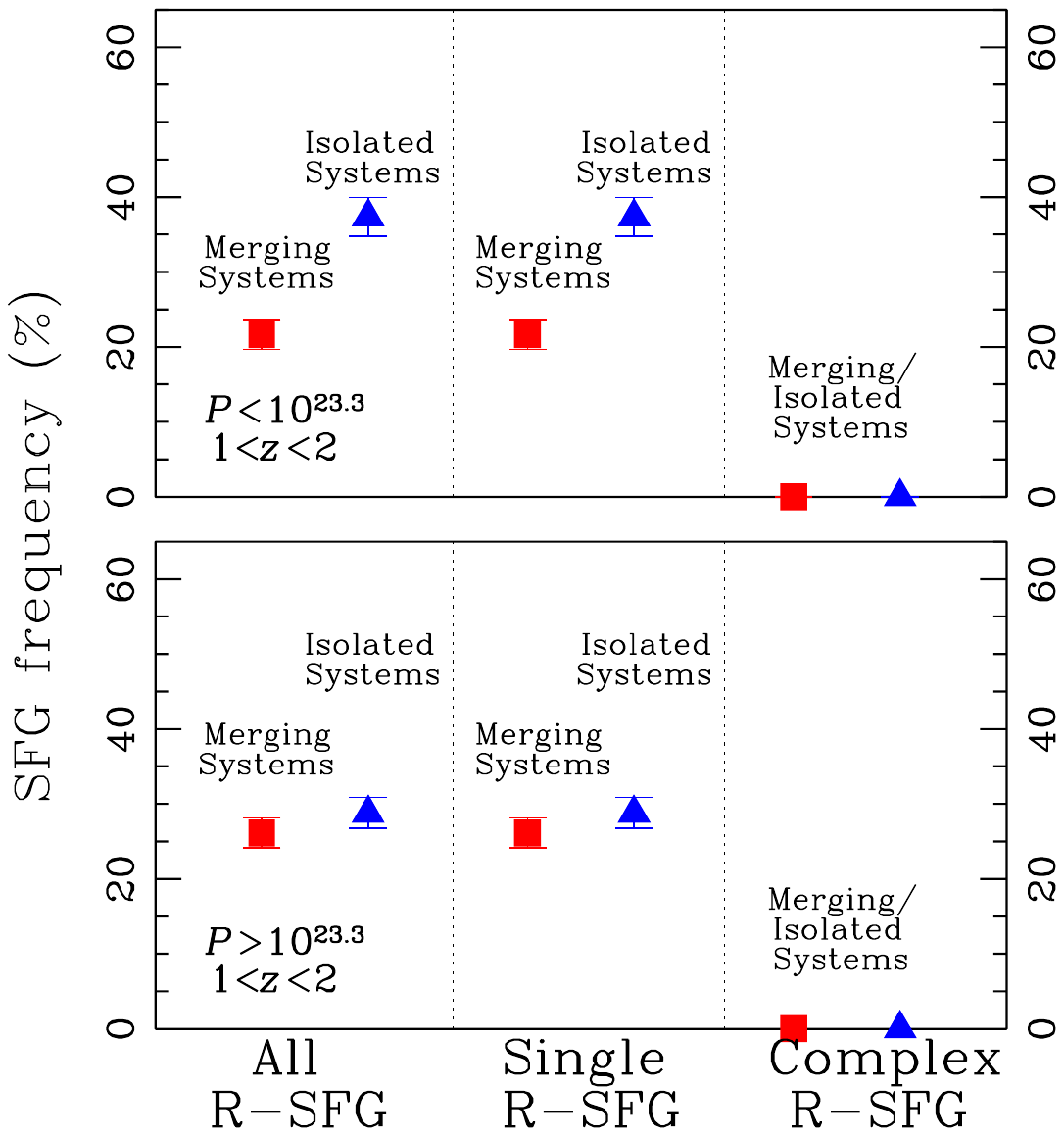}
\caption{Similar to Fig.~6, except that now different panels present different redshift/radio-luminosity (expressed in W\,Hz$^{-1}$\,sr$^{-1}$) combinations as shown (see text for details). 
\label{fig:hist_SF_mer_z_P}}
\end{center}
\end{figure*}

\section{Results for radio AGN and star-forming galaxies \label{results}}
In order to assess the relative role of mergers in the radio activity of galaxies, in Fig.~\ref{fig:hist_mer} we show the fraction of radio-selected AGN (left-hand panel) and star-forming galaxies (right-hand panel) respectively associated with isolated galaxies (\lq Isolated Systems\rq) or galaxy-galaxy mergers (\lq Merging Systems\rq).
The values plotted are obtained from those presented in Tables \ref{table1} and \ref{table2} and we also present a subdivision based on radio morphology for complex/extended or compact emission. Note that the figure does not include LOFAR sources associated with \Euclid galaxies that remain unclassified in the analysis of \citetalias{Q1-SP013}. A discussion of these sources is presented in Sect.~\ref{sys}.

A number of features can be appreciated here. The first, most striking one is the remarkably opposite behaviour observed for radio-selected AGN and SFG. Indeed, regardless of their radio morphology (i.e., whether compact or extended), AGN are preferentially associated with merging systems ($\SI {\sim 40}\percent$ versus $\SI {\sim 25}\percent$ in isolated galaxies for the entire population of radio AGN), while the opposite is true for star-forming galaxies ($\SI {\sim 25}\percent$ vs. $\SI {\sim 40}\percent$). This clear preference for radio-selected AGN to reside within merging systems becomes even more pronounced if we focus on those that present a complex radio morphology. In this case, we have that about $\SI {50}\percent$ of them are found associated with merging systems while only $\SI {\sim 15}\percent$ reside within isolated galaxies. Incidentally, we also note that the sample of SFG with available optical morphological information does not contain any object with complex radio emission, so the values in the rightmost section of the right-hand panel are both zero.\\
In the following we will separately analyse the cases for AGN and star-forming galaxies.

\subsection {AGN \label{AGN}}
The exquisite statistics provided by both LOFAR and \Euclid observations of the EDF-N allow us to investigate the behaviour presented in Fig.~\ref{fig:hist_mer} in greater detail as a function of redshift and radio luminosity, in order to study possible evolutionary trends.
In the case of radio AGN this is done in Fig.~\ref{fig:hist_AGN_mer}, left-hand panel for redshift and right-hand panel for radio luminosity. In all the considered cases (i.e., $0.5<z<1$, top-left; $1<z<2$, bottom-left; $P_{144\, \rm MHz}<10^{24}$\,W\,Hz$^{-1}$\,sr$^{-1}$ -- top-right; $P_{144\, \rm MHz}>10^{24}$\,W\,Hz$^{-1}$\,sr$^{-1}$ -- bottom-right) we note a preference for radio AGN to be associated with galaxy mergers, with a more marked tendency observed for AGN with extended radio morphologies, which are very seldom found within isolated galaxies. 
However, while such a behaviour is very evident in the case of relatively low redshift ($0.5<z<1$) or high luminosity ($P_{144\, \rm MHz}> 10^{24}$\,W\,Hz$^{-1}$\,sr$^{-1}$) AGN, this becomes less so for sources of relatively low radio luminosities ($P_{144\,\rm MHz}< 10^{24}$\,W\,Hz$^{-1}$\,sr$^{-1}$) or redshifts in the range $1<z<2$. Indeed, in these latter cases the small spread between the still higher fraction of radio AGN within merging systems and that of those residing within isolated galaxies is basically only driven by the subpopulation of complex/extended AGN (cf. bottom-left and top-right panels of Fig.~\ref{fig:hist_AGN_mer}).

In order to disentangle redshift and radio-luminosity dependence and understand whether the observed trends are mainly due to evolutionary effects (i.e., vary with cosmic time) or are rather more connected with radio activity, we have then considered the four subsets: 
\begin{enumerate}
\item $0.5<z<1$; $10^{23.35}< P_{144\, \rm MHz}< 10^{24.2}$ (top-left panel of Fig.~\ref{fig:hist_AGN_mer_z_P});
\item $1<z<2$; $10^{23.35}< P_{144\, \rm MHz}< 10^{24.2}$ (top-right panel of Fig.~\ref{fig:hist_AGN_mer_z_P});
\item $0.5<z<1$; $P_{144\, \rm MHz}> 10^{24.2}$  (bottom-left panel of Fig.~\ref{fig:hist_AGN_mer_z_P});
\item $1<z<2$; $P_{144\, \rm MHz}> 10^{24.2}$  (bottom-right panel of Fig.~\ref{fig:hist_AGN_mer_z_P}).
\end{enumerate}
Here all the radio luminosities are in W\,Hz$^{-1}$\,sr$^{-1}$ units. 

As in the previous case (see Fig.~\ref{fig:hist_AGN_mer}), we chose the redshift intervals to have a similar number of sources in the two low-$z$ and high-$z$ ranges and also to differentiate the regime where the redshift distribution of LOFAR sources with optical morphological information is independent of redshift from that where a dependence is instead observed (although merely driven by the optical $\IE=23.5$ cut -- see Fig.~\ref{fig:hist_z} and Sect.~\ref{combined}). The radio luminosity intervals were instead chosen to minimise possible incompleteness effects in the radio-luminosity distribution of LOFAR sources, while still allowing for a direct comparison between AGN of similar brightness at high and low redshift. Indeed, as shown in Fig.~\ref{fig:L_z}, in the range $0.5<z<1$ the AGN sample is complete above $P_{144\, \rm MHz}= 10^{23.35}$\,W\,Hz$^{-1}$\,sr$^{-1}$, while between $z=1$ and $z=2$ this is only true above $P_{144\, \rm MHz}= 10^{24.2}$\,W\,Hz$^{-1}$\,sr$^{-1}$.

Figure~\ref{fig:hist_AGN_mer_z_P} presents the results of our analysis, where it should be kept in mind that all the panels are complete in radio luminosity except for the top-right one (low luminosities and high redshifts). It appears clear that the observed global trend for radio AGN to be preferentially hosted within merging systems (see left-hand panel of Fig.~\ref{fig:hist_mer}) is mainly driven by high-luminosity sources at low redshifts. Indeed, in this case we find that $\SI{\sim 60}\percent$ of the radio AGN considered are associated with merging systems, while just a mere $\SI{\sim 10}\percent$ is hosted by isolated galaxies. Note that this result is roughly independent of the radio morphology of the sources. A smaller, but still significant difference  is observed in the low-luminosity/low-redshift regime (top-left panel of Fig.~\ref{fig:hist_AGN_mer_z_P}), where this time though the difference between the fractions of radio AGN associated with merging versus isolated systems is mainly due to AGN that present complex radio morphologies. The situation changes at high redshifts, where we do not notice any significant preference for radio AGN to appear within merging rather than isolated systems and -- if there is any hint (see bottom-right panel of Fig.~\ref{fig:hist_AGN_mer_z_P}) --  this seems to be only driven by AGN with extended or complex radio emission. 

As will be discussed at greater length in Sect.~\ref{conclusions}, we interpret the observed behaviour of radio AGN in the relatively local, $z<1$ Universe, with their need to accrete the gas necessary to trigger (radio) AGN activity from galaxy-galaxy encounters. This is even more mandatory in the case of intense radio activity (i.e., high luminosities) and/or extended morphologies, when radio emission permeates most of the surrounding environment. 
Indeed, in the low-redshift Universe galaxies are more gas-deprived than their $z>1$ counterparts (e.g., \citealt{genzel}), so it seems that the main if not almost the only (like in the case of bright AGN with complex radio morphologies) way to accrete the gas that then triggers radio activity is via local encounters. 

The situation appears different in the more distant Universe, since there does not seem to be any strong preference for radio AGN activity to happen in merging versus isolated systems, apart from possibly in the case of complex radio morphologies. This is likely due to a higher abundance of available gas in the galaxies (e.g., \citealt{tacconi}) which can be accreted by the AGN even without the need to rely on close encounters such as mergers. However, in this latter case we recall that we might just be seeing the tip of the \lq optically-bright AGN\rq \,iceberg since -- as already discussed before -- due to the \IE magnitude cut (cf. Fig.~\ref{fig:morph}), only a small fraction (from about $\SI{30}\percent$ at $z>1$ to about $\SI{15}\percent$ at $z\simeq 2$) of the general population of \Euclid galaxies is considered in the analysis of \citetalias{Q1-SP013} and therefore by our work. Additionally, as recently shown by \cite{degraaff}, at higher redshifts the observed trends may be weaker than the true ones because the identification of mergers becomes harder.

\subsection {Star-forming galaxies \label{SF}}
An analysis very similar to that presented in Sect.~\ref{AGN} can be performed for the population of radio-emitting star-forming galaxies. This is done in Figs.~\ref{fig:hist_SF_mer} and \ref{fig:hist_SF_mer_z_P}. In more detail, Fig.~\ref{fig:hist_SF_mer} presents the fraction of star-forming galaxies associated with either an isolated or a merging system in the two redshift intervals $0.5<z<1$ (top-left panel) and $1<z<2$ (bottom-left panel) and in the two luminosity ranges, $P_{144\, \rm MHz}< 10^{23}$\,W\,Hz$^{-1}$\,sr$^{-1}$ (top-right panel) and $P_{144\, \rm MHz}> 10^{23}$\,W\,Hz$^{-1}$\,sr$^{-1}$ (bottom-right panel). What clearly emerges from investigation of the trends is a net preference for star-forming galaxies to be hosted within isolated systems in all considered cases. However, we find that such a preference is more marked in the low-redshift regime and for low radio luminosities. Indeed, in both cases the percentage of star-forming galaxies associated with isolated systems is as high as $\SI{\sim 45}\percent$. Is this a cosmic evolutionary effect (outer driver) or is the observed trend due to different levels of the sources' radio activity (inner driver)?

In order to answer the above question, in the same fashion as in Sect.~\ref{AGN}, we have then subdivided our sample into four intervals: 
\begin{enumerate}
\item $0.5<z<1$; $10^{22.7}< P_{144\, \rm MHz}< 10^{23}$ (top-left panel of Fig.~\ref{fig:hist_SF_mer_z_P});
\item $1<z<2$; $10^{22.7}< P_{144\, \rm MHz}< 10^{23.3}$ (top-right panel of Fig.~\ref{fig:hist_SF_mer_z_P});
\item $0.5<z<1$; $P_{144\, \rm MHz}> 10^{23}$  (bottom-left panel of Fig.~\ref{fig:hist_SF_mer_z_P});
\item $1<z<2$; $P_{144\, \rm MHz}> 10^{23.3}$  (bottom-right panel of Fig.~\ref{fig:hist_SF_mer_z_P}). 
\end{enumerate}
Here all the radio luminosities are in W\,Hz$^{-1}$\,sr$^{-1}$ units. 

The choice for the redshift intervals follows the same argument presented in Sect.~\ref{AGN}.
The radio luminosity intervals were instead chosen to minimise possible incompleteness effects in the radio-luminosity distribution. Indeed, as shown in Fig.~\ref{fig:L_z}, in the range $0.5<z<1$ the SFG sample is complete above $P_{144\, \rm MHz}= 10^{22.7}$ \,W\,Hz$^{-1}$\,sr$^{-1}$, while between $z=1$ and $z=2$ this is only true above $P_{144\, \rm MHz}= 10^{23.3}$\,W\,Hz$^{-1}$\,sr$^{-1}$. The cut $P_{144\, \rm MHz}= 10^{23}$\,W\,Hz$^{-1}$\,sr$^{-1}$ in the low-redshift regime was instead chosen because above $P_{144\, \rm MHz}= 10^{23.3}$\,W\,Hz$^{-1}$\,sr$^{-1}$ 
no star-forming galaxy is found. 

Figure~\ref{fig:hist_SF_mer_z_P} presents the results of our analysis. As for the AGN case, all the panels are complete in radio luminosity except for the top-right one (low luminosities and high redshifts). Low-luminosity star-forming galaxies clearly prefer to be hosted by isolated systems. This is even more true in the low-redshift regime. Such a preference decreases up to vanishing at high redshifts for the radio-brighter sample. Interestingly, we notice that the fraction of SFG associated with merging systems remains the same ($\SI{\sim 20}\percent$) in all considered cases: it is the fraction of radio SFG residing within isolated galaxies that varies, going from $\SI{\sim 45}\percent$ for $0.5<z<1$ and $10^{22.7}$ W\,Hz$^{-1}$\,sr$^{-1}$ $< P_{144\, \rm MHz}< 10^{23.3}$  W\,Hz$^{-1}$\,sr$^{-1}$ to $\SI{\sim 30}\percent$ in the case of $1<z<2$ and  $P_{144\, \rm MHz}> 10^{23.3}$ W\,Hz$^{-1}$\,sr$^{-1}$. 

Given that the SFG sample is a mixed bag of pure star-forming galaxies and other low radio luminosity sources (see Sect.~\ref{AGNvsSF}), it is unclear whether the observed trend for a preference of faint radio SFG to reside within isolated systems is due to a real radio luminosity (i.e., star-forming activity since these two quantities are intimately connected, see e.g. \citealt{condon}) effect, in the sense that galaxies that are forming fewer stars are mostly isolated systems, or if there is a selection effect, since the higher-luminosity SFG sample is more contaminated not only by radio-quiet AGN but also by the low-luminosity tail of radio AGN (\citealt{maglio2002,mcalpine}) which, as seen in Sect.~\ref{AGN}, definitely prefer merging systems. 

Despite this uncertainty, the result that clearly emerges from our analysis is that --  at variance with the radio AGN case -- radio-emitting star-forming galaxies are preferentially isolated galaxies. This implies that local encounters, especially at low redshifts (see left-hand panel of Fig.~\ref{fig:hist_SF_mer_z_P}), {\it are not} the main source of gas availability or, in other words, that in the majority of cases reservoirs of gas already present within the galaxy suffice to fuel star formation. 

\subsection{Possible systematics \label{sys}}
The present work has been subject to extensive tests for possible systematics that could hamper the significance of our results. We took into account radio luminosity completeness effects (Sects.~\ref{AGN} and \ref{SF}), removed all those sources whose photometric redshifts were uncertain (Sect.~\ref{AGNvsSF}), worked with an optical catalogue complete in $\IE$ magnitude (Sect.~\ref{combined} and Appendix~\ref{morph}) and discussed possible contaminants in the sample of radio-selected star-forming galaxies. The next step is to investigate further systematics in the parent optical catalogue of \citetalias{Q1-SP013}. 

Indeed, as described in Sect.~\ref{combined}, this work excludes by construction all \Euclid galaxies that show point-like structures in VIS imaging. Since these objects are most likely isolated galaxies or quasars, their exclusion from the considered catalogue is expected to introduce biases in our analysis. In order to test for this effect, in the right-hand panel of Fig.~\ref{fig:morph} we plot the distribution of the values for the quantity $\texttt{MUMAX\_MINUS\_MAG}$, which characterises visual extension, for all \Euclid galaxies with magnitudes $\IE\leq 23.5$ as a function of redshift. As it is possible to appreciate, only a handful of sources (those below the dashed line) have been excluded from the \citetalias{Q1-SP013} analysis, and this exclusion does not depend on redshift. In more detail, only 61\,512 objects --   corresponding to $\SI{\sim 5.7}\percent$ of the parent optical sample -- were discarded. It follows that, even if we assume all these galaxies/quasars to be {\it bona-fide} isolated systems, their addition to the analysis presented in the previous sections does not affect any of the results obtained. 

As a further point, we have to consider all those \Euclid galaxies from the \citetalias{Q1-SP013} work that have classifier scores between 0.35 and 0.59 and for this reason are considered dubious or unclassified cases (see Sect.~{\ref{combined})}. Extensive tests are presented in the above paper (see their Section~5.1) to assess the properties of these sources and their effect on the mergers-non mergers statistics, with the conclusion that they are an almost even mixture of merger and non-merger systems and, as such, their addition is not expected to modify the results of either their or our study.
However, since the above tests have been performed for the whole population of \Euclid galaxies, while the present study concentrates on a subset of radio-detected sources for which different results might emerge (especially in the case radio-selected AGN) we have also tackled this issue. In order to do so, we have then visually investigated the \Euclid images for all the 310 radio AGN in our sample associated with an unclassified host. We were able to classify morphologies for 247 of them because some images were corrupted. 
In the case of compact radio AGN, we find that 110 out of a total of 198 (i.e., $\SI{\sim 56}\percent$) are associated with an isolated galaxy, while 88 (i.e., $\SI{\sim 44}\percent$) with a merging system. Concerning extended radio AGN, instead we 
find that their hosts are evenly split into mergers and non-mergers (respectively 24 mergers and 25 non-mergers).
These results confirm what found by \citetalias{Q1-SP013} about the even mixture of merger and non-merger systems in unclassified cases and ultimately assess the robustness of our analysis and conclusions.

Lastly, the impact of the classifier's imperfect performance on derived results was investigated by \citetalias{Q1-SP013} using Monte Carlo simulations. These show variations of about 6 percentage points for merging systems. It follows that our key findings remain qualitatively robust also in spite of potential CNN misclassifications.

\section{Discussion \label{conclusions}}
Thanks to the exquisite statistics provided by both LOFAR and \Euclid, we have presented the first large-scale study
of the connection between radio emission and its morphology (i.e., whether compact versus complex or extended) in radio sources and the merging properties of their host galaxies. By dividing the radio sample into AGN and star-forming galaxies, the data clearly indicates a net preference for radio AGN to reside within galaxies undergoing a merging event. This preference is more marked for AGN that show signatures of extended or complex radio emission: indeed, about $\SI{ 50}\percent$ of them are found to be associated with merging systems, while only $\SI{\sim 15}\percent$ are hosted by an isolated galaxy. The observed trend is primarily driven by AGN residing in the relatively local Universe ($z<1$), especially in the case of high  -- $P_{144\, \rm MHz}>10^{24}$\,W\,Hz$^{-1}$\,sr$^{-1}$ -- radio luminosities, for which we find $\SI{\sim 60}\percent$ in merging systems versus $\SI{\sim 10}\percent$ in isolated galaxies, regardless of radio morphology. This dichotomic behaviour is instead observed to disappear in the more distant Universe for redshifts $1<z<2$, where only bright AGN with extended radio emission still seem to prefer merging systems. 

The situation is totally reversed in the case of radio-emitting star-forming galaxies, which instead are mostly associated with isolated systems. This preference is more pronounced for low radio-luminosity/star-formation objects ($\SI{\sim 40}\percent$ versus $\SI{\sim 20}\percent$ in mergers for $P_{144\, \rm MHz}<10^{23}$\,W\,Hz$^{-1}$\,sr$^{-1}$) regardless of redshift, but also in the case of brighter galaxies in the relatively local, $0.5<z<1$, Universe. 
Incidentally, we also note that the fraction of mergers in non-active galaxies reported by \citetalias{Q1-SP013} -- $\SI{\sim 18}\percent$ in all considered cases -- is remarkably similar to the $\SI{\sim 20}\percent$ we find for our sample of radio-emitting star-forming galaxies at all luminosities and in all redshift intervals. This provides us with a reassuring check on the reliability of the results obtained also for this population.

Selection effects (discussed in Sect.~\ref{sys}) aside, we interpret the above result for AGN with their need to accrete outer gas from local encounters in order to trigger (radio) activity, especially in the case of extended radio emission such as hot-spots, lobes, and jets. This is mostly observed at $z<1$, since in the local Universe galaxies are more gas deprived with respect to epochs approaching cosmic noon (e.g., \citealt{genzel}) and an external supply might be needed to power radio emission of AGN origin (e.g., \citealt{heckman2}). Internal gas reservoirs instead seem sufficient to trigger star-formation within the majority of galaxies, which indeed prefer to be associated with isolated systems at all redshifts, more so for moderate-to-low star-formation rates (see, e.g., \citealt{gurkan} for the conversion between 144\,MHz radio luminosities and star-formation rates).    

Results similar to ours for AGN candidates selected in ways different from radio emission are presented in \citetalias{Q1-SP013}. These authors consider four main categories: extragalactic point-like X-rays sources from 4XMM-DR13, CSC2, and eROSITA surveys (\citealt{Q1-SP003}); spectroscopically-identified DESI QSOs (\citealt{siudek}); AGN candidates detected via Deep Learning-based image decomposition methods (\citealt{Q1-SP015}); and AGN candidates selected via their MIR colours (\citealt{assef}).

Bearing in mind that their merger and non-merger fractions are defined in a different way from ours,\footnote{They consider $f_{\rm merg/iso}=\frac{N_{\rm merger/isolated}}{N_{\rm merger}+{N_{\rm isolated}}}$, while we adopt $f_{\rm merg/iso}=\frac{N_{\rm merger/isolated}}{N_{\rm TOT}}$ which also takes into account the $\SI{36}\percent$ of non-classified cases in the parent \Euclid catalogue. Here we call $N_{\rm merger}$ and $N_{\rm isolated}$, respectively, the number of galaxies associated with a merging event or isolated, while $N_{\rm TOT}$ denotes the total number of galaxies.} 
and therefore have to be renormalised to be comparable with our results, we find that the percentage of radio AGN associated with merging systems found in this work is in very good agreement with the values reported by \citetalias{Q1-SP013}. More specifically, they obtain around $\SI{ 40}\percent$ (a little less for X-ray-selected AGN) in all analysed cases except for the AllWISE AGN sample with the lowest completeness level ($\SI{\sim 26}\percent$ -- see \citealt{assef} for more detail). 

We stress that the above agreement holds for all radio AGN considered in this work, regardless of their radio appearance while -- as already discussed at length -- the fraction of radio AGN with extended emission associated with merging systems is substantially higher, reaching values as high as $\SI{60}\percent$. Putting together all the pieces of information, this implies that {\it in order to have complex and/or extended radio emission more close encounters are needed with respect to the case of unresolved AGN emission, regardless of the observed wavelength}. The likely explanation for this can be found in the connection between availability of large amounts of gas and the ability of an AGN to produce radio emission that extends beyond the black hole neighborhood.

Is the above result only driven by radio morphology or does it hide an underlying dependence on the properties of gas accretion onto the central black hole (i.e., whether efficient or inefficient)? As we have seen in the Introduction, efficient accretion is thought to produce the population of HERGs that present strong emission lines in their optical spectra and radiate at all wavelengths including the radio band. On the other hand, inefficient accretion -- which only generates radio emission --  is expected to produce the population of LERGs that present weak or no emission lines in their optical spectra (e.g., \citealt{best2}).

A thorough study of the populations of LERGs and HERGs would require investigation of the spectra for the LOFAR AGN considered in this work, which unfortunately we do not yet have. However, we can still draw some conclusions by relying once again on radio morphology. Indeed, it is now assessed that there is an almost one-to-one correspondence between LERGs and FR\,I radio sources, in the sense that there are very few known
examples of FR\,I HERGs (e.g., \citealt{mingo19,gurkan21}). On the other hand, FR\,II's can come in both HERG and LERG types (e.g., \citealt{hine,chiaberge1,croston}), with FR\,II LERGs being much more frequent at lower radio luminosities (e.g., \citealt{mingo}).

With the aim of characterising our sample of radio AGN, two members of our team have individually visually investigated the LOFAR images of all 69 extended sources associated with merging systems (see Table~\ref{table1}). As can be seen in Figs.~\ref{fig:FRI_radio} and \ref{fig:FRII_radio}, where we show radio cutouts for a random selection of 48 of them, the AGN with extended radio emission considered in this work are an even mixture of FR\,I and FR\,II morphologies. Furthermore, the work of \cite{mingo} clearly shows that below $P_{144\, \rm MHz}\simeq 10^{25}$\,W\,Hz$^{-1}$\,sr$^{-1}$ the contribution of HERGs to the FR\,II population is negligible. Our sample only includes a handful of sources brighter than the above limit. Based on both radio appearance and radio luminosity, we can then safely conclude that the vast majority of our AGN (including those with extended radio emission) are indeed LERGs. 

The above discussion then indicates that -- at least within the population of LERGs -- the observed enhanced frequency of galaxy mergers associated with extended radio emission of AGN origin is a result that only depends on radio morphology. 
Some combined LOFAR and \Euclid images for FR\,I and FR\,II radio galaxies associated with merging systems are presented in Figs.~\ref{fig:FRI_mer_1} to \ref{fig:FRII_mer_2}.

For completeness, in Fig.~\ref{fig:extended_single} we also show LOFAR images for 20 out of the 21 extended radio AGN hosted by isolated galaxies (see Table~\ref{table1}). As in the previous case, here we also have that the sample is an even mixture of FR\,I and FR\,II morphologies. The only point worth mentioning is that three of the four most extended radio galaxies in our sample appear to be associated with isolated systems. However, the statistics is too poor to draw any solid conclusion out of this finding. 

One last point to notice is that the LOFAR images used in this work have a resolution of 6$\arcsec$. In the redshift range considered and for the adopted cosmology, this corresponds to physical scales between 36 kpc (at $z=0.5$) and 50 kpc (at $z=2$) and implies that the resolved radio emission we observe originates from the outskirts of a typical-sized galaxy and extends to larger scales. In the near future it will be interesting to compare the present results with those obtained using the LOFAR international stations, which will allow us to zoom-in on radio emission down to a resolution as high as $0\arcsecf3$ (e.g., \citealt{morabito}), capable to pierce right into the galaxy cores.

\begin{acknowledgements}
\AckQone
  \AckEC 

LB acknowledges support from the INAF Large Grant \lq AGN \& \Euclid: a
close entanglement\rq Ob. Fu. 01.05.23.01.14. Part of the research activities described in this paper were carried out with contribution of the Next Generation EU funds within the National Recovery and Resilience Plan (PNRR), Mission 4 -- Education and Research, Component 2 -- From Research to Business (M4C2), Investment Line 3.1 - Strengthening and creation of Research Infrastructures, Project IR0000034 – \lq STILES -- Strengthening the Italian Leadership in ELT and SKA\rq.
MM, LB, MB, and IP acknowledge support from INAF under the Large Grant 2022 funding scheme (project \lq Euclid and LOFAR Team up: a Unique Radio Window on Galaxy/AGN co-evolution\rq ).
\end{acknowledgements}

%
%

\bibliography{my, Euclid, Q1}

\begin{thebibliography}{91}
\expandafter\ifx\csname natexlab\endcsname\relax\def\natexlab#1{#1}\fi

\bibitem[{{Assef} {et~al.}(2018){Assef}, {Stern}, {Noirot}, {Jun}, {Cutri}, \& {Eisenhardt}}]{assef}
{Assef}, R.~J., {Stern}, D., {Noirot}, G., {et~al.} 2018, \apjs, 234, 23

\bibitem[{{Best} \& {Heckman}(2012)}]{best2}
{Best}, P.~N. \& {Heckman}, T.~M. 2012, \mnras, 421, 1569

\bibitem[{{Best} {et~al.}(2005){Best}, {Kauffmann}, {Heckman}, {Brinchmann}, {Charlot}, {Ivezi{\'c}}, \& {White}}]{best2005}
{Best}, P.~N., {Kauffmann}, G., {Heckman}, T.~M., {et~al.} 2005, \mnras, 362, 25

\bibitem[{{Best} {et~al.}(2023){Best}, {Kondapally}, {Williams}, {Cochrane}, {Duncan}, {Hale}, {Haskell}, {Ma{\l}ek}, {McCheyne}, {Smith}, {Wang}, {Botteon}, {Bonato}, {Bondi}, {Calistro Rivera}, {Gao}, {G{\"u}rkan}, {Hardcastle}, {Jarvis}, {Mingo}, {Miraghaei}, {Morabito}, {Nisbet}, {Prandoni}, {R{\"o}ttgering}, {Sabater}, {Shimwell}, {Tasse}, \& {van Weeren}}]{best23}
{Best}, P.~N., {Kondapally}, R., {Williams}, W.~L., {et~al.} 2023, \mnras, 523, 1729

\bibitem[{{Bichang'a} {et~al.}(2024){Bichang'a}, {Kaviraj}, {Lazar}, {Jackson}, {Das}, {Smith}, {Watkins}, \& {Martin}}]{bichang}
{Bichang'a}, B., {Kaviraj}, S., {Lazar}, I., {et~al.} 2024, \mnras, 532, 613

\bibitem[{{Bickley} {et~al.}(2024){Bickley}, {Ellison}, {Salvato}, {Salim}, {Patton}, {Merloni}, {Byrne-Mamahit}, {Ferreira}, \& {Wilkinson}}]{bickley}
{Bickley}, R.~W., {Ellison}, S.~L., {Salvato}, M., {et~al.} 2024, \mnras, 533, 3068

\bibitem[{Bisigello {et~al.}(2025)Bisigello, Giulietti, Prandoni, Bondi, Bonato, Magliocchetti, Rottgering, Morabito, \& White}]{bisigello}
Bisigello, L., Giulietti, M., Prandoni, I., {et~al.} 2025, The Open Journal of Astrophysics, 8

\bibitem[{{Bondi} {et~al.}(2024){Bondi}, {Scaramella}, {Zamorani}, {Ciliegi}, {Vitello}, {Arias}, {Best}, {Bonato}, {Botteon}, {Brienza}, {Brunetti}, {Hardcastle}, {Magliocchetti}, {Massaro}, {Morabito}, {Pentericci}, {Prandoni}, {R{\"o}ttgering}, {Shimwell}, {Tasse}, {van Weeren}, \& {White}}]{bondi}
{Bondi}, M., {Scaramella}, R., {Zamorani}, G., {et~al.} 2024, \aap, 683, A179

\bibitem[{{Breiding} {et~al.}(2024){Breiding}, {Chiaberge}, {Lambrides}, {Meyer}, {Willner}, {Hilbert}, {Haas}, {Miley}, {Perlman}, {Barthel}, {O'Dea}, {Capetti}, {Wilkes}, {Baum}, {Macchetto}, {Sparks}, {Tremblay}, \& {Norman}}]{breiding}
{Breiding}, P., {Chiaberge}, M., {Lambrides}, E., {et~al.} 2024, \apj, 963, 91

\bibitem[{{Buat} {et~al.}(2018){Buat}, {Boquien}, {Ma{\l}ek}, {Corre}, {Salas}, {Roehlly}, {Shirley}, \& {Efstathiou}}]{buat}
{Buat}, V., {Boquien}, M., {Ma{\l}ek}, K., {et~al.} 2018, \aap, 619, A135

\bibitem[{{Calistro Rivera} {et~al.}(2024){Calistro Rivera}, {Alexander}, {Harrison}, {Fawcett}, {Best}, {Williams}, {Hardcastle}, {Rosario}, {Smith}, {Arnaudova}, {Escott}, {G{\"u}rkan}, {Kondapally}, {Miley}, {Morabito}, {Petley}, {Prandoni}, {R{\"o}ttgering}, \& {Yue}}]{calistro2024}
{Calistro Rivera}, G., {Alexander}, D.~M., {Harrison}, C.~M., {et~al.} 2024, \aap, 691, A191

\bibitem[{{Calistro Rivera} {et~al.}(2017){Calistro Rivera}, {Williams}, {Hardcastle}, {Duncan}, {R{\"o}ttgering}, {Best}, {Br{\"u}ggen}, {Chy{\.z}y}, {Conselice}, {de Gasperin}, {Engels}, {G{\"u}rkan}, {Intema}, {Jarvis}, {Mahony}, {Miley}, {Morabito}, {Prandoni}, {Sabater}, {Smith}, {Tasse}, {van der Werf}, \& {White}}]{calistro}
{Calistro Rivera}, G., {Williams}, W.~L., {Hardcastle}, M.~J., {et~al.} 2017, \mnras, 469, 3468

\bibitem[{{Capetti} {et~al.}(2022){Capetti}, {Brienza}, {Balmaverde}, {Best}, {Baldi}, {Drabent}, {G{\"u}rkan}, {Rottgering}, {Tasse}, \& {Webster}}]{capetti}
{Capetti}, A., {Brienza}, M., {Balmaverde}, B., {et~al.} 2022, \aap, 660, A93

\bibitem[{{Chiaberge} {et~al.}(2000){Chiaberge}, {Capetti}, \& {Celotti}}]{chiaberge1}
{Chiaberge}, M., {Capetti}, A., \& {Celotti}, A. 2000, \aap, 355, 873

\bibitem[{{Chiaberge} {et~al.}(2015){Chiaberge}, {Gilli}, {Lotz}, \& {Norman}}]{chiaberge}
{Chiaberge}, M., {Gilli}, R., {Lotz}, J.~M., \& {Norman}, C. 2015, \apj, 806, 147

\bibitem[{{Comerford} {et~al.}(2024){Comerford}, {Nevin}, {Negus}, {Barrows}, {Eracleous}, {M{\"u}ller-S{\'a}nchez}, {Roy}, {Stemo}, {Storchi-Bergmann}, \& {Wylezalek}}]{comerford}
{Comerford}, J.~M., {Nevin}, R., {Negus}, J., {et~al.} 2024, \apj, 963, 53

\bibitem[{Condon(1992)}]{condon}
Condon, J.~J. 1992, ARA\&A, 30, 575

\bibitem[{{Croston} {et~al.}(2018){Croston}, {Ineson}, \& {Hardcastle}}]{croston}
{Croston}, J.~H., {Ineson}, J., \& {Hardcastle}, M.~J. 2018, \mnras, 476, 1614

\bibitem[{{de Graaff} {et~al.}(2025){de Graaff}, {Margalef-Bentabol}, {Wang}, {La Marca}, {Pearson}, {Rodriguez-Gomez}, \& {Walmsley}}]{degraaff}
{de Graaff}, R., {Margalef-Bentabol}, B., {Wang}, L., {et~al.} 2025, \aap, 697, A207

\bibitem[{{Delhaize} {et~al.}(2017){Delhaize}, {Smol{\v{c}}i{\'c}}, {Delvecchio}, {Novak}, {Sargent}, {Baran}, {Magnelli}, {Zamorani}, {Schinnerer}, {Murphy}, {Aravena}, {Berta}, {Bondi}, {Capak}, {Carilli}, {Ciliegi}, {Civano}, {Ilbert}, {Karim}, {Laigle}, {Le F{\`e}vre}, {Marchesi}, {McCracken}, {Salvato}, {Seymour}, \& {Tasca}}]{delhaize2017}
{Delhaize}, J., {Smol{\v{c}}i{\'c}}, V., {Delvecchio}, I., {et~al.} 2017, \aap, 602, A4

\bibitem[{{DESI Collaboration} {et~al.}(2024){DESI Collaboration}, {Adame}, {Aguilar}, {Ahlen}, {Alam}, {Aldering}, {Alexander}, {Alfarsy}, {Allende Prieto}, {Alvarez}, {Alves}, {Anand}, {Andrade-Oliveira}, {Armengaud}, {Asorey}, {Avila}, {Aviles}, {Bailey}, {Balaguera-Antol{\'\i}nez}, {Ballester}, {Baltay}, {Bault}, {Bautista}, {Behera}, {Beltran}, {BenZvi}, {Beraldo e Silva}, {Bermejo-Climent}, {Berti}, {Besuner}, {Beutler}, {Bianchi}, {Blake}, {Blum}, {Bolton}, {Brieden}, {Brodzeller}, {Brooks}, {Brown}, {Buckley-Geer}, {Burtin}, {Cabayol-Garcia}, {Cai}, {Canning}, {Cardiel-Sas}, {Carnero Rosell}, {Castander}, {Cervantes-Cota}, {Chabanier}, {Chaussidon}, {Chaves-Montero}, {Chen}, {Chen}, {Chuang}, {Claybaugh}, {Cole}, {Cooper}, {Cuceu}, {Davis}, {Dawson}, {de Belsunce}, {de la Cruz}, {de la Macorra}, {Della Costa}, {de Mattia}, {Demina}, {Demirbozan}, {DeRose}, {Dey}, {Dey}, {Dhungana}, {Ding}, {Ding}, {Doel}, {Doshi}, {Douglass}, {Edge}, {Eftekharzadeh}, {Eisenstein}, {Elliott}, {Ereza}, {Escoffier},
  {Fagrelius}, {Fan}, {Fanning}, {Fawcett}, {Ferraro}, {Flaugher}, {Font-Ribera}, {Forero-Romero}, {Forero-S{\'a}nchez}, {Frenk}, {G{\"a}nsicke}, {Garc{\'\i}a}, {Garc{\'\i}a-Bellido}, {Garcia-Quintero}, {Garrison}, {Gil-Mar{\'\i}n}, {Golden-Marx}, {Gontcho A Gontcho}, {Gonzalez-Morales}, {Gonzalez-Perez}, {Gordon}, {Graur}, {Green}, {Gruen}, {Guy}, {Hadzhiyska}, {Hahn}, {Han}, {Hanif}, {Herrera-Alcantar}, {Honscheid}, {Hou}, {Howlett}, {Huterer}, {Ir{\v{s}}i{\v{c}}}, {Ishak}, {Jacques}, {Jana}, {Jiang}, {Jimenez}, {Jing}, {Joudaki}, {Joyce}, {Jullo}, {Juneau}, {Kara{\c{c}}ayl{\i}}, {Karim}, {Kehoe}, {Kent}, {Khederlarian}, {Kim}, {Kirkby}, {Kisner}, {Kitaura}, {Kizhuprakkat}, {Kneib}, {Koposov}, {Kov{\'a}cs}, {Kremin}, {Krolewski}, {L'Huillier}, {Lahav}, {Lambert}, {Lamman}, {Lan}, {Landriau}, {Lang}, {Lange}, {Lasker}, {Leauthaud}, {Le Guillou}, {Levi}, {Li}, {Linder}, {Lyons}, {Magneville}, {Manera}, {Manser}, {Margala}, {Martini}, {McDonald}, {Medina}, {Medina-Varela}, {Meisner}, {Mena-Fern{\'a}ndez},
  {Meneses-Rizo}, {Mezcua}, {Miquel}, {Montero-Camacho}, {Moon}, {Moore}, {Moustakas}, {Mueller}, {Mundet}, {Mu{\~n}oz-Guti{\'e}rrez}, {Myers}, {Nadathur}, {Napolitano}, {Neveux}, {Newman}, {Nie}, {Nikutta}, {Niz}, {Norberg}, {Noriega}, {Paillas}, {Palanque-Delabrouille}, {Palmese}, {Pan}, {Parkinson}, {Penmetsa}, {Percival}, {P{\'e}rez-Fern{\'a}ndez}, {P{\'e}rez-R{\`a}fols}, {Pieri}, {Poppett}, {Porredon}, \& {Pothier}}]{desi}
{DESI Collaboration}, {Adame}, A.~G., {Aguilar}, J., {et~al.} 2024, \aj, 168, 58

\bibitem[{{Duncan} {et~al.}(2021){Duncan}, {Kondapally}, {Brown}, {Bonato}, {Best}, {R{\"o}ttgering}, {Bondi}, {Bowler}, {Cochrane}, {G{\"u}rkan}, {Hardcastle}, {Jarvis}, {Kunert-Bajraszewska}, {Leslie}, {Ma{\l}ek}, {Morabito}, {O'Sullivan}, {Prandoni}, {Sabater}, {Shimwell}, {Smith}, {Wang}, {Wo{\l}owska}, \& {Tasse}}]{Duncan}
{Duncan}, K.~J., {Kondapally}, R., {Brown}, M.~J.~I., {et~al.} 2021, \aap, 648, A4

\bibitem[{{Ellison} {et~al.}(2015){Ellison}, {Patton}, \& {Hickox}}]{ellison}
{Ellison}, S.~L., {Patton}, D.~R., \& {Hickox}, R.~C. 2015, \mnras, 451, L35

\bibitem[{{Ellison} {et~al.}(2019){Ellison}, {Viswanathan}, {Patton}, {Bottrell}, {McConnachie}, {Gwyn}, \& {Cuillandre}}]{Ellison19}
{Ellison}, S.~L., {Viswanathan}, A., {Patton}, D.~R., {et~al.} 2019, \mnras, 487, 2491

\bibitem[{{Er{\'o}stegui} {et~al.}(2025){Er{\'o}stegui}, {Mezcua}, {Siudek}, {Dom{\'\i}nguez S{\'a}nchez}, \& {Rodr{\'\i}guez Morales}}]{erostegui}
{Er{\'o}stegui}, A., {Mezcua}, M., {Siudek}, M., {Dom{\'\i}nguez S{\'a}nchez}, H., \& {Rodr{\'\i}guez Morales}, V. 2025, arXiv e-prints, arXiv:2503.03742

\bibitem[{{Euclid Collaboration: Aussel} {et~al.}(2025){Euclid Collaboration: Aussel}, {Tereno}, {Schirmer}, {et~al.}}]{Q1-TP001}
{Euclid Collaboration: Aussel}, H., {Tereno}, I., {Schirmer}, M., {et~al.} 2025, A\&A, submitted (Euclid Q1 SI), arXiv:2503.15302

\bibitem[{{Euclid Collaboration: Cropper} {et~al.}(2025){Euclid Collaboration: Cropper}, {Al-Bahlawan}, {Amiaux}, {et~al.}}]{EuclidSkyVIS}
{Euclid Collaboration: Cropper}, M., {Al-Bahlawan}, A., {Amiaux}, J., {et~al.} 2025, A\&A, 697, A2

\bibitem[{{Euclid Collaboration: Huertas-Company} {et~al.}(2025){Euclid Collaboration: Huertas-Company}, {Walmsley}, {Siudek}, {et~al.}}]{Q1-SP043}
{Euclid Collaboration: Huertas-Company}, M., {Walmsley}, M., {Siudek}, M., {et~al.} 2025, A\&A, submitted (Euclid Q1 SI), arXiv:2503.15311

\bibitem[{{Euclid Collaboration: La Marca} {et~al.}(2025){Euclid Collaboration: La Marca}, {Wang}, {Margalef-Bentabol}, {et~al.}}]{Q1-SP013}
{Euclid Collaboration: La Marca}, A., {Wang}, L., {Margalef-Bentabol}, B., {et~al.} 2025, A\&A, submitted (Euclid Q1 SI), arXiv:2503.15317

\bibitem[{{Euclid Collaboration: Margalef-Bentabol} {et~al.}(2025){Euclid Collaboration: Margalef-Bentabol}, {Wang}, {La Marca}, {et~al.}}]{Q1-SP015}
{Euclid Collaboration: Margalef-Bentabol}, B., {Wang}, L., {La Marca}, A., {et~al.} 2025, A\&A, submitted (Euclid Q1 SI), arXiv:2503.15318

\bibitem[{{Euclid Collaboration: Matamoro Zatarain} {et~al.}(2025){Euclid Collaboration: Matamoro Zatarain}, {Fotopoulou}, {Ricci}, {et~al.}}]{Q1-SP027}
{Euclid Collaboration: Matamoro Zatarain}, T., {Fotopoulou}, S., {Ricci}, F., {et~al.} 2025, A\&A, submitted (Euclid Q1 SI), arXiv:2503.15320

\bibitem[{{Euclid Collaboration: McCracken} {et~al.}(2025){Euclid Collaboration: McCracken}, {Benson}, {Dolding}, {et~al.}}]{Q1-TP002}
{Euclid Collaboration: McCracken}, H.~J., {Benson}, K., {Dolding}, C., {et~al.} 2025, A\&A, submitted (Euclid Q1 SI), arXiv:2503.15303

\bibitem[{{Euclid Collaboration: Mellier} {et~al.}(2025){Euclid Collaboration: Mellier}, {Abdurro'uf}, {Acevedo~Barroso}, {et~al.}}]{EuclidSkyOverview}
{Euclid Collaboration: Mellier}, Y., {Abdurro'uf}, {Acevedo~Barroso}, J., {et~al.} 2025, A\&A, 697, A1

\bibitem[{{Euclid Collaboration: Moneti} {et~al.}(2022){Euclid Collaboration: Moneti}, {McCracken}, {Shuntov}, {Kauffmann}, {Capak}, {Davidzon}, {Ilbert}, {Scarlata}, {Toft}, {Weaver}, {Chary}, {Cuby}, {Faisst}, {Masters}, {McPartland}, {Mobasher}, {Sanders}, {Scaramella}, {Stern}, {Szapudi}, {Teplitz}, {Zalesky}, {Amara}, {Auricchio}, {Bodendorf}, {Bonino}, {Branchini}, {Brau-Nogue}, {Brescia}, {Brinchmann}, {Capobianco}, {Carbone}, {Carretero}, {Castander}, {Castellano}, {Cavuoti}, {Cimatti}, {Cledassou}, {Congedo}, {Conselice}, {Conversi}, {Copin}, {Corcione}, {Costille}, {Cropper}, {Da Silva}, {Degaudenzi}, {Douspis}, {Dubath}, {Duncan}, {Dupac}, {Dusini}, {Farrens}, {Ferriol}, {Fosalba}, {Frailis}, {Franceschi}, {Fumana}, {Garilli}, {Gillis}, {Giocoli}, {Granett}, {Grazian}, {Grupp}, {Haugan}, {Hoekstra}, {Holmes}, {Hormuth}, {Hudelot}, {Jahnke}, {Kermiche}, {Kiessling}, {Kilbinger}, {Kitching}, {Kohley}, {K{\"u}mmel}, {Kunz}, {Kurki-Suonio}, {Ligori}, {Lilje}, {Lloro}, {Maiorano}, {Mansutti},
  {Marggraf}, {Markovic}, {Marulli}, {Massey}, {Maurogordato}, {Meneghetti}, {Merlin}, {Meylan}, {Moresco}, {Moscardini}, {Munari}, {Niemi}, {Padilla}, {Paltani}, {Pasian}, {Pedersen}, {Pires}, {Poncet}, {Popa}, {Pozzetti}, {Raison}, {Rebolo}, {Rhodes}, {Rix}, {Roncarelli}, {Rossetti}, {Saglia}, {Schneider}, {Secroun}, {Seidel}, {Serrano}, {Sirignano}, {Sirri}, {Stanco}, {Tallada-Cresp{\'\i}}, {Taylor}, {Tereno}, {Toledo-Moreo}, {Torradeflot}, {Wang}, {Welikala}, {Weller}, {Zamorani}, {Zoubian}, {Andreon}, {Bardelli}, {Camera}, {Graci{\'a}-Carpio}, {Medinaceli}, {Mei}, {Polenta}, {Romelli}, {Sureau}, {Tenti}, {Vassallo}, {Zacchei}, {Zucca}, {Baccigalupi}, {Balaguera-Antol{\'\i}nez}, {Bernardeau}, {Biviano}, {Bolzonella}, {Bozzo}, {Burigana}, {Cabanac}, {Cappi}, {Carvalho}, {Casas}, {Castignani}, {Colodro-Conde}, {Coupon}, {Courtois}, {Di Ferdinando}, {Farina}, {Finelli}, {Flose-Reimberg}, {Fotopoulou}, {Galeotta}, {Ganga}, {Garcia-Bellido}, {Gaztanaga}, {Gozaliasl}, {Hook}, {Joachimi}, {Kansal}, {Keihanen},
  {Kirkpatrick}, {Lindholm}, {Mainetti}, {Maino}, {Maoli}, {Martinelli}, {Martinet}, {Maturi}, {Metcalf}, {Morgante}, {Morisset}, {Nucita}, {Patrizii}, {Potter}, {Renzi}, {Riccio}, {S{\'a}nchez}, {Sapone}, {Schirmer}, {Schultheis}, {Scottez}, {Sefusatti}, {Teyssier}, {Tubio}, {Tutusaus}, {Valiviita}, {Viel}, \& {Hildebrandt}}]{moneti}
{Euclid Collaboration: Moneti}, A., {McCracken}, H.~J., {Shuntov}, M., {et~al.} 2022, \aap, 658, A126

\bibitem[{{Euclid Collaboration: Quilley} {et~al.}(2025){Euclid Collaboration: Quilley}, {Damjanov}, {de Lapparent}, {et~al.}}]{Q1-SP040}
{Euclid Collaboration: Quilley}, L., {Damjanov}, I., {de Lapparent}, V., {et~al.} 2025, A\&A, submitted (Euclid Q1 SI), arXiv:2503.15309

\bibitem[{{Euclid Collaboration: Roster} {et~al.}(2025){Euclid Collaboration: Roster}, {Salvato}, {Buchner}, {et~al.}}]{Q1-SP003}
{Euclid Collaboration: Roster}, W., {Salvato}, M., {Buchner}, J., {et~al.} 2025, A\&A, submitted (Euclid Q1 SI), arXiv:2503.15316

\bibitem[{{Euclid Collaboration: Siudek} {et~al.}(2025){Euclid Collaboration: Siudek}, {Huertas-Company}, {Smith}, {et~al.}}]{Q1-SP049}
{Euclid Collaboration: Siudek}, M., {Huertas-Company}, M., {Smith}, M., {et~al.} 2025, A\&A, submitted (Euclid Q1 SI), arXiv:2503.15312

\bibitem[{{Euclid Collaboration: Stevens} {et~al.}(2025){Euclid Collaboration: Stevens}, {Fotopoulou}, {Bremer}, {et~al.}}]{Q1-SP009}
{Euclid Collaboration: Stevens}, G., {Fotopoulou}, S., {Bremer}, M.~N., {et~al.} 2025, A\&A, submitted (Euclid Q1 SI), arXiv:2503.15321

\bibitem[{{Euclid Collaboration: Tucci} {et~al.}(2025){Euclid Collaboration: Tucci}, {Paltani}, {Hartley}, {et~al.}}]{Q1-TP005}
{Euclid Collaboration: Tucci}, M., {Paltani}, S., {Hartley}, W.~G., {et~al.} 2025, A\&A, accepted (Euclid Q1 SI), arXiv:2503.15306

\bibitem[{{Euclid Collaboration: Walmsley} {et~al.}(2025){Euclid Collaboration: Walmsley}, {Huertas-Company}, {Quilley}, {et~al.}}]{Q1-SP047}
{Euclid Collaboration: Walmsley}, M., {Huertas-Company}, M., {Quilley}, L., {et~al.} 2025, A\&A, submitted (Euclid Q1 SI), arXiv:2503.15310

\bibitem[{{Euclid Quick Release Q1}(2025)}]{Q1cite}
{Euclid Quick Release Q1}. 2025, \url{https://doi.org/10.57780/esa-2853f3b}

\bibitem[{{Fanaroff} \& {Riley}(1974)}]{FR}
{Fanaroff}, B.~L. \& {Riley}, J.~M. 1974, \mnras, 167, 31P

\bibitem[{{Gao} {et~al.}(2020){Gao}, {Wang}, {Pearson}, {Gordon}, {Holwerda}, {Hopkins}, {Brown}, {Bland-Hawthorn}, \& {Owers}}]{gao}
{Gao}, F., {Wang}, L., {Pearson}, W.~J., {et~al.} 2020, \aap, 637, A94

\bibitem[{{Gehrels}(1986)}]{gehrels}
{Gehrels}, N. 1986, \apj, 303, 336

\bibitem[{{Genzel} {et~al.}(2015){Genzel}, {Tacconi}, {Lutz}, {Saintonge}, {Berta}, {Magnelli}, {Combes}, {Garc{\'\i}a-Burillo}, {Neri}, {Bolatto}, {Contini}, {Lilly}, {Boissier}, {Boone}, {Bouch{\'e}}, {Bournaud}, {Burkert}, {Carollo}, {Colina}, {Cooper}, {Cox}, {Feruglio}, {F{\"o}rster Schreiber}, {Freundlich}, {Gracia-Carpio}, {Juneau}, {Kovac}, {Lippa}, {Naab}, {Salome}, {Renzini}, {Sternberg}, {Walter}, {Weiner}, {Weiss}, \& {Wuyts}}]{genzel}
{Genzel}, R., {Tacconi}, L.~J., {Lutz}, D., {et~al.} 2015, \apj, 800, 20

\bibitem[{{Gordon} {et~al.}(2019){Gordon}, {Pimbblet}, {Kaviraj}, {Owers}, {O'Dea}, {Walmsley}, {Baum}, {Crossett}, {Fraser-McKelvie}, {Lintott}, \& {Pierce}}]{gordon}
{Gordon}, Y.~A., {Pimbblet}, K.~A., {Kaviraj}, S., {et~al.} 2019, \apj, 878, 88

\bibitem[{{G{\"u}rkan} {et~al.}(2021){G{\"u}rkan}, {Croston}, {Hardcastle}, {Mahatma}, {Mingo}, \& {Williams}}]{gurkan21}
{G{\"u}rkan}, G., {Croston}, J., {Hardcastle}, M.~J., {et~al.} 2021, Galaxies, 10, 2

\bibitem[{Gürkan {et~al.}(2018)Gürkan, Hardcastle, Smith, Best, Bourne, Calistro-Rivera, Heald, Jarvis, Prandoni, Röttgering, Sabater, Shimwell, Tasse, \& Williams}]{gurkan}
Gürkan, G., Hardcastle, M.~J., Smith, D. J.~B., {et~al.} 2018, MNRAS, 475, 3010

\bibitem[{{Hardcastle} \& {Croston}(2020)}]{hardcastle}
{Hardcastle}, M.~J. \& {Croston}, J.~H. 2020, \nar, 88, 101539

\bibitem[{{Heckman} {et~al.}(2024){Heckman}, {Roy}, {Best}, \& {Kondapally}}]{heckman2}
{Heckman}, T.~M., {Roy}, N., {Best}, P.~N., \& {Kondapally}, R. 2024, \apj, 977, 125

\bibitem[{{Heckman} {et~al.}(1986){Heckman}, {Smith}, {Baum}, {van Breugel}, {Miley}, {Illingworth}, {Bothun}, \& {Balick}}]{heckman}
{Heckman}, T.~M., {Smith}, E.~P., {Baum}, S.~A., {et~al.} 1986, \apj, 311, 526

\bibitem[{{Hickox} {et~al.}(2009){Hickox}, {Jones}, {Forman}, {Murray}, {Kochanek}, {Eisenstein}, {Jannuzi}, {Dey}, {Brown}, {Stern}, {Eisenhardt}, {Gorjian}, {Brodwin}, {Narayan}, {Cool}, {Kenter}, {Caldwell}, \& {Anderson}}]{hickox}
{Hickox}, R.~C., {Jones}, C., {Forman}, W.~R., {et~al.} 2009, \apj, 696, 891

\bibitem[{{Hine} \& {Longair}(1979)}]{hine}
{Hine}, R.~G. \& {Longair}, M.~S. 1979, \mnras, 188, 111

\bibitem[{{Janssen} {et~al.}(2012){Janssen}, {R{\"o}ttgering}, {Best}, \& {Brinchmann}}]{Janssen2012}
{Janssen}, R.~M.~J., {R{\"o}ttgering}, H.~J.~A., {Best}, P.~N., \& {Brinchmann}, J. 2012, \aap, 541, A62

\bibitem[{{Koulouridis} {et~al.}(2024){Koulouridis}, {Gkini}, \& {Drigga}}]{koulouridis}
{Koulouridis}, E., {Gkini}, A., \& {Drigga}, E. 2024, \aap, 684, A111

\bibitem[{{La Marca} {et~al.}(2024){La Marca}, {Margalef-Bentabol}, {Wang}, {Gao}, {Goulding}, {Martin}, {Rodriguez-Gomez}, {Trager}, {Yang}, {Dav{\'e}}, \& {Dubois}}]{lamarca24}
{La Marca}, A., {Margalef-Bentabol}, B., {Wang}, L., {et~al.} 2024, \aap, 690, A326

\bibitem[{Lilly \& Longair(1984)}]{lilly1}
Lilly, S.~J. \& Longair, M.~S. 1984, MNRAS, 211, 833

\bibitem[{Magliocchetti(2022)}]{maglio22}
Magliocchetti, M. 2022, A\&ARv, 30

\bibitem[{Magliocchetti {et~al.}(2014)Magliocchetti, {Lutz}, {Rosario}, {Berta}, \& et~al.}]{maglio13}
Magliocchetti, M., {Lutz}, D., {Rosario}, D., {Berta}, S., \& et~al. 2014, MNRAS, 442, 682

\bibitem[{Magliocchetti {et~al.}(2016)Magliocchetti, {Lutz}, {Santini}, {Salvato}, {Popesso}, {Berta}, \& {Pozzi}}]{maglio14}
Magliocchetti, M., {Lutz}, D., {Santini}, P., {et~al.} 2016, MNRAS, 456, 431

\bibitem[{{Magliocchetti} {et~al.}(2002){Magliocchetti}, {Maddox}, {Jackson}, {Bland-Hawthorn}, {Bridges}, {Cannon}, {Cole}, {Colless}, {Collins}, {Couch}, {Dalton}, {de Propris}, {Driver}, {Efstathiou}, {Ellis}, {Frenk}, {Glazebrook}, {Lahav}, {Lewis}, {Lumsden}, {Peacock}, {Peterson}, {Sutherland}, \& {Taylor}}]{maglio2002}
{Magliocchetti}, M., {Maddox}, S.~J., {Jackson}, C.~A., {et~al.} 2002, \mnras, 333, 100

\bibitem[{Magliocchetti {et~al.}(2020)Magliocchetti, {Pentericci}, {Cirasuolo}, {Zamorani}, \& et~al.}]{maglio18}
Magliocchetti, M., {Pentericci}, L., {Cirasuolo}, M., {Zamorani}, G., \& et~al. 2020, MNRAS, 493, 3838

\bibitem[{Magliocchetti {et~al.}(2018)Magliocchetti, {Popesso}, {Brusa}, \& {Salvato}}]{maglio15}
Magliocchetti, M., {Popesso}, P., {Brusa}, M., \& {Salvato}, M. 2018, MNRAS, 473, 2493

\bibitem[{Magliocchetti {et~al.}(2017)Magliocchetti, Popesso, Brusa, Salvato, Laigle, {McCracken}, \& Ilbert}]{maglio16}
Magliocchetti, M., Popesso, P., Brusa, M., {et~al.} 2017, MNRAS, 464, 3271

\bibitem[{Mauch \& Sadler(2007)}]{mauch}
Mauch, T. \& Sadler, E.~M. 2007, MNRAS, 375, 931

\bibitem[{McAlpine {et~al.}(2013)McAlpine, Jarvis, \& Bonfield}]{mcalpine}
McAlpine, K., Jarvis, M.~J., \& Bonfield, D.~G. 2013, MNRAS, 436, 1084

\bibitem[{{Mingo} {et~al.}(2022){Mingo}, {Croston}, {Best}, {Duncan}, {Hardcastle}, {Kondapally}, {Prandoni}, {Sabater}, {Shimwell}, {Williams}, {Baldi}, {Bonato}, {Bondi}, {Dabhade}, {G{\"u}rkan}, {Ineson}, {Magliocchetti}, {Miley}, {Pierce}, \& {R{\"o}ttgering}}]{mingo}
{Mingo}, B., {Croston}, J.~H., {Best}, P.~N., {et~al.} 2022, \mnras, 511, 3250

\bibitem[{{Mingo} {et~al.}(2019){Mingo}, {Croston}, {Hardcastle}, {Best}, {Duncan}, {Morganti}, {Rottgering}, {Sabater}, {Shimwell}, {Williams}, {Brienza}, {Gurkan}, {Mahatma}, {Morabito}, {Prandoni}, {Bondi}, {Ineson}, \& {Mooney}}]{mingo19}
{Mingo}, B., {Croston}, J.~H., {Hardcastle}, M.~J., {et~al.} 2019, \mnras, 488, 2701

\bibitem[{{Morabito} {et~al.}(2025){Morabito}, {Jackson}, {de Jong}, {Escott}, {Groeneveld}, {Mahatma}, {Petley}, {Sweijen}, {Timmerman}, \& {van Weeren}}]{morabito}
{Morabito}, L.~K., {Jackson}, N., {de Jong}, J., {et~al.} 2025, \apss, 370, 19

\bibitem[{{Noll} {et~al.}(2009){Noll}, {Burgarella}, {Giovannoli}, {Buat}, {Marcillac}, \& {Mu{\~n}oz-Mateos}}]{noll}
{Noll}, S., {Burgarella}, D., {Giovannoli}, E., {et~al.} 2009, \aap, 507, 1793

\bibitem[{Padovani(2016)}]{padovani}
Padovani, P. 2016, A\&ARev, 24, 13

\bibitem[{{Pierce} {et~al.}(2023){Pierce}, {Tadhunter}, {Ramos Almeida}, {Bessiere}, {Heaton}, {Ellison}, {Speranza}, {Gordon}, {O'Dea}, {Grimmett}, \& {Makrygianni}}]{Pierce23}
{Pierce}, J.~C.~S., {Tadhunter}, C., {Ramos Almeida}, C., {et~al.} 2023, \mnras, 522, 1736

\bibitem[{{Pierce} {et~al.}(2022){Pierce}, {Tadhunter}, {Gordon}, {Ramos Almeida}, {Ellison}, {O'Dea}, {Grimmett}, {Makrygianni}, {Bessiere}, \& {Do{\~n}a Gir{\'o}n}}]{pierce22}
{Pierce}, J.~C.~S., {Tadhunter}, C.~N., {Gordon}, Y., {et~al.} 2022, \mnras, 510, 1163

\bibitem[{{Pierce} {et~al.}(2019){Pierce}, {Tadhunter}, {Ramos Almeida}, {Bessiere}, \& {Rose}}]{pierce19}
{Pierce}, J.~C.~S., {Tadhunter}, C.~N., {Ramos Almeida}, C., {Bessiere}, P.~S., \& {Rose}, M. 2019, \mnras, 487, 5490

\bibitem[{{Ramos Almeida} {et~al.}(2011){Ramos Almeida}, {Tadhunter}, {Inskip}, {Morganti}, {Holt}, \& {Dicken}}]{ramos11}
{Ramos Almeida}, C., {Tadhunter}, C.~N., {Inskip}, K.~J., {et~al.} 2011, \mnras, 410, 1550

\bibitem[{{Ricci} {et~al.}(2021){Ricci}, {Privon}, {Pfeifle}, {Armus}, {Iwasawa}, {Torres-Alb{\`a}}, {Satyapal}, {Bauer}, {Treister}, {Ho}, {Aalto}, {Ar{\'e}valo}, {Barcos-Mu{\~n}oz}, {Charmandaris}, {Diaz-Santos}, {Evans}, {Gao}, {Inami}, {Koss}, {Lansbury}, {Linden}, {Medling}, {Sanders}, {Song}, {Stern}, {U}, {Ueda}, \& {Yamada}}]{ricci}
{Ricci}, C., {Privon}, G.~C., {Pfeifle}, R.~W., {et~al.} 2021, \mnras, 506, 5935

\bibitem[{{Sabater} {et~al.}(2019){Sabater}, {Best}, {Hardcastle}, {Shimwell}, {Tasse}, {Williams}, {Br{\"u}ggen}, {Cochrane}, {Croston}, {de Gasperin}, {Duncan}, {G{\"u}rkan}, {Mechev}, {Morabito}, {Prandoni}, {R{\"o}ttgering}, {Smith}, {Harwood}, {Mingo}, {Mooney}, \& {Saxena}}]{sabater19}
{Sabater}, J., {Best}, P.~N., {Hardcastle}, M.~J., {et~al.} 2019, \aap, 622, A17

\bibitem[{{Schlafly} {et~al.}(2019){Schlafly}, {Meisner}, \& {Green}}]{schlafly}
{Schlafly}, E.~F., {Meisner}, A.~M., \& {Green}, G.~M. 2019, \apjs, 240, 30

\bibitem[{{Siudek} {et~al.}(2024){Siudek}, {Pucha}, {Mezcua}, {Juneau}, {Aguilar}, {Ahlen}, {Brooks}, {Circosta}, {Claybaugh}, {Cole}, {Dawson}, {de la Macorra}, {Dey}, {Dey}, {Doel}, {Font-Ribera}, {Forero-Romero}, {Gazta{\~n}aga}, {Gontcho A Gontcho}, {Gutierrez}, {Honscheid}, {Howlett}, {Ishak}, {Kehoe}, {Kirkby}, {Kisner}, {Kremin}, {Lambert}, {Landriau}, {Le Guillou}, {Manera}, {Martini}, {Meisner}, {Miquel}, {Moustakas}, {Newman}, {Niz}, {Pan}, {Percival}, {Poppett}, {Prada}, {Rossi}, {Saintonge}, {Sanchez}, {Schlegel}, {Scholte}, {Schubnell}, {Seo}, {Speranza}, {Sprayberry}, {Tarl{\'e}}, {Weaver}, \& {Zou}}]{siudek}
{Siudek}, M., {Pucha}, R., {Mezcua}, M., {et~al.} 2024, \aap, 691, A308

\bibitem[{{Smol{\v{c}}i{\'c}} {et~al.}(2017){Smol{\v{c}}i{\'c}}, {Delvecchio}, {Zamorani}, {Baran}, {Novak}, {Delhaize}, {Schinnerer}, {Berta}, {Bondi}, {Ciliegi}, {Capak}, {Civano}, {Karim}, {Le Fevre}, {Ilbert}, {Laigle}, {Marchesi}, {McCracken}, {Tasca}, {Salvato}, \& {Vardoulaki}}]{smolcic2017}
{Smol{\v{c}}i{\'c}}, V., {Delvecchio}, I., {Zamorani}, G., {et~al.} 2017, \aap, 602, A2

\bibitem[{{Spinrad} {et~al.}(1985){Spinrad}, {Djorgovski}, {Marr}, \& {Aguilar}}]{spinrad}
{Spinrad}, H., {Djorgovski}, S., {Marr}, J., \& {Aguilar}, L. 1985, \pasp, 97, 932

\bibitem[{{Sutherland} \& {Saunders}(1992)}]{sutherland}
{Sutherland}, W. \& {Saunders}, W. 1992, \mnras, 259, 413

\bibitem[{{Tacconi} {et~al.}(2010){Tacconi}, {Genzel}, {Neri}, {Cox}, {Cooper}, {Shapiro}, {Bolatto}, {Bouch{\'e}}, {Bournaud}, {Burkert}, {Combes}, {Comerford}, {Davis}, {F{\"o}rster Schreiber}, {Garcia-Burillo}, {Gracia-Carpio}, {Lutz}, {Naab}, {Omont}, {Shapley}, {Sternberg}, \& {Weiner}}]{tacconi}
{Tacconi}, L.~J., {Genzel}, R., {Neri}, R., {et~al.} 2010, \nat, 463, 781

\bibitem[{{Tadhunter} {et~al.}(1993){Tadhunter}, {Morganti}, {di Serego Alighieri}, {Fosbury}, \& {Danziger}}]{tadhunter}
{Tadhunter}, C.~N., {Morganti}, R., {di Serego Alighieri}, S., {Fosbury}, R.~A.~E., \& {Danziger}, I.~J. 1993, \mnras, 263, 999

\bibitem[{{Tanaka} {et~al.}(2023){Tanaka}, {Koike}, {Naito}, {Shibata}, {Usuda-Sato}, {Yamaoka}, {Ando}, {Ito}, {Kobayashi}, {Kofuji}, {Kuwata}, {Nakano}, {Shimakawa}, {Tadaki}, {Takebayashi}, {Tsuchiya}, {Umemoto}, \& {Bottrell}}]{tanaka}
{Tanaka}, M., {Koike}, M., {Naito}, S., {et~al.} 2023, \pasj, 75, 986

\bibitem[{{Toba} {et~al.}(2019){Toba}, {Yamashita}, {Nagao}, {Wang}, {Ueda}, {Ichikawa}, {Kawaguchi}, {Akiyama}, {Hsieh}, {Kajisawa}, {Lee}, {Matsuoka}, {Noboriguchi}, {Onoue}, {Schramm}, {Tanaka}, \& {Komiyama}}]{toba}
{Toba}, Y., {Yamashita}, T., {Nagao}, T., {et~al.} 2019, \apjs, 243, 15

\bibitem[{{van Haarlem} {et~al.}(2013){van Haarlem}, {Wise}, {Gunst}, {Heald}, {McKean}, {Hessels}, {de Bruyn}, {Nijboer}, {Swinbank}, {Fallows}, {Brentjens}, {Nelles}, {Beck}, {Falcke}, {Fender}, {H{\"o}randel}, {Koopmans}, {Mann}, {Miley}, {R{\"o}ttgering}, {Stappers}, {Wijers}, {Zaroubi}, {van den Akker}, {Alexov}, {Anderson}, {Anderson}, {van Ardenne}, {Arts}, {Asgekar}, {Avruch}, {Batejat}, {B{\"a}hren}, {Bell}, {Bell}, {van Bemmel}, {Bennema}, {Bentum}, {Bernardi}, {Best}, {B{\^\i}rzan}, {Bonafede}, {Boonstra}, {Braun}, {Bregman}, {Breitling}, {van de Brink}, {Broderick}, {Broekema}, {Brouw}, {Br{\"u}ggen}, {Butcher}, {van Cappellen}, {Ciardi}, {Coenen}, {Conway}, {Coolen}, {Corstanje}, {Damstra}, {Davies}, {Deller}, {Dettmar}, {van Diepen}, {Dijkstra}, {Donker}, {Doorduin}, {Dromer}, {Drost}, {van Duin}, {Eisl{\"o}ffel}, {van Enst}, {Ferrari}, {Frieswijk}, {Gankema}, {Garrett}, {de Gasperin}, {Gerbers}, {de Geus}, {Grie{\ss}meier}, {Grit}, {Gruppen}, {Hamaker}, {Hassall}, {Hoeft}, {Holties},
  {Horneffer}, {van der Horst}, {van Houwelingen}, {Huijgen}, {Iacobelli}, {Intema}, {Jackson}, {Jelic}, {de Jong}, {Juette}, {Kant}, {Karastergiou}, {Koers}, {Kollen}, {Kondratiev}, {Kooistra}, {Koopman}, {Koster}, {Kuniyoshi}, {Kramer}, {Kuper}, {Lambropoulos}, {Law}, {van Leeuwen}, {Lemaitre}, {Loose}, {Maat}, {Macario}, {Markoff}, {Masters}, {McFadden}, {McKay-Bukowski}, {Meijering}, {Meulman}, {Mevius}, {Middelberg}, {Millenaar}, {Miller-Jones}, {Mohan}, {Mol}, {Morawietz}, {Morganti}, {Mulcahy}, {Mulder}, {Munk}, {Nieuwenhuis}, {van Nieuwpoort}, {Noordam}, {Norden}, {Noutsos}, {Offringa}, {Olofsson}, {Omar}, {Orr{\'u}}, {Overeem}, {Paas}, {Pandey-Pommier}, {Pandey}, {Pizzo}, {Polatidis}, {Rafferty}, {Rawlings}, {Reich}, {de Reijer}, {Reitsma}, {Renting}, {Riemers}, {Rol}, {Romein}, {Roosjen}, {Ruiter}, {Scaife}, {van der Schaaf}, {Scheers}, {Schellart}, {Schoenmakers}, {Schoonderbeek}, {Serylak}, {Shulevski}, {Sluman}, {Smirnov}, {Sobey}, {Spreeuw}, {Steinmetz}, {Sterks}, {Stiepel}, {Stuurwold},
  {Tagger}, {Tang}, {Tasse}, {Thomas}, {Thoudam}, {Toribio}, {van der Tol}, {Usov}, {van Veelen}, {van der Veen}, {ter Veen}, {Verbiest}, {Vermeulen}, {Vermaas}, {Vocks}, {Vogt}, {de Vos}, {van der Wal}, {van Weeren}, {Weggemans}, {Weltevrede}, {White}, {Wijnholds}, {Wilhelmsson}, {Wucknitz}, {Yatawatta}, {Zarka}, \& {Zensus}}]{van}
{van Haarlem}, M.~P., {Wise}, M.~W., {Gunst}, A.~W., {et~al.} 2013, \aap, 556, A2

\bibitem[{{Villforth}(2023)}]{villforth23}
{Villforth}, C. 2023, The Open Journal of Astrophysics, 6, 34

\bibitem[{{Villforth} {et~al.}(2019){Villforth}, {Herbst}, {Hamann}, {Hamilton}, {Bertemes}, {Efthymiadou}, \& {Hewlett}}]{villforth19}
{Villforth}, C., {Herbst}, H., {Hamann}, F., {et~al.} 2019, \mnras, 483, 2441

\bibitem[{{Wang} {et~al.}(2025){Wang}, {De Breuck}, {Wylezalek}, {Vernet}, {Lehnert}, {Stern}, {Rupke}, {Nesvadba}, {Vayner}, {Zakamska}, {Lin}, {Kukreti}, {Dall'Agnol de Oliveira}, \& {Groth}}]{wang}
{Wang}, W., {De Breuck}, C., {Wylezalek}, D., {et~al.} 2025, \aap, 696, A88

\bibitem[{{Whittam} {et~al.}(2022){Whittam}, {Jarvis}, {Hale}, {Prescott}, {Morabito}, {Heywood}, {Adams}, {Afonso}, {An}, {Ao}, {Bowler}, {Collier}, {Deane}, {Delhaize}, {Frank}, {Glowacki}, {Hatfield}, {Maddox}, {Marchetti}, {Matthews}, {Prandoni}, {Randriamampandry}, {Randriamanakoto}, {Smith}, {Taylor}, {Thomas}, \& {Vaccari}}]{whittam2022}
{Whittam}, I.~H., {Jarvis}, M.~J., {Hale}, C.~L., {et~al.} 2022, \mnras, 516, 245

\end{thebibliography}

\onecolumn
\begin{appendix}
\section{Radio Luminosity Distributions \label{Lums}}
\begin{figure*}
\begin{center}
\vspace{0 cm}  
\includegraphics[scale=0.30, viewport=120 180 600 760] {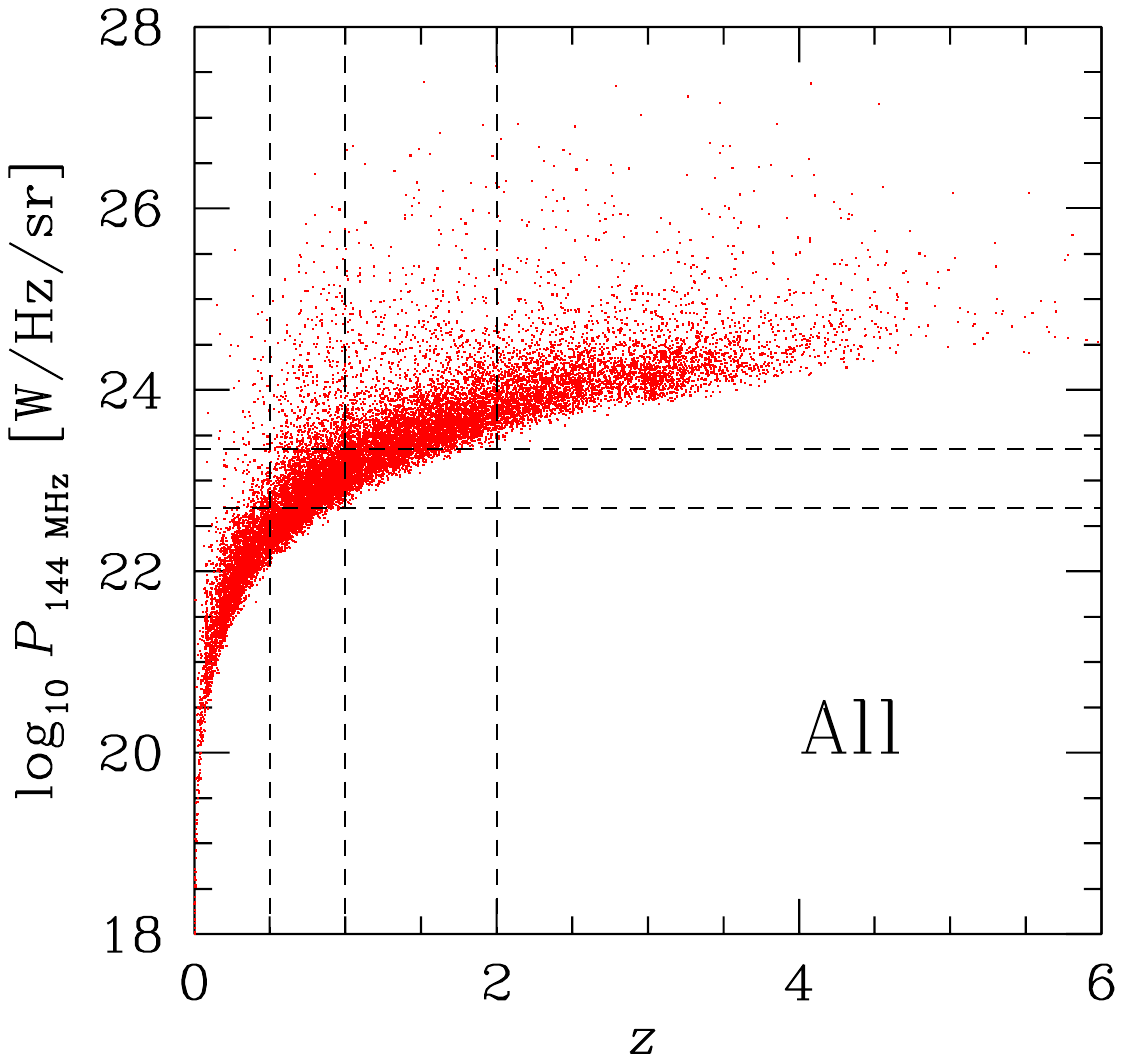}
\includegraphics[scale=0.30, viewport=10 180 600 760] {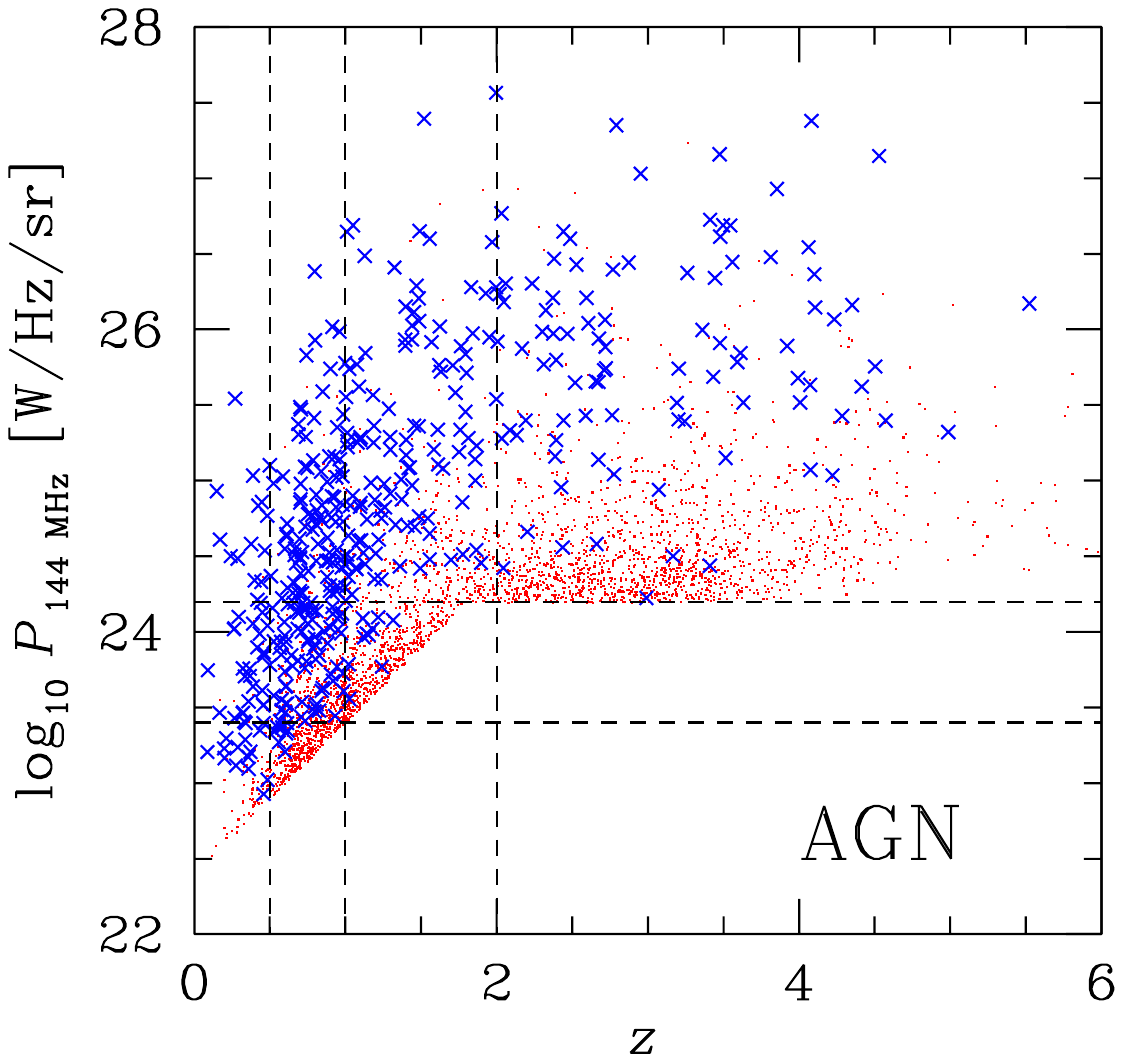}
\includegraphics[scale=0.30, viewport=-10 180 600 760] {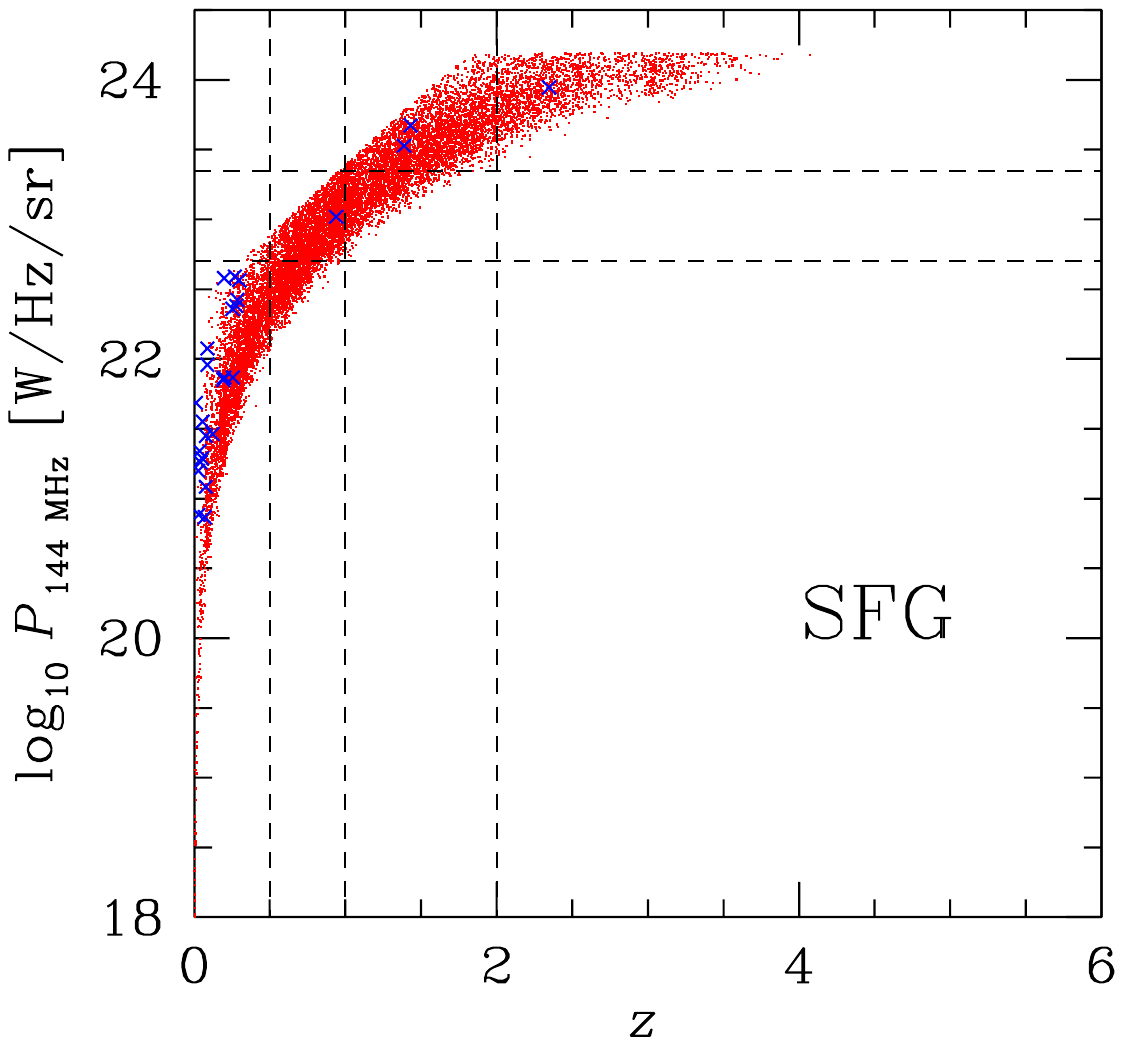}
\caption{144\,MHz luminosity distribution of LOFAR sources in the EDF-N region as a function of redshift. The left-hand panel refers to the whole \citetalias{bisigello} sample, the middle panel to the sub-class of radio AGN, and the right-hand panel to star-forming galaxies. The distinction between these two populations has been obtained via Eq.~(\ref{eq:P}). Crosses indicate objects with a complex or extended radio morphology. The horizontal dashed lines indicate the completeness limits of the whole (left-hand panel), AGN (middle panel), and SFG (right-hand panel) samples in the two redshift intervals $0.5<z<1$ and $1<z<2$ discussed in Sect.~\ref{results}.
\label{fig:L_z}}
\end{center}
\end{figure*}

We show here in Fig.~\ref{fig:L_z} the 144\,MHz luminosities for radio-selected sources within the EDF-N.

\section{Properties of the optical catalogue \label{morph}}
\begin{figure*}[htbp!]
\begin{center}
\vspace{-3 cm}  
\includegraphics[scale=0.45, viewport=50 160 600 900] {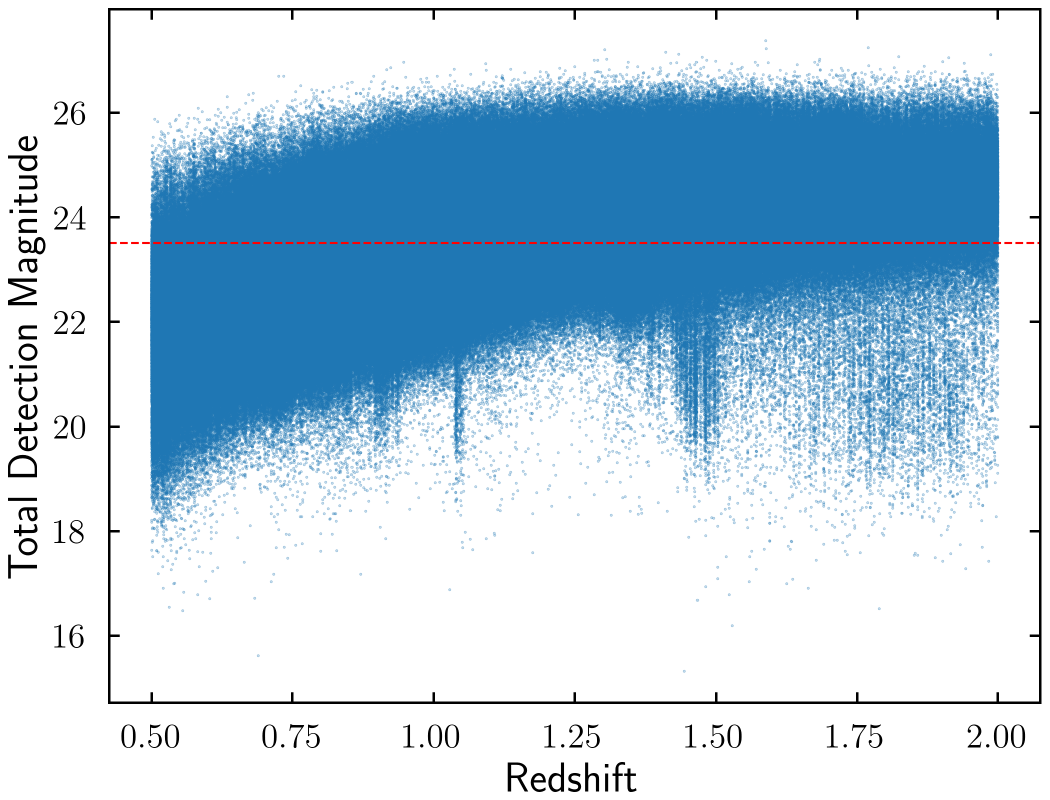}
\includegraphics[scale=0.45, viewport=50 160 600 600] {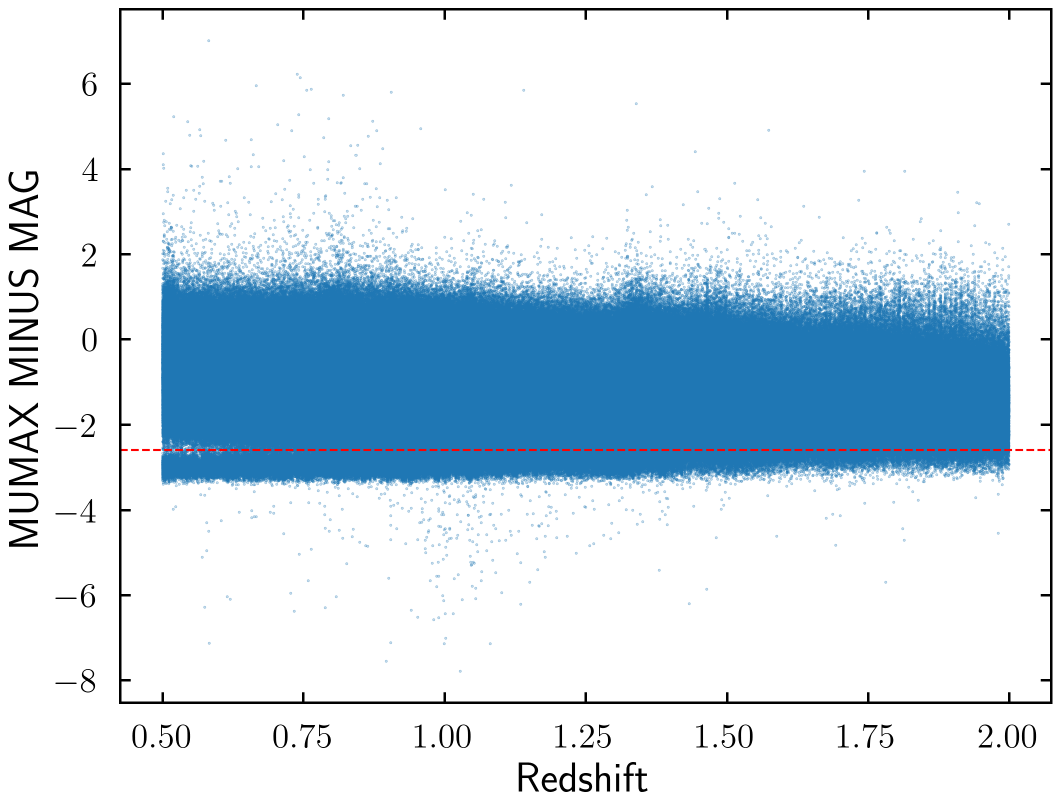}
\caption{Left panel: distribution of total detection magnitudes, $\IE$, for all \Euclid galaxies in the range $0.5<z<2$. The dashed line indicates the cut at $\IE=23.5$ applied by \citetalias{Q1-SP013} for the production of their catalogue. 
Right panel: distribution of values for the quantity $\texttt{MUMAX\_MINUS\_MAG}$ which characterises the  visual extension of \Euclid galaxies. The dashed line represents the value of $-2.6$ adopted by \citetalias{Q1-SP013} to exclude point-like sources.
\label{fig:morph}}
\end{center}
\end{figure*}

In Fig.~\ref{fig:morph} we show trends for all \Euclid galaxies observed within the Q1 fields as a function of redshift.

\newpage

\section{Radio and optical images for AGN with extended radio morphology \label{images}}
\begin{figure*}
\begin{center}
\vspace{0 cm}  
\includegraphics[scale=0.5] 
{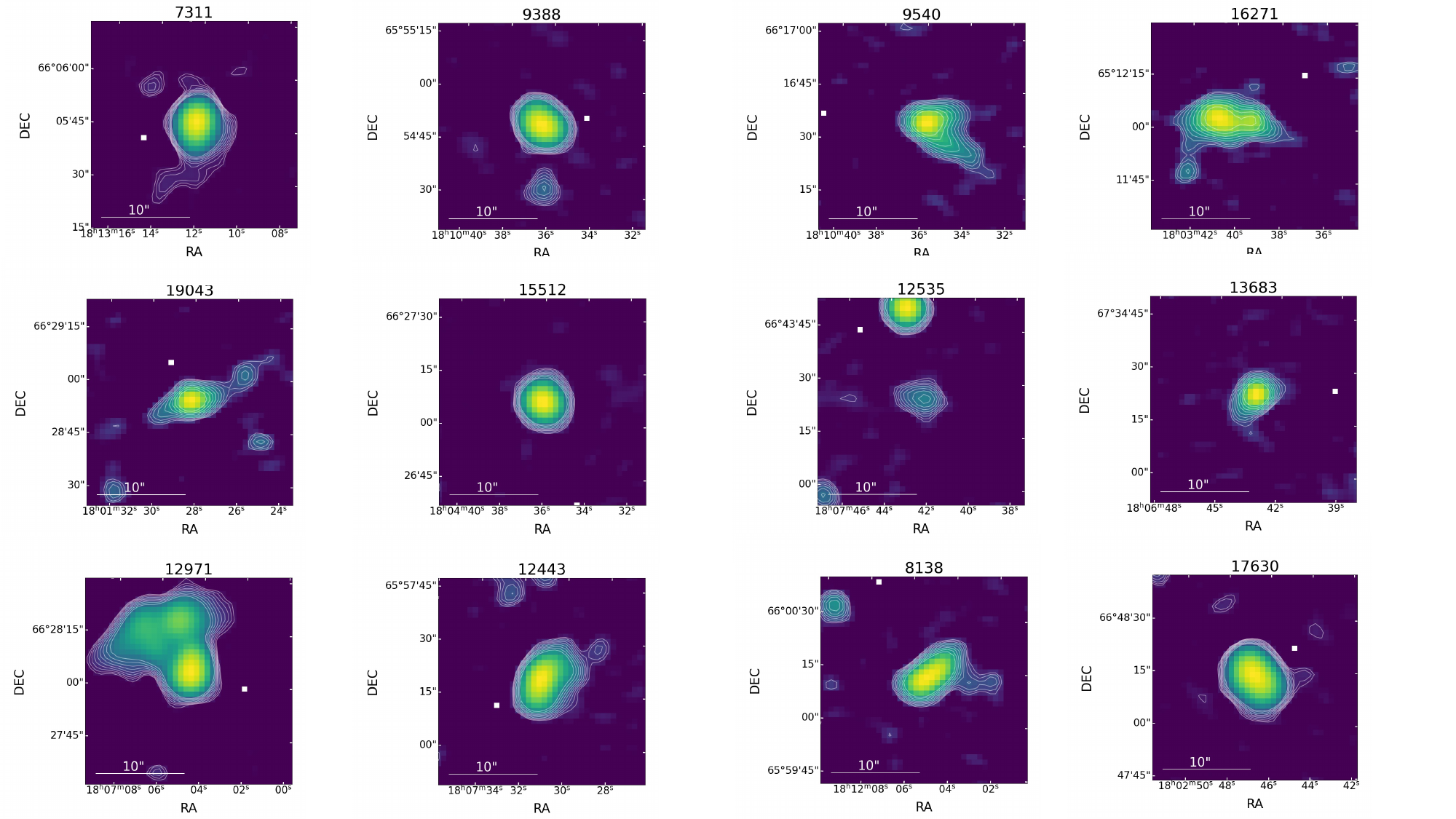}
\includegraphics[scale=0.5] 
{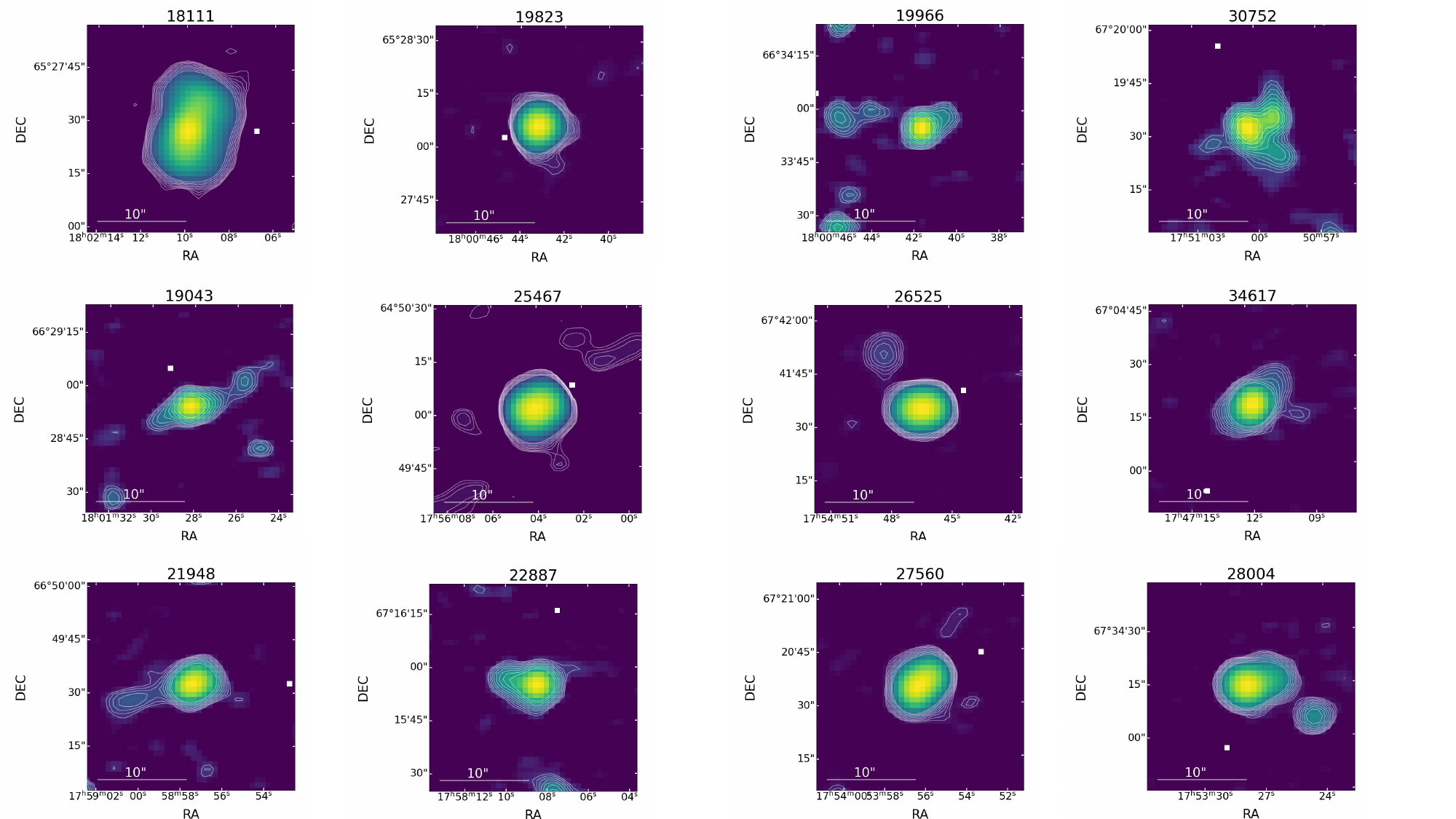}
\caption{LOFAR images for a random selection of extended radio AGN with FR\,I morphology.
\label{fig:FRI_radio}}
\end{center}
\end{figure*}

\begin{figure*}[htbp!]
\begin{center}
\vspace{0 cm}  
\includegraphics[scale=0.5]
{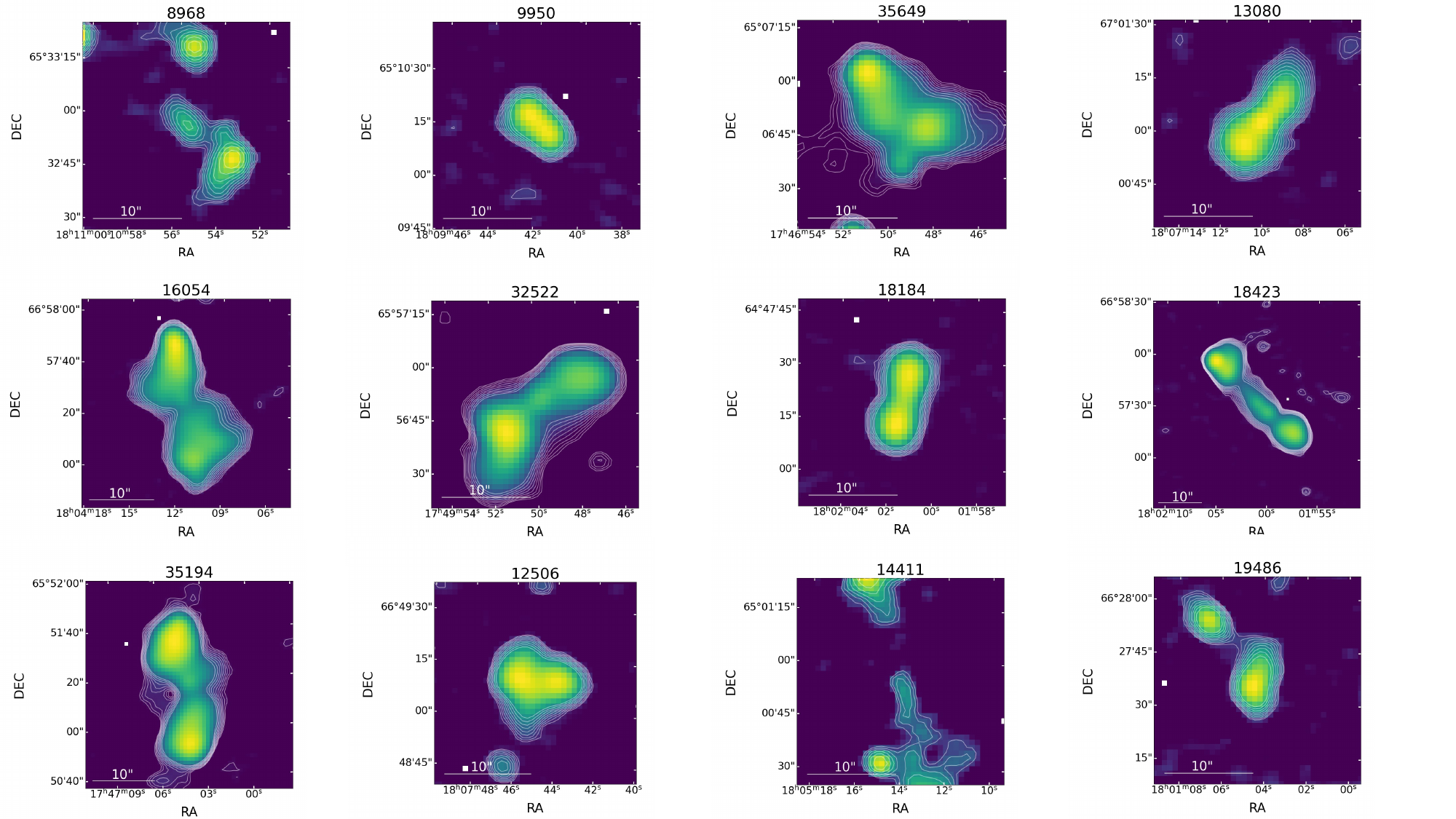}
\includegraphics[scale=0.5]
{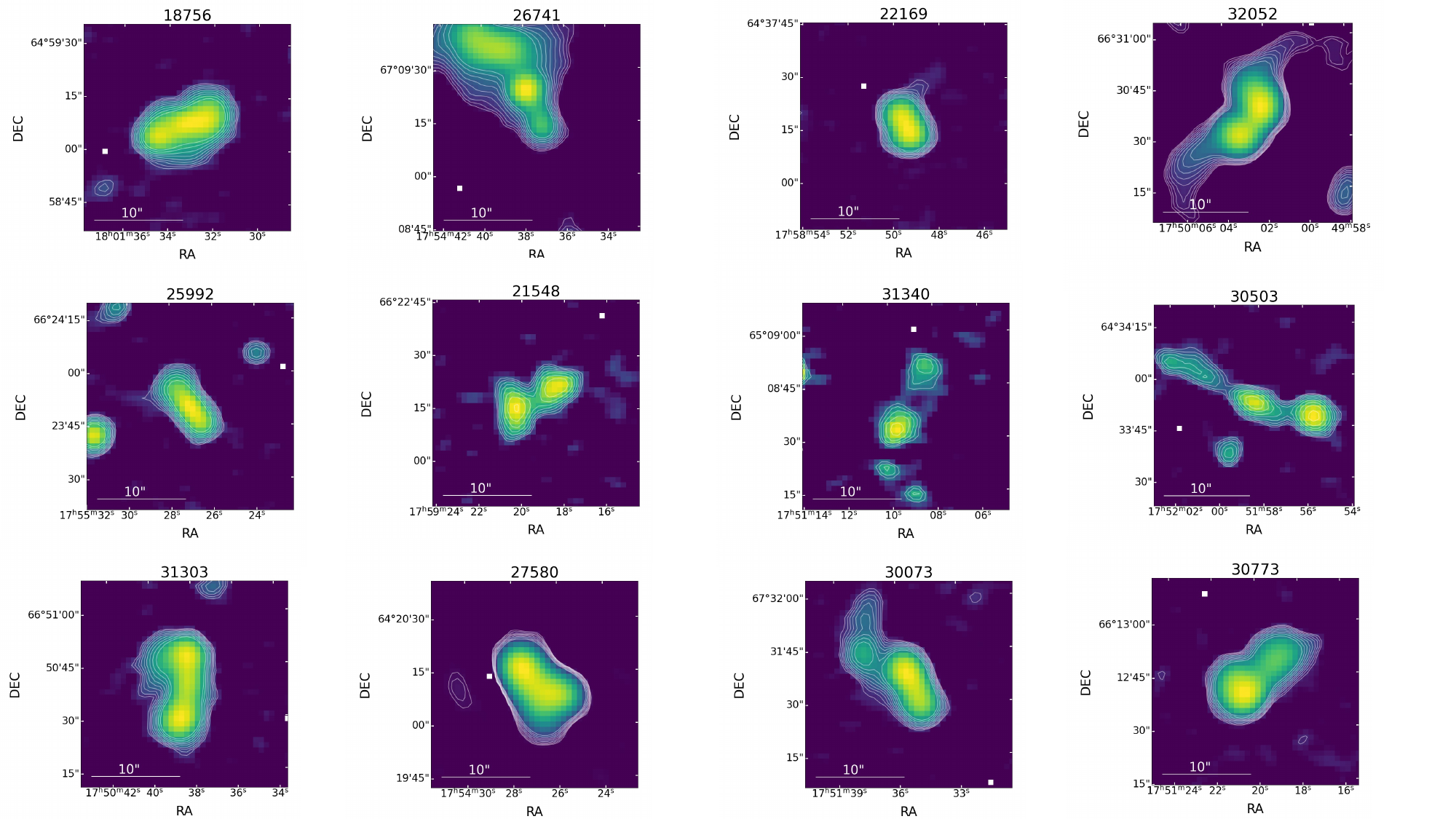}
\caption{LOFAR images for a random selection of extended radio AGN with FR\,II morphology.
\label{fig:FRII_radio}}
\end{center}
\end{figure*}

\begin{figure*}[htbp!]
\begin{center}
\vspace{0 cm}  
\includegraphics[scale=0.75]
{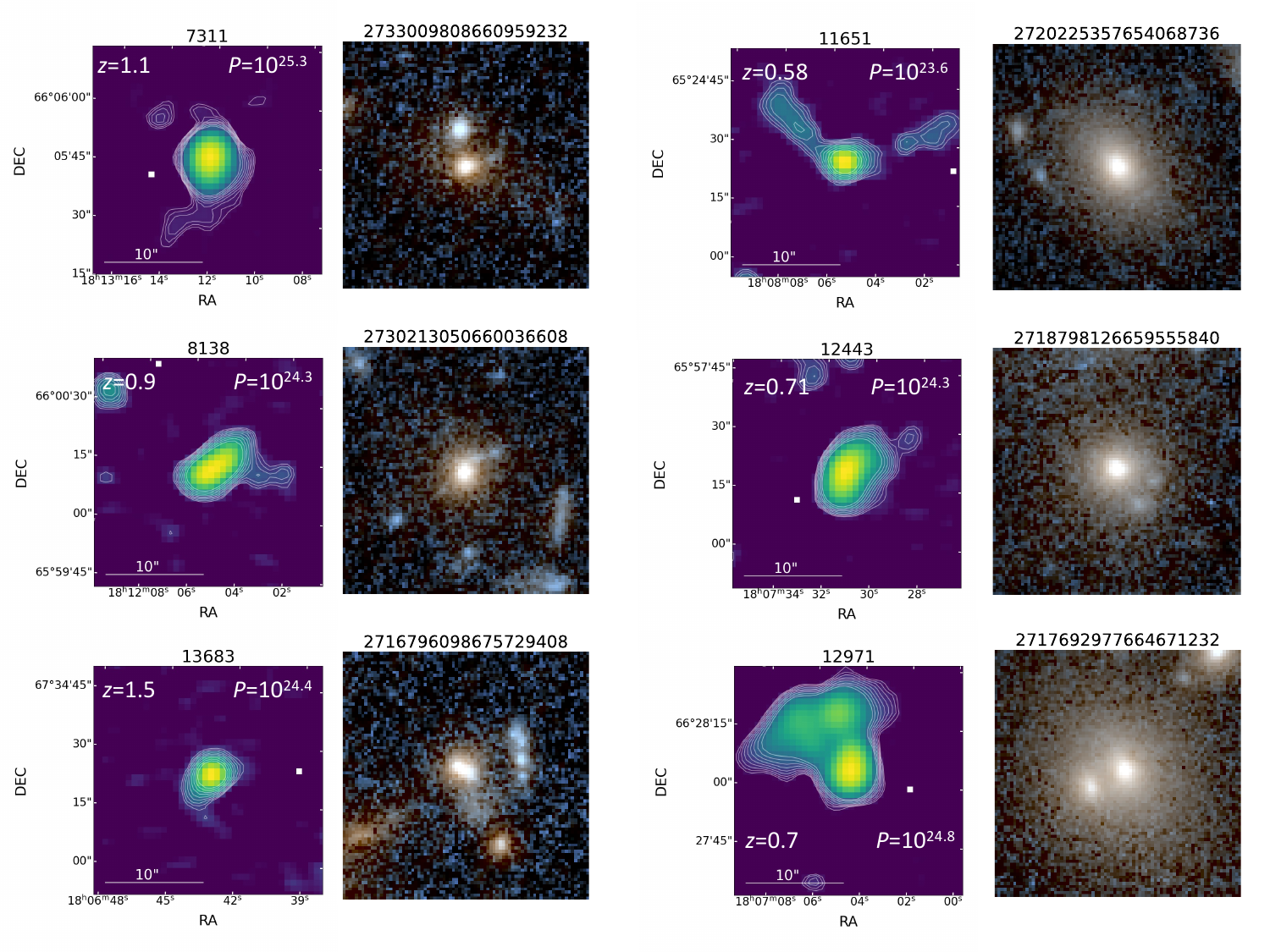}
\caption{Some examples of LOFAR (left) and \Euclid (right) images of AGN with extended/FR\,I-like radio emission hosted by galaxies undergoing close encounters. Approximate redshifts and 144\,MHz radio luminosities (in units W\,Hz$^{-1}$\,sr$^{-1}$) are also reported for all sources. The optical cutouts are $8\arcsec\times 8\arcsec$, and the radio ones are as indicated. We choose different scales to be able to appreciate both the details of galaxy close encounters and the generally more extended radio emission.
\label{fig:FRI_mer_1}}
\end{center}
\end{figure*}

\begin{figure*}[htbp!]
\begin{center}
\vspace{0 cm}  
\includegraphics[scale=0.75]
{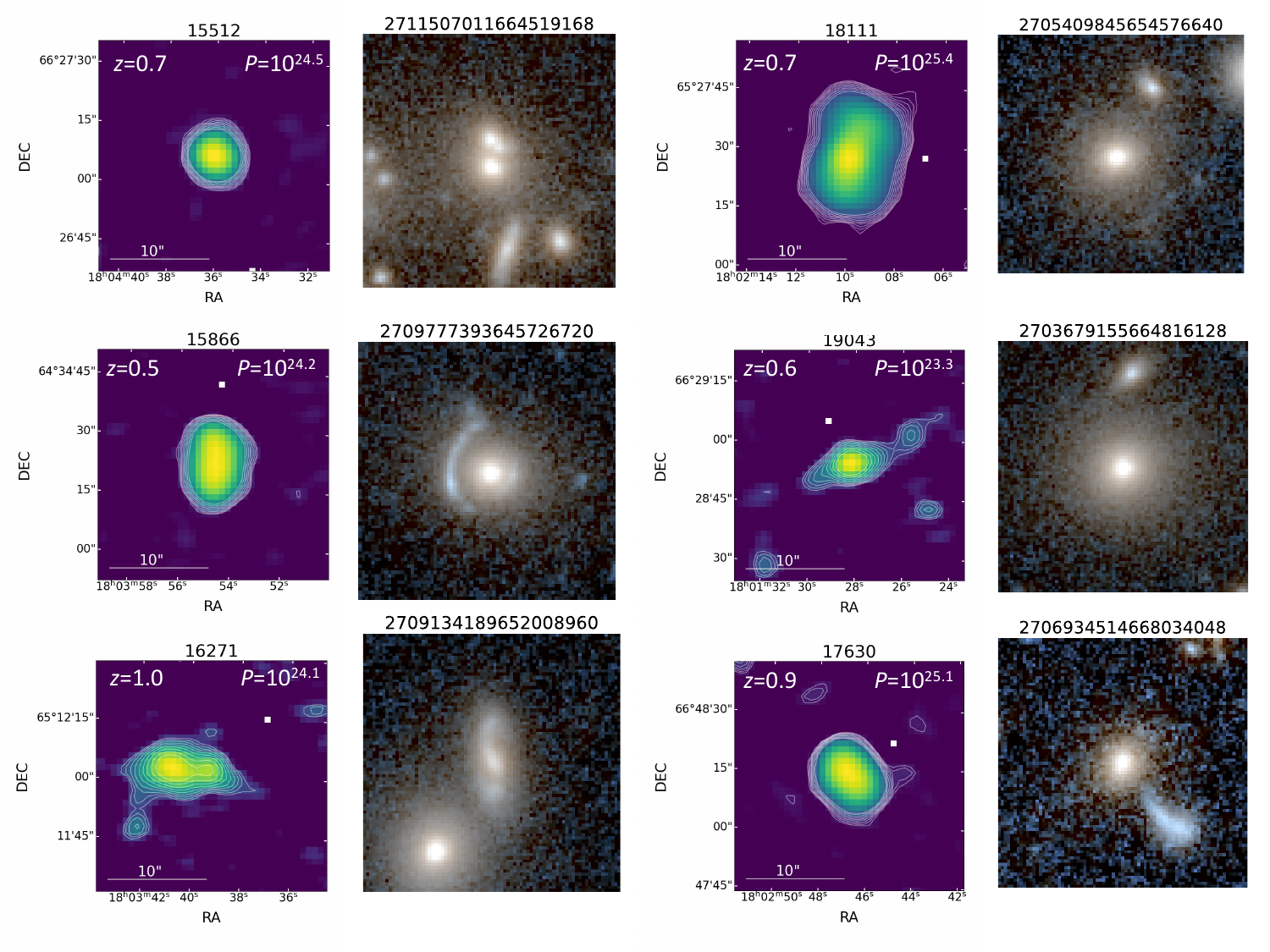}
\caption{As in Fig.~\ref{fig:FRI_mer_1}.
\label{fig:FRI_mer_2}}
\end{center}
\end{figure*}

\begin{figure*}[htbp!]
\begin{center}
\vspace{0 cm}  
\includegraphics[scale=0.75]
{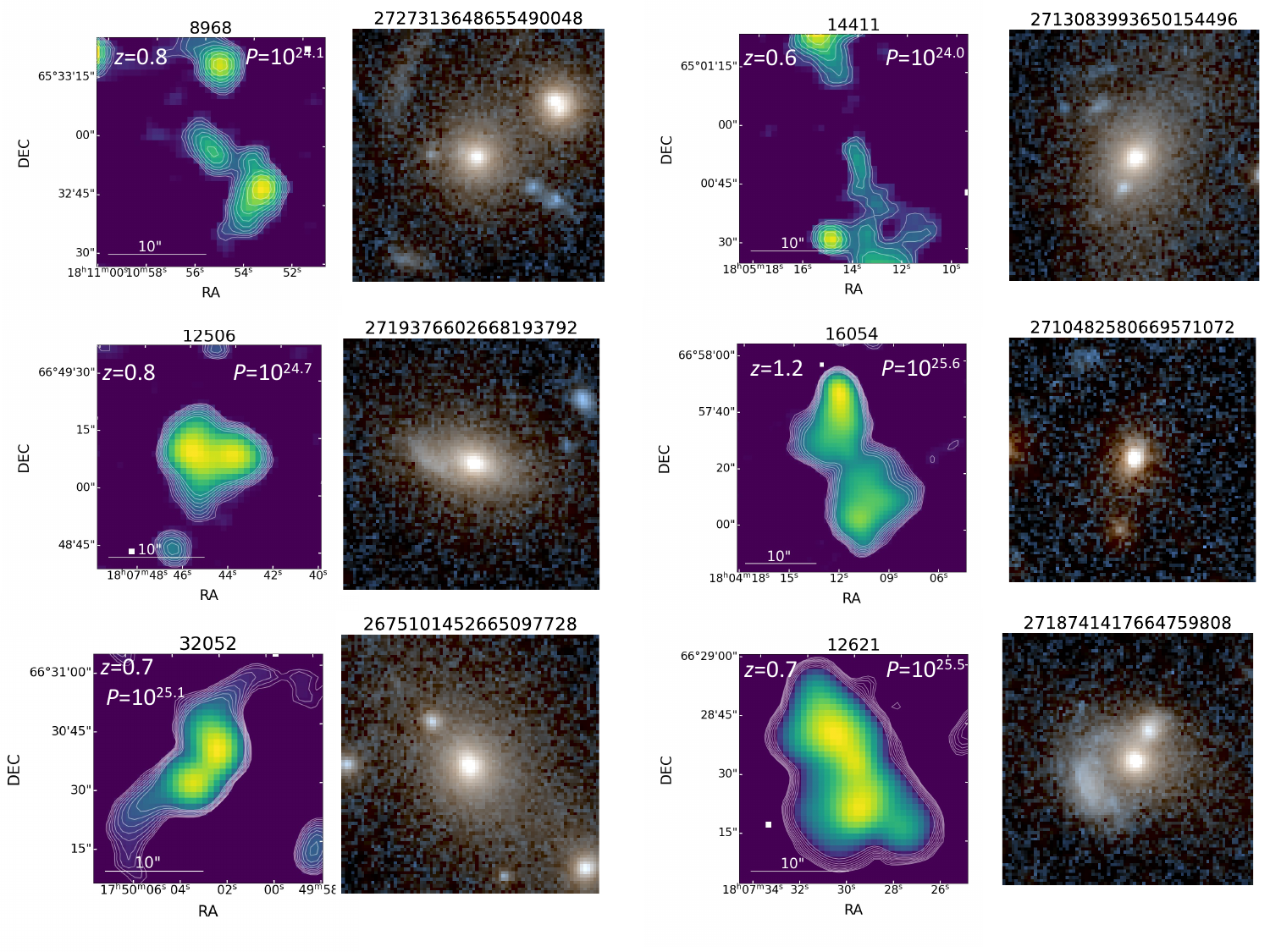}
\caption{Some examples of LOFAR (left) and \Euclid (right) images of AGN with extended/FR\,II-like radio emission hosted by galaxies undergoing close encounters. Approximate redshifts and 144\,MHz radio luminosities (in units W\,Hz$^{-1}$\,sr$^{-1}$) are also reported for all sources. The optical cutouts are $8\arcsec\times 8\arcsec$, and the radio ones are as indicated. We choose different scales to be able to appreciate both the details of galaxy close encounters and the generally more extended radio emission.
\label{fig:FRII_mer_1}}
\end{center}
\end{figure*}

\begin{figure*}[htbp!]
\begin{center}
\vspace{0 cm}  
\includegraphics[scale=0.75]
{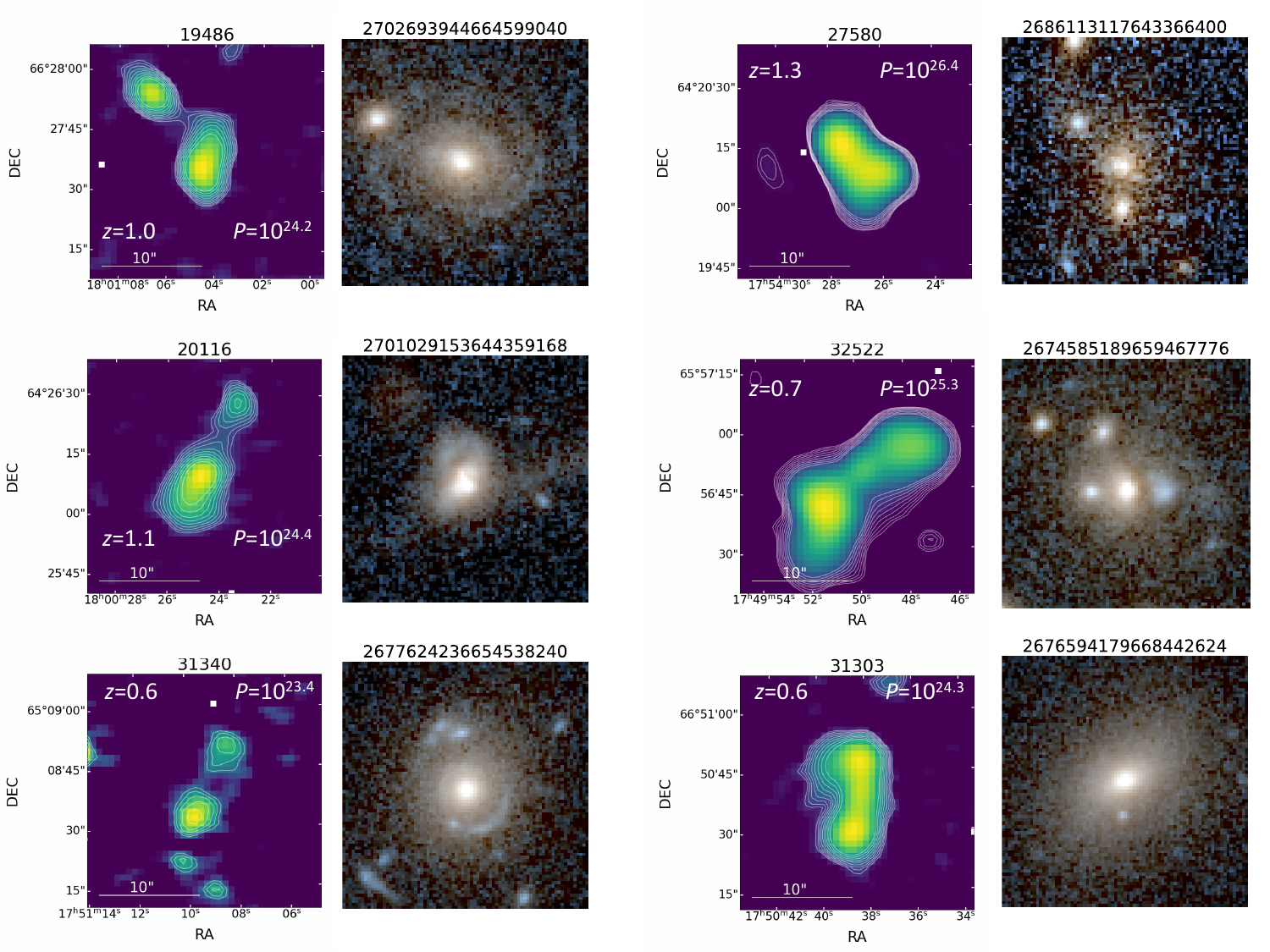}
\caption{As in Fig.~\ref{fig:FRII_mer_1}.
\label{fig:FRII_mer_2}}
\end{center}
\end{figure*}

\begin{figure*}[htbp!]
\begin{center}
\vspace{0 cm}  
\includegraphics[scale=0.75]
{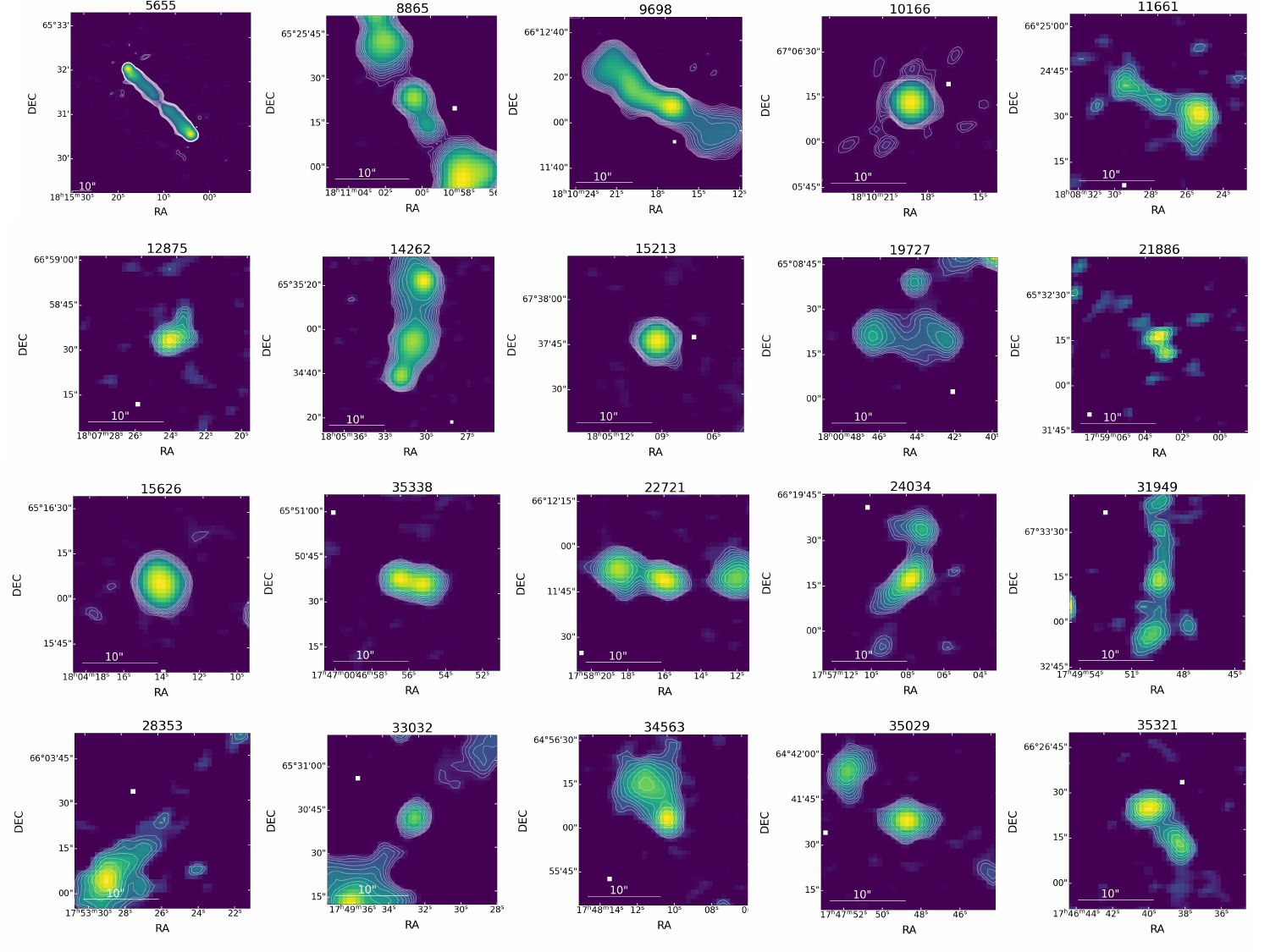}
\caption{LOFAR images for 20 out of the 21 AGN with extended radio emission hosted by isolated systems (cf. Table~\ref{table1}).
\label{fig:extended_single}
\label{LastPage}}
\end{center}
\end{figure*}

Here we show some LOFAR and \Euclid images for extended radio AGN associated with either isolated or merging galaxies (Figs.~\ref{fig:FRI_radio} to \ref{fig:extended_single}).

\end{appendix}

\end{document}